\documentclass[12pt]{article}
\pdfoutput=1

\usepackage{booktabs}
\usepackage{tabulary}
\usepackage{multirow}
\usepackage{rotating}
\usepackage{epsfig}
\usepackage{psfrag}
\usepackage{latexsym}
\usepackage{indentfirst}
\usepackage{fancyhdr}
\usepackage{dsfont}
\usepackage{amsmath}
\usepackage{amssymb}
\usepackage{amsfonts}
\usepackage{mathrsfs}
\usepackage{amsthm}
\usepackage{pifont}
\usepackage{dsfont}
\usepackage{multirow}
\usepackage{array}
\usepackage{chngpage}
\usepackage{longtable}
\usepackage{cite}
\usepackage{bbold}
\usepackage{color}
\usepackage{braket}
\usepackage{colordvi}
\usepackage{makecell}
\usepackage{fancybox}
\usepackage[footnotesize]{caption2}
\usepackage{graphicx}
\usepackage[center,footnotesize,hang]{subfigure}
\usepackage{bbm}
\usepackage{url}
\usepackage[colorlinks, linkcolor=red, anchorcolor=black, citecolor=green]{hyperref}
\usepackage{multirow}
\usepackage[integrals]{wasysym}
\usepackage{harpoon}
\usepackage{textcomp}
\usepackage{enumitem}
\usepackage{mathrsfs}
\usepackage{adjustbox}
\newcommand{\PreserveBackslash}[1]{\let\temp=\\#1\let\\=\temp}
\newcolumntype{C}[1]{>{\PreserveBackslash\centering}p{#1}}
\newcolumntype{R}[1]{>{\PreserveBackslash\raggedleft}p{#1}}
\newcolumntype{L}[1]{>{\PreserveBackslash\raggedright}p{#1}}
\allowdisplaybreaks

\newcommand{\bq}{\begin{eqnarray}}
\newcommand{\nq}{\end{eqnarray}}

\newcommand\harp[1]{\mathstrut\mkern2.5mu#1\mkern-11mu\raise1.7ex \hbox{\scalebox{.3}{\bf(}$\kern 0.4em$\scalebox{.3}{\bf)}}}

\begin{document}
\title{
\begin{flushright}
\hfill\mbox{{\small\tt USTC-ICTS/PCFT-20-18}}\\[5mm]
\begin{minipage}{0.2\linewidth}
\normalsize
\end{minipage}
\end{flushright}
{\Large \bf Modular Invariant Quark and Lepton Models in Double Covering of $S_4$ Modular Group \\[2mm]}
}
\date{}

\author{
Xiang-Gan~Liu$^{1,2}$\footnote{E-mail: {\tt
hepliuxg@mail.ustc.edu.cn}}, \
Chang-Yuan~Yao$^{3}$\footnote{E-mail: {\tt
yaocy@nankai.edu.cn}},  \
Gui-Jun~Ding$^{1,2}$\footnote{E-mail: {\tt
dinggj@ustc.edu.cn}}
\\*[20pt]
\centerline{
\begin{minipage}{\linewidth}
\begin{center}
$^1${\it\small Peng Huanwu Center for Fundamental Theory, Hefei, Anhui 230026, China} \\[2mm]
$^2${\it \small
Interdisciplinary Center for Theoretical Study and  Department of Modern Physics,\\
University of Science and Technology of China, Hefei, Anhui 230026, China}\\[2mm]
$^3${\it \small
School of Physics, Nankai University, Tianjin 300071, China}\\
\end{center}
\end{minipage}}
\\[10mm]}
\maketitle
\thispagestyle{empty}

\begin{abstract}
We perform a comprehensive analysis of the homogeneous finite modular group $\Gamma'_4\equiv S'_4$ which is the double covering of $S_4$ group. The weight 1 modular forms of level 4 are constructed in terms of Dedekind eta function, and they transform as a triplet $\mathbf{\hat{3}'}$ of $S'_4$. The integral weight modular forms until weight 6 are built from the tensor products of weight 1 modular forms. We perform a systematical classification of $S'_4$ modular models for lepton masses and mixing with/without generalized CP, where the left-handed leptons are assigned to triplet of $S'_4$ and right-handed charged leptons transform as singlets under $S'_4$, and we consider both scenarios where the neutrino masses arise from Weinberg operator or type I seesaw mechanism. The phenomenological implications of the minimal models for lepton masses, mixing angles, CP violation phases and neutrinoless double decay are discussed. The $S'_4$ modular symmetry is extended to quark sector, we present several predictive models which use nine or ten free parameters including real and imaginary parts of $\tau$ to describe  quark masses and Cabibbo-Kobayashi-Maskawa mixing matrix. We give a quark-lepton unified model which can explain the flavor structure of quarks and leptons simultaneously for a common value of $\tau$.

\end{abstract}
\newpage

\section{\label{sec:introduction}Introduction}

The origin of fermion masses and the mixing matrices is one of the greatest challenges for modern particle physics. Neutrino oscillation provides new clues for the understanding of the flavor problem. It is known that neutrino mixing angles show a pattern which is completely different than that of quark mixing: all quarks mixing angles are small, while for the lepton sector two mixing angles $\theta_{12}$, $\theta_{23}$ are large, the third one $\theta_{13}$ is small and it is comparable to the size of the quark Cabibbo mixing angle~\cite{Tanabashi:2018oca}. The evidence of CP violation in neutrino oscillation is reported recently~\cite{Abe:2019vii}. Given the successful use of symmetries in various fields of physics, it was conceived that the flavor structure of quarks and leptons is dictated by certain flavor symmetry, and different kinds of flavor symmetry groups (abelian, non-abelian, continuous, discrete, global, local, linearly or non-linearly realized) have been considered so far. In particular, it turns that the discrete non-abelian flavor symmetry is quite suitable to reproduce the large lepton mixing angles, a huge number of models have been constructed, see~\cite{Feruglio:2019ktm} for recent review. If discrete flavor symmetry is combined with generalized CP symmetry~\cite{Feruglio:2012cw,Holthausen:2012dk}, one can predict leptonic CP violation phase. It is notable that a unified description of the observed structure of the quark and lepton mixing can be achieved if the flavor and CP symmetries are broken down to $Z_2\times CP$ in  neutrino,  charged  lepton,  up  quark  and  down  quark  sectors,  and the minimal flavor group is the dihedral group $D_{14}$~\cite{Lu:2016jit,Li:2017abz,Lu:2018oxc,Lu:2019gqp}.

In any realistic model based on discrete flavor symmetry, the flavor symmetry is spontaneously broken by the vacuum expectation values (VEVs) of a set of scalar fields called flavons which are standard model singlet albeit transforming non-trivially under the flavor symmetry group. The VEVs of flavon are typically aligned along certain directions in flavor space, and the vacuum alignment determines the flavor structure of quarks and leptons. One has to intelligently design the flavon energy density to achieve the required vacuum alignment as the global minimum of the scalar potential. In most models, discrete flavor symmetry is accompanied by additional symmetries, either discrete like $Z_N$ or continuous like $U(1)$, to ensure the needed vacuum alignment and to reproduce the observed mass hierarchies. Hence the flavor symmetry breaking sector introduces many independent parameters, makes the flavor model rather complicated. Moreover, high dimensional operators compatible with symmetry the model can lead to corrections to leading order results such the predictability of the model is spoiled in some sense.

Recently modular invariance as flavor symmetry has suggested to understand the neutrino masses and lepton flavor mixing~\cite{Feruglio:2017spp}. Modular symmetry naturally appears in torus and orbifold compactifications of string theory. In this approach, flavon fields are not absolute requirement, the flavor symmetry can be uniquely broken by the VEV of the modulus $\tau$. Hence the vacuum alignment problem is simplified considerably although a moduli stabilization mechanism is needed. In modular invariant models, the Yukawa couplings transform nontrivially under the modular symmetry and they are just modular forms which are holomorphic functions of $\tau$. In the limit of exact supersymmetry, the superpotenial is completely fixed by modular symmetry. Furthermore, modular invariant models can be quite predictive, typical minimal modular models describe the neutrino masses, mixing angles and CP violating phases in terms of five free real parameters including the real and imaginary parts of $\tau$.

The finite modular group $\Gamma_N=SL(2,Z)/\Gamma(N)$ arising from the quotient of the $SL(2,Z)$ modular group by congruence subgroups $\Gamma(N)$ have been utilized for the flavor symmetry of quarks and leptons. Some models for lepton masses and flavor mixing have been constructed at level $N=2$\cite{Kobayashi:2018vbk,Kobayashi:2018wkl,Kobayashi:2019rzp,Okada:2019xqk}, level $N=3$~\cite{Feruglio:2017spp,Criado:2018thu,Kobayashi:2018vbk,Kobayashi:2018scp,deAnda:2018ecu,
Okada:2018yrn,Kobayashi:2018wkl,Novichkov:2018yse,Nomura:2019jxj,Okada:2019uoy,Nomura:2019yft,Liu:2019khw,
Ding:2019zxk,Okada:2019mjf,Nomura:2019lnr,Kobayashi:2019xvz,Asaka:2019vev,Gui-JunDing:2019wap,
Zhang:2019ngf,Nomura:2019xsb,Kobayashi:2019gtp,Lu:2019vgm,King:2020qaj,Okada:2020rjb}, level $N=4$~\cite{Penedo:2018nmg,Novichkov:2018ovf,deMedeirosVarzielas:2019cyj,Kobayashi:2019mna,King:2019vhv,Criado:2019tzk,Wang:2019ovr,Gui-JunDing:2019wap}, level $N=5$~\cite{Novichkov:2018nkm,Ding:2019xna,Criado:2019tzk} and level $N=7$~\cite{Ding:2020msi}. The quark masses and mixing parameters can also be addressed by using modular symmetry~\cite{Lu:2019vgm,Okada:2018yrn,Okada:2019uoy,Okada:2020rjb}, and the fermion mass hierarchies can naturally arise as a result of a weighton which is a standard model singlet field with non-zero modular weight~\cite{King:2020qaj}. Modular symmetry has been discussed in the
context of $SU(5)$ grand unification theory~\cite{deAnda:2018ecu,Kobayashi:2019rzp}. It is notable that the dynamics of modular symmetry could be tested at present and future neutrino
oscillation experiments~\cite{Ding:2020yen}. The modular symmetry has been extended to consistently include generalized CP symmetry under which the complex modulus $\tau$ transforms as $\tau\rightarrow-\tau^{*}$~\cite{Novichkov:2019sqv,Baur:2019kwi,Acharya:1995ag,Dent:2001cc,Giedt:2002ns}. The interplay between flavor symmetry, CP symmetry and modular invariance was recently analyzed in string theory~\cite{Baur:2019kwi,Baur:2019iai}. Extension to the direct product of multiple modular symmetry has been proposed~\cite{deMedeirosVarzielas:2019cyj,King:2019vhv}. We have generalized the modular invariance approach to include the odd weight modular forms which can be organized into irreducible representations of the homogeneous finite modular group $\Gamma'_{N}$~\cite{Liu:2019khw}. $\Gamma'_N$ is generally the double covering of the inhomogeneous finite modular group $\Gamma_N$. Texture zeros of fermion mass matrices can be naturally obtained from $\Gamma'_N$, the masses and mixing of quarks and leptons can be addressed in $\Gamma'_3\cong T'$~\cite{Lu:2019vgm}. There are many papers on modular symmetry $\Gamma_3\cong A_4$, $\Gamma_4\cong S_4$ and $\Gamma_5\cong A_5$, nevertheless the double covering modular groups $\Gamma'_N$ are less well studied except few papers on $\Gamma'_3\cong T'$~\cite{Liu:2019khw,Lu:2019vgm} although $\Gamma'_N$ can naturally appears in top-down string constructions~\cite{Baur:2019kwi,Baur:2019iai}. The modular symmetry $\Gamma'_{N}$ provides new ingredient for modular invariance approach, it could help us to further understand the possible role of modular symmetry in addressing the standard model flavor puzzle. In the present work, we shall consider the next-to-minimal homogeneous finite modular group $\Gamma'_4\equiv S'_4$ which is of order 48.

We emphasise that there are good motivations to study the $S'_4$ modular symmetry. It is known that interesting lepton mixing patterns can arise from the modular groups $\Gamma_N$ with $N=3,4,5,7$. However, light neutrino masses are usually predicted to quasi-degenerate such that the sum of neutrino masses are rather close to or beyond the upper limit of Planck collaboration. The $S'_4$ modular group opens up new model building possibilities, it is notable that light neutrino masses can be very tiny and all the experimental bounds from neutrino oscillations and cosmology can be satisfied, as shown below. So far most papers in the literature use modular symmetry to understand the neutrino masses and mixing. In order to incorporate the quarks and obtain a complete flavor theory, the lesson learned from convention flavor symmetry tells us that it is highly advantageous to extend the group to its double covering group which allows for spinorial representations. The most prominent example is the tetrahedral $A_4$ group, the successful $U(2)$ quark textures can be obtained together with the successful $A_4$ predictions for lepton sector by considering its double cover $T'$~\cite{Aranda:1999kc,Feruglio:2007uu,Chen:2007afa,Ding:2008rj}. Similar features are observed to hold true in the modular $T'$ group~\cite{Liu:2019khw,Lu:2019vgm}. The modular group $\Gamma_4\cong S_4$ has been considered as a family symmetry group for the leptons~\cite{Penedo:2018nmg,Novichkov:2018ovf,deMedeirosVarzielas:2019cyj,Kobayashi:2019mna,King:2019vhv,Criado:2019tzk,Wang:2019ovr,Gui-JunDing:2019wap}. In this paper, we investigate the extension of $S_4$ to its double covering $S'_4$\footnote{As discussed in section~\ref{sec:ModSym-ModF}, $S'_4$ is the double covering of $S_4$ in the sense of $S_4\cong S'_4/\{1, R\}$. $S_4$ is the symmetry group of the regular octahedron, consequently it is a subgroup of $SO(3)$. However, $S'_4$ is not a subgroup of $SU(2)$, and thus it is not the inverse image of $S_4$ under the homomorphism from $SU(2)$ to $SO(3)$, see the footnote~\ref{footnote2} for details.
Note that the $SU(2)$ analogue of $S_4$ is the binary octahedral group with \texttt{GAP} id $[48, 28]$.}, and apply $S'_4$ to explain the masses and mixing patterns of both leptons and quarks.

We intend to perform a systematical analysis of lepton and quark models based on $S'_4$ modular symmetry with/without generalized CP. For normal ordering neutrino masses, we find that fifteen viable models which can describe the neutrino masses, mixing angles and CP violation phases in terms of five real parameters  $|g_2/g_1|$, $\arg{(g_2/g_1)}$, $\texttt{Re}(\tau)$, $\texttt{Im}(\tau)$, and $g^2_1v_u^2/\Lambda$. After imposing the  generalized CP symmetry, the phase $\arg{(g_2/g_1)}$ is constrained to be $0$ or $\pi$, seven out of the fifteen models can produce a good fit to the data. The neutrino mass spectrum tends to be quasi-degenerate in previous models based on inhomogeneous finite modular group $\Gamma_N$, nevertheless the neutrino masses are much lighter in these $S'_4$ models. Moreover, we extend the $S'_4$ modular symmetry to the quark sector. The rich structure of the integral weight modular forms at level 4 allows many possibilities to accommodate the experimental on quark masses and CKM matrix. The modular models at level 3 use ten~\cite{Lu:2019vgm} or more free parameters~\cite{Okada:2018yrn,Okada:2019uoy,Okada:2020rjb} including $\tau$ to describe quark masses and mixing. The first five benchmark models constructed in this work involve only nine parameters and can be regarded as minimal. Aiming at minimal and predictive model for quarks and leptons, we impose both $S'_4$ modular symmetry and the generalized CP symmetry which are spontaneously broken by the VEV of the modulus field $\tau$. After comprehensively scanning the possible weight and representation assignments for the quark and lepton fields, we find a model which can describe the flavor structure of quarks and leptons simultaneously for a common value of $\tau$. This model has fifteen real parameters to explain the twenty-two observables: six quark masses, three quark mixing angles, one quark CP violation phases, three charged lepton masses, three neutrino masses, three lepton mixing angles and three leptonic CP violation phases. It is the most predictive modular models for quarks and leptons so far.

The remainder of the paper is organized as follows. In section~\ref{sec:ModSym-ModF}, we briefly review the basic aspects of modular symmetry, we show that the inhomogeneous finite modular group $\Gamma_N$ is isomorphic to the quotient of the homogeneous finite modular group $\Gamma'_N$ over the center $\{1, R\}$, i.e. $\Gamma_{N}\cong\Gamma'_N/\{1, R\}$, where $R$ is related to $-I\in SL(2, \mathbb{Z})$. The integral weight modular forms at level 4 are constructed up to weight 6 in section~\ref{sec:modularform_of_N=4}, and they are arranged into different irreducible representations of $S'_4$. The generalized CP symmetry compatible with $S'_4$ modular symmetry is discussed in section~\ref{sec:gCP-S4DC}. We find that the generalized CP symmetry requires all the coupling constants real in our working basis.
In section~\ref{sec:lepton-models}, we perform a systematical classification of $S'_4$ modular models for lepton masses and mixing, where the left-handed leptons are assigned to triplet of $S'_4$ and right-handed charged leptons transform as singlets under $S'_4$, and the neutrino masses are described
are described by the Weinberg operator or through the type I seesaw
mechanism. The $S'_4$ modular symmetry is utilized to address the flavor problem of quark masses hierarchies and CKM mixing matrix, and several models with small number of free parameters are presented in section~\ref{sec:quark-models}. We give a quark-lepton unification model in section~\ref{sec:quark-lepton-unification}, which can explain the masses and mixing patterns of quark and lepton for a common value of $\tau$. Section~\ref{sec:conclusion} concludes the paper. Appendix~\ref{sec:S4DC-app} gives the necessary group theory of $S'_4$ as well as the Clebsch-Gordan (CG) coefficients. We present the explicit forms of the modular forms for higher weight in Appendix~\ref{sec:higher weight-app}. The models based on another two possible assignments of right-handed charged leptons are discussed in Appendix~\ref{sec:lepton sector-app}.

\section{\label{sec:ModSym-ModF}Modular symmetry and finite modular group}

The modular group $\overline{\Gamma}$ is isomorphic to the projective special linear group $PSL(2, \mathbb{Z})$ of $2\times2$ matrices with integer coefficients and unit determinant,
\begin{equation}
\overline{\Gamma}\cong PSL(2, \mathbb{Z}) =\left\{\left.\pm\begin{pmatrix}
a  &  b \\
c  &  d
\end{pmatrix}\right| a,b,c,d\in\mathbb{Z}, ad-bc=1
\right\}\,,
\end{equation}
where the pairs of matrices $A$ and $-A$ are identified. Hence $PSL(2, \mathbb{Z})$ is the quotient of the two-dimensional special linear group $\Gamma\equiv SL(2, \mathbb{Z})$ over the integers by its center $\{I, -I\}$, i.e., $\overline{\Gamma}=PSL(2, \mathbb{Z})\cong SL(2, \mathbb{Z})/\{I, -I\}$, where $I$ is two dimensional unit matrix. The modular group acts on the upper-half complex plane $\mathcal{H}=\{\tau\in\mathbb{C}|\texttt{Im}(\tau)>0\}$ by fractional linear transformations,
\begin{equation}
\tau\mapsto\gamma\tau=\gamma(\tau)=\frac{a\tau+b}{c\tau+d},~~~\texttt{Im}(\tau) >0\,,
\end{equation}
which implies
\begin{eqnarray}
\nonumber&&I(\tau)=\tau\,,\\
\nonumber&&\texttt{Im}(\gamma(\tau))=\frac{\texttt{Im}\tau}{|c\tau+d|^2}>0\,,\\
&&(\gamma\gamma')(\tau)=\gamma(\gamma'(\tau)),~~\text{for}~~\forall \gamma, \gamma'\in\overline{\Gamma}\,.
\end{eqnarray}
Hence every fractional linear transformation corresponds to a modular group element $\begin{pmatrix}a  &  b \\
c  &  d
\end{pmatrix}$, and $\begin{pmatrix}a  &  b \\
c  &  d
\end{pmatrix}$ and $-\begin{pmatrix}a  &  b \\
c  &  d
\end{pmatrix}$ represent the same fractional linear transformation.
The modular group $\overline{\Gamma}$ has infinity group elements which can be obtained as a combination of the two fundamental transformations
\begin{equation}
S(\tau)=-\frac{1}{\tau},~~~~~T(\tau)=\tau+1\,,
\end{equation}
with the corresponding matrices
\begin{equation}
S=\begin{pmatrix}
0  ~& 1 \\
-1 ~& 0
\end{pmatrix},~~~~T=\begin{pmatrix}
1  ~&  1 \\
0  ~&  1
\end{pmatrix}\,.
\end{equation}
We check immediately that in $\overline{\Gamma}$ we have the relations
\begin{equation}
S^2=(ST)^3=I\,,
\end{equation}
and also $(TS)^3=I$ which is equivalent to $(ST)^3=I$ if $S^2=I$. The corresponding relations in $\Gamma$ are $S^2=-I, (ST^3)=I$ so that $S^4=(ST)^3=I$. The $\overline{\Gamma}$ orbit of every $\tau\in\mathcal{H}$ has a unique representative in the standard fundamental domain $\mathcal{D}$,
\begin{equation}
\mathcal{D}=\{\tau|\texttt{Im}(\tau)>0, |\texttt{Re}(\tau)|<\frac{1}{2}, |\tau|>1 \}\,,
\end{equation}
which is bounded by the vertical lines $\texttt{Re}(\tau)=-\frac{1}{2}$, $\texttt{Re}(\tau)=\frac{1}{2}$ and the circle $|\tau|=1$ in the upper half plane $\mathcal{H}$. The transformations $S$ and $T$ can map any point in $\mathcal{H}$ into the fundamental domain $\mathcal{D}$, and no two points inside $\mathcal{D}$ differ by a linear fraction transformation. The transformation $T$ pairs the two vertical lines $\texttt{Re}(\tau)=\pm\frac{1}{2}$, and the transformation $S$ maps the arc of $|\tau|=1$ from $i$ to $e^{\pi i/3}$ into the arc from $i$ to $e^{2\pi i/3}$. The principal congruence subgroup of level $N$ is defined as
\begin{eqnarray}
\nonumber \Gamma(N)&=&\left\{\begin{pmatrix}
a  &  b \\
c   & d
\end{pmatrix}\in\Gamma, \begin{pmatrix}
a  &  b \\
c   & d
\end{pmatrix}=\begin{pmatrix}
1  &  0 \\
0   & 1
\end{pmatrix}~~(\texttt{mod}~N) \right\}\,, \\
\overline{\Gamma}(N)&=&\left\{\begin{pmatrix}
a  &  b \\
c   & d
\end{pmatrix}\in\overline{\Gamma}, \begin{pmatrix}
a  &  b \\
c   & d
\end{pmatrix}=\pm\begin{pmatrix}
1  &  0 \\
0   & 1
\end{pmatrix}~~(\texttt{mod}~N) \right\}\,,
\end{eqnarray}
which are normal subgroups of $\Gamma$ and $\overline{\Gamma}$ respectively. Obviously we have $T^{N}\in\Gamma(N)$, $\Gamma=\Gamma(1)$, $\overline{\Gamma}=\overline{\Gamma}(1)$, $\overline{\Gamma}=\Gamma/\{I, -I\}$, $\overline{\Gamma}(2)=\Gamma(2)/\{I, -I\}$. For $N>2$, we have $-I\neq I~(\texttt{mod}~N)$ and thus $I\notin\Gamma(N)$, consequently $\Gamma(N)=\overline{\Gamma}(N)$. The finite modular group is the quotient of modular group over its principal congruence subgroup~\cite{deAdelhartToorop:2011re,Liu:2019khw},
\begin{equation}
\begin{aligned}
\textit{inhomogeneous finite modular group}~:~&  \Gamma_N\equiv \overline{\Gamma}/\overline{\Gamma}(N)\,,\\
\textit{homogeneous finite modular group}~:~&  \Gamma'_N\equiv \Gamma/\Gamma(N)\,.
\end{aligned}
\end{equation}
We see $\Gamma_2\cong\Gamma'_2$, and $\Gamma_N$ for $N>2$ is isomorphic to the quotient of $\Gamma'_N$ over its center $\{I, -I\}$, i.e., $\Gamma_{N}\cong\Gamma'_N/\{I, -I\}$ in matrix form. Hence $\Gamma'_N$ has double the number of group elements as $\Gamma_N$ with $|\Gamma'_N|=2|\Gamma_N|$. We broadly call $\Gamma'_N$ is the double covering\footnote{\label{footnote2}It is known that $SU(2)$ is the double covering of $SO(3)$. There is group homomorphisms that maps two distinct elements of $SU(2)$ into the same set of Euler angles of $SO(3)$. For every real vector $(x_1, x_2, x_3)^T\in \mathbb{R}^3$, we identify a Hermitian matrix $X=\sum^{3}_ix_i\sigma^i$ where $\sigma^i$ are the Pauli matrices. If $U$ is an element of $SU(2)$, the transformation $X\rightarrow UXU^{\dagger}=\sum^{3}_ix'_i\sigma^i$ induces a $SO(3)$ transformation $(x_1, x_2, x_3)^T\rightarrow (x'_1, x'_2, x'_3)=\mathcal{R}(x_1, x_2, x_3)^T$ with $\mathcal{R}\in SO(3)$. In this way, each element $SO(3)$ element $\mathcal{R}$ is mapped into two different elements $U$ and $-U$ of $SU(2)$. More precisely, $SO(3)$ is the isomorphic to $SU(2)/\{I, -I\}$, where $\{I, -I\}$ is the center of $SU(2)$. Quite similarly $\Gamma_N$ is isomorphic to $\Gamma'_N/\{1, R\}$, in this sense we call $\Gamma'_N$ as the double covering of $\Gamma_N$. Although $\Gamma'_3=T'$ and $\Gamma'_5=A'_5$ can be regarded as the inverse images of the group $\Gamma_3=A_4$ and $\Gamma_5=A_5$ respectively under the map from $SU(2)$ to $SO(3)$, $\Gamma'_4=S'_4$ is not the double cover of $\Gamma_4=S_4$ in $SU(2)$. In particular, the inhomogeneous finite modular group $\Gamma_N$ for $N>5$ is not a subgroup of $SO(3)$, thus the usual concepts of double covering learned from $SU(2)$ and $SO(3)$ groups don't hold true.} of $\Gamma_N$ in the present work. The homogeneous finite modular group $\Gamma'_N$ can be obtained from $\Gamma_N$ by including another generator $R$ which commutes with all elements of the $SL(2,\mathbb{Z})$ group. For $N\leq5$, the multiplication rules of the finite modular groups are\footnote{The multiplication rules of $\Gamma'_N$ can also be written as $S^4=(ST)^3=T^N=1, S^2T=TS^2$.}~\cite{Liu:2019khw},
\begin{equation}
\begin{aligned}
\Gamma_N~:~& S^2=(ST)^3=T^N=1\,,\\
\Gamma'_N~:~& S^2=R,~~(ST)^3=T^N=R^2=1,~~RT=TR\,,
\end{aligned}
\label{eq:mult-rules}
\end{equation}
where $R$ is related to $-I \in SL(2,\mathbb{Z})$. Obviously $\{1, R\}$ is the center of $\Gamma'_N$, and $\Gamma_N$ is isomorphic to the quotient group of $\Gamma'_N$ over this center group, i.e., $\Gamma_N\cong \Gamma'_N/\{1, R\}$. It is remarkable that $\Gamma_N$ and $\Gamma'_N$ for $N\leq 5$ is isomorphic to permutation groups and their double coverings, e.g., $\Gamma_2=\Gamma'_2\cong S_3$, $\Gamma_3\cong A_4$, $\Gamma'_3\cong T'$, $\Gamma_4\cong S_4$, $\Gamma_4\cong S'_4$, $\Gamma_5\cong A_5$, $\Gamma'_5\cong A'_5$. For $N>5$, additional relations besides those in Eq.~\eqref{eq:mult-rules} are needed to render the groups $\Gamma_N$ and $\Gamma'_N$ finite~\cite{deAdelhartToorop:2011re,Ding:2020msi}.

The modular form $f(\tau)$ of level $N$ and weight $k$ is a holomorphic function on $\mathcal{H}$ and at all cusps, and it is required to satisfy the following modular transformation property
\begin{equation}
f\left(\gamma\tau\right)=(c\tau+d)^{k}f(\tau)~~\text{for all}~~\gamma=\begin{pmatrix}
a  &  b \\
c   & d
\end{pmatrix}\in\Gamma(N)\,.
\end{equation}
The modular forms of level $N$ and weight $k$ span a linear space denoted by $M_k(\Gamma(N))$, and the dimension formula is~\cite{Bruinier2008The,schultz2015notes}
\begin{equation}
\label{eq:dim-formula}\text{dim}M_k(\Gamma(N))=\frac{(k-1)N+6}{24}N^2\prod_{p|N}(1-\frac{1}{p^2})\,,
\end{equation}
for $N>2$, where the product is over the prime divisors $p$ of $N$. For level $N=4$, we have $\text{dim}M_k(\Gamma(N))=2k+1$. As has been proved in~\cite{Liu:2019khw}, one can always find a basis of $M_k(\Gamma(N))$ such that the weight $k$ modular forms of level $N$
can be decomposed into different irreducible representations of $\Gamma'_N$ up to the automorphy factor $(c\tau+d)^k$. To be more specific, the transformation of the weight $k$ modular form multiplet $Y^{(k)}_{\mathbf{r}}(\tau)=(f_1(\tau),\,f_2(\tau),\,\dots)^T$ at level $N$ can be described by an irreducible representation $\rho_{\mathbf{r}}$ of $\Gamma'_{N}$,
\begin{equation}
\label{eq:MF-decomp}Y^{(k)}_{\mathbf{r}}(\gamma\tau)=(c\tau+d)^{k}\rho_{\mathbf{r}}(\gamma)Y^{(k)}_{\mathbf{r}}(\tau)\,,
\end{equation}
where $\gamma=\begin{pmatrix}
a  & b \\
c  & d
\end{pmatrix}$ is a representative element of $\Gamma'_{N}$. In a given representation basis of $\Gamma'_{N}$, the modular multiplet  $Y^{(k)}_{\mathbf{r}}$ can be fixed up to an overall irrelevant constant by applying Eq.~\eqref{eq:MF-decomp} for the generators $S$ and $T$.

\section{\label{sec:modularform_of_N=4}Modular forms of level $N=4$ }

Applying the general dimension formula in Eq.~\eqref{eq:dim-formula} for $N=4$, we find the modular space $M_{k}(\Gamma(4))$ has dimension $2k+1$. The modular space $M_{k}(\Gamma(4))$ has been constructed explicitly by
making use of Dedekind eta function~\cite{schultz2015notes},
\begin{equation}
\label{eq:Mk_Gamma4}M_{k}(\Gamma(4))=\bigoplus_{a+b=2k,\,a,b\ge0} \mathbb{C} \frac{\eta^{2b-2a}(4\tau)\eta^{5a-b}(2\tau)}{\eta^{2a}(\tau)}\,,
\end{equation}
where $\eta(\tau)$ is the famous Dedekind eta function defined by
\begin{equation}
\eta(\tau)=q^{1/24}\prod_{n=1}^\infty \left(1-q^n \right),~~~q= e^{i 2 \pi\tau}\,.
\end{equation}
The Dedekind eta function is a crucial example of a half-integral weight modular form, having weight $1/2$ and level $1$. The eta function satisfies the well-known transformation formulas~\cite{diamond2005first,Bruinier2008The,lang2012introduction},
\begin{equation}
\label{eq:eta-identities}\begin{aligned}
& \eta(\tau)\stackrel{S}{\longmapsto}\eta(-1/\tau)=\sqrt{-i \tau}~\eta(\tau),\\
&\eta(\tau)\stackrel{T}{\longmapsto}\eta(\tau+1)=e^{i \pi/12}\eta(\tau)\,.
\end{aligned}
\end{equation}
As shown in Eq.~\eqref{eq:Mk_Gamma4}, we can choose the three linearly independent basis vectors of the weight $1$ modular space of level $4$ as
\begin{eqnarray}
\label{eq:original_basis}
e_1(\tau)=\frac{\eta^{4}(4\tau)}{\eta^{2}(2\tau)},~~~ e_{2}(\tau)=\frac{\eta^{10}(2\tau)}{\eta^{4}(4\tau)\eta^{4}(\tau)},~~~ e_3(\tau)=\frac{\eta^{4}(2\tau)}{\eta^{2}(2\tau)}\,.
\end{eqnarray}
The $q-$expansion of $e_1(\tau),e_2(\tau),e_3(\tau)$ reads:
\begin{align}
\nonumber &e_1(\tau)=\sqrt{q} \left(1+2 q^2+q^4+2 q^6+2 q^8+3 q^{12}+2 q^{14}+2 q^{18}+2 q^{20}+\dots \right),\\
\nonumber &e_2(\tau)=1+4 q+4 q^2+4 q^4+8 q^5+4 q^8+4 q^9+8 q^{10}+8 q^{13}+4 q^{16} +\dots,  \\
\label{eq:ei-expansion}&e_3(\tau)=q^{1/4} \left(1+2 q+q^2+2 q^3+2 q^4+3 q^6+2 q^7+2 q^9+2 q^{10} +\dots\right)\,.
\end{align}
From the identities of the eta function in Eq.~\eqref{eq:eta-identities}, we know that $e_{1,2,3}(\tau)$ transform under the actions of $S$ and $T$ as follow,
\begin{eqnarray}
\nonumber&& e_1(\tau)\stackrel{T}{\longmapsto}-e_1(\tau),~~ e_{2}(\tau)\stackrel{T}{\longmapsto} e_2 ,~~ e_{3}(\tau)\stackrel{T}{\longmapsto} i e_3\,. \\
\nonumber&& e_1(\tau)\stackrel{S}{\longmapsto}\frac{1}{8}(-i\tau)(4 e_1+e_2-4 e_3),\\
\nonumber&&e_{2}(\tau)\stackrel{S}{\longmapsto} \frac{1}{8}(-i\tau)(16 e_1 +4 e_2 +16 e_3),\\
&& e_{3}(\tau)\stackrel{S}{\longmapsto} \frac{1}{8}(-i\tau)(-8 e_1 +2 e_2)\,.
\end{eqnarray}
As shown in Eq.~\eqref{eq:MF-decomp}, it is always possible to choose a set of basis in $M_{k}(\Gamma(4))$ such that the basis vectors can be arranged into several modular multiplets which transform in irreducible representations of $\Gamma'_4\equiv S'_4$. Thus for the weight 1 modular forms of level 4, solving the condition of Eq.~\eqref{eq:MF-decomp}, we find the original basis $e_{1,2,3}(\tau)$ can be arranged into triplet modular form $Y^{(1)}_{\mathbf{\hat{3}'}}$ transforming as a triplet $\mathbf{\hat{3}'}$ of $S'_4$,
\begin{equation}
\label{eq:modular_space}
Y^{(1)}_{\mathbf{\hat{3}'}}(\tau) \equiv \begin{pmatrix}
Y_1(\tau) \\
Y_2(\tau) \\
Y_3(\tau)
\end{pmatrix}\,,
\end{equation}
where $Y_{1,2,3}(\tau)$ are linear combinations of $e_{1,2,3}(\tau)$ as follows
\begin{align}
\nonumber
Y_1(\tau)&=4\sqrt{2}\,e_1(\tau)+\sqrt{2}\,ie_2(\tau)+2\sqrt{2}(1-i) e_3(\tau),\\
\nonumber
Y_2(\tau)&=-2\sqrt{2}(1+\sqrt{3})\omega^2e_1(\tau)-\frac{1-\sqrt{3}}{\sqrt{2}}i\omega^2
e_2(\tau)+2\sqrt{2}(1-i)\omega^2 e_3(\tau), \\
Y_3(\tau)&=2\sqrt{2}(\sqrt{3}-1)\omega e_1(\tau)-\frac{1+\sqrt{3}}{\sqrt{2}}i\omega e_2(\tau)+2\sqrt{2}(1-i)\omega e_3(\tau).
\end{align}
It is straightforward to check that $Y^{(1)}_{\mathbf{\hat{3}'}}(\tau)$ transforms under $S$ and $T$ as
\begin{equation}
\label{eq:MF_decomp_ST}
Y^{(1)}_{\mathbf{\hat{3}'}}(-1/\tau)=-\tau\rho_\mathbf{\hat{3}'}(S)Y^{(1)}_{\mathbf{\hat{3}'}}(\tau),\qquad Y^{(1)}_{\mathbf{\hat{3}'}}(\tau+1)=\rho_\mathbf{\hat{3}'}(T)Y^{(1)}_{\mathbf{\hat{3}'}}(\tau)\,,
\end{equation}
where the representation matrices $\rho_\mathbf{\hat{3}'}(S)$ and $\rho_\mathbf{\hat{3}'}(T)$ in our working basis are summarized in table~\ref{tab:Rep_matrix}.

The modular forms $Y_{1,2,3}$ satisfy the following constraint,
\begin{equation}
\label{eq:MF-constraint}\left(Y^{(1)}_{\mathbf{\hat{3}'}}Y^{(1)}_{\mathbf{\hat{3}'}}\right)_{\mathbf{1'}}
=Y_1^2+2Y_2Y_3=0\,.
\end{equation}
The higher weight modular forms can be constructed from the tensor products of lower weight modular forms with the help of the CG coefficients of $S'_4$ in Appendix~\ref{sec:S4DC-app}, and they are
homogeneous polynomials of $Y_{1,2,3}$. Using the contraction rules for $\mathbf{\hat{3}'} \otimes \mathbf{\hat{3}'}\rightarrow \mathbf{1'} \oplus \mathbf{2} \oplus \mathbf{3} \oplus \mathbf{3'}$, we find that the weight 2 modular forms of level 4 decompose $\mathbf{2}\oplus\mathbf{3}$ under $S'_4$
\begin{eqnarray}
\nonumber&& Y^{(2)}_{\mathbf{2}}=\left(Y^{(1)}_{\mathbf{\hat{3}'}}Y^{(1)}_{\mathbf{\hat{3}'}}\right)_{\mathbf{2}}=\begin{pmatrix} -Y_2^2- 2Y_1Y_3 \\  Y_3^2 + 2 Y_1Y_2 \end{pmatrix}, \\
\nonumber&& \\[+0.05in]
&& Y^{(2)}_{\mathbf{3}}=\left(Y^{(1)}_{\mathbf{\hat{3}'}}Y^{(1)}_{\mathbf{\hat{3}'}}\right)_{\mathbf{3}}=\begin{pmatrix}
2Y_1^2-2Y_2Y_3 \\ 2 Y_3^2-2Y_1Y_2 \\ 2Y_2^2-2Y_1Y_3
\end{pmatrix}\,.
\end{eqnarray}
Note $\left(Y^{(1)}_{\mathbf{\hat{3}'}}Y^{(1)}_{\mathbf{\hat{3}'}}\right)_{\mathbf{3'}}=(0,\,0,\,0)^T$ which arises from the antisymmetric CG coefficients. Likewise, the weight 3 modular forms can be obtained from the tensor products of $Y^{(1)}_{\mathbf{\hat{3}'}}$ with $Y^{(2)}_{\mathbf{2}}$ and $Y^{(2)}_{\mathbf{3}}$, and they are arranged into a singlet $\mathbf{\hat{1}'}$ and two triplets $\mathbf{\hat{3}}$ and $\mathbf{\hat{3}'}$ under $S'_4$,
\begin{eqnarray}
\nonumber&& Y^{(3)}_{\mathbf{\hat{1}'}}=\left(Y^{(2)}_{\mathbf{3}}Y^{(1)}_{\mathbf{\hat{3}'}}\right)_{\mathbf{\hat{1}'}}=2(Y_1^3+Y_2^3+Y_3^3-3 Y_1Y_2Y_3 ) ,\\
\nonumber&& Y^{(3)}_{\mathbf{\hat{3}}}=\left(Y^{(2)}_{\mathbf{3}}Y^{(1)}_{\mathbf{\hat{3}'}}\right)_{\mathbf{\hat{3}}}=\begin{pmatrix}2(2Y_1^3-Y_2^3-Y_3^3)\\ 6Y_3(Y_2^2-Y_1Y_3)\\ 6Y_2(Y_3^2-Y_1Y_2) \end{pmatrix} , \\
\nonumber&& \\[+0.05in]
&& Y^{(3)}_{\mathbf{\hat{3}'}}=\left(Y^{(2)}_{\mathbf{3}}Y^{(1)}_{\mathbf{\hat{3}'}}\right)_{\mathbf{\hat{3}'}}=\begin{pmatrix}2(Y_2^3-Y_3^3)\\ 2(-2 Y_1^2Y_2+Y_2^2Y_3+Y_1Y_3^2)\\ 2(2Y_1^2Y_3-Y_1Y_2^2-Y_2Y_3^2)
\end{pmatrix}\,.
\end{eqnarray}
We have three additional contractions between weight 1 and 2 modular forms, nevertheless they are not independent from $Y^{(3)}_{\mathbf{\hat{1}'}}$, $Y^{(3)}_{\mathbf{\hat{3}}}$ and $Y^{(3)}_{\mathbf{\hat{3}'}}$,
\begin{subequations}
\begin{align}
\label{eq:wt3-ad1}&\left(Y^{(2)}_{\mathbf{3}}Y^{(1)}_{\mathbf{\hat{3}'}}\right)_{\mathbf{\hat{2}}}=\begin{pmatrix}
0\\
0
\end{pmatrix}\,, \\
\label{eq:wt3-ad2}&\left(Y^{(2)}_{\mathbf{2}}Y^{(1)}_{\mathbf{\hat{3}'}}\right)_{\mathbf{\hat{3}'}} = \begin{pmatrix}
-Y_2^3+Y_3^3 \\
2Y_1^2Y_2-Y_2^2Y_3-Y_1Y_3^2) \\
-2Y_1^2Y_3+Y_1Y_2^2+Y_2Y_3^2
\end{pmatrix} =-\frac{1}{2} Y^{(3)}_{\mathbf{\hat{3}'}}\,, \\
\label{eq:wt3-ad3}&\left(Y^{(2)}_{\mathbf{2}}Y^{(1)}_{\mathbf{\hat{3}'}}\right)_{\mathbf{\hat{3}}} = \begin{pmatrix}
Y_2^3+Y_3^3+4Y_1Y_2Y_3 \\ 2Y_1^2Y_2+Y_2^2Y_3+3Y_1Y_3^2 \\ 3Y_1Y_2^2+2Y_1^2Y_3+Y_2Y_3^2
\end{pmatrix} =-\frac{1}{2} Y^{(3)}_{\mathbf{\hat{3}}}\,.
\end{align}
\end{subequations}
The last relation follows from the constraint in Eq.~\eqref{eq:MF-constraint}. In a similar manner, we can find out the linearly independent modular forms of higher weights and corresponding constraints. The expressions of the higher weight modular multiplets with $k=4, 5, 6$ are given in Appendix~\ref{sec:higher weight-app}. We summarize the modular forms of level 4 up to weight 6 in table~\ref{Tab:Level4_MM}. We notice that all the odd weight modular forms are in hatted irreducible representations of $S'_4$ while the even weight modular forms are in unhatted irreducible representations of $S'_4$. Note that generator $R$ is represented by unit matrix and the $S'_4$ group can not be distinguished from $S_4$ in unhatted irreducible representations. For notation simplicity of model construction in the following, we denote the components of modular multiplets as follows,
\begin{eqnarray}
\nonumber&&
Y^{(2)}_\mathbf{2}\equiv\begin{pmatrix} Y^{(2)}_1 \\Y^{(2)}_2 \end{pmatrix}, ~
Y^{(2)}_\mathbf{3}\equiv\begin{pmatrix} Y^{(2)}_3 \\Y^{(2)}_4 \\ Y^{(2)}_5 \end{pmatrix},~
Y^{(3)}_\mathbf{\hat{1}'}\equiv Y^{(3)}_1, ~
Y^{(3)}_\mathbf{\hat{3}}\equiv \begin{pmatrix} Y^{(3)}_2 \\Y^{(3)}_3 \\ Y^{(3)}_4 \end{pmatrix}, ~ Y^{(3)}_\mathbf{\hat{3}'}\equiv \begin{pmatrix} Y^{(3)}_5 \\Y^{(3)}_6 \\ Y^{(3)}_7 \end{pmatrix}\,, \\
\nonumber
&& Y^{(4)}_\mathbf{1}\equiv Y^{(4)}_1, ~~
Y^{(4)}_\mathbf{2}\equiv\begin{pmatrix} Y^{(4)}_2 \\Y^{(4)}_3  \end{pmatrix}, ~~ Y^{(4)}_\mathbf{3}\equiv\begin{pmatrix} Y^{(4)}_4 \\Y^{(4)}_5 \\Y^{(4)}_6 \end{pmatrix}, ~~ Y^{(4)}_\mathbf{3'}\equiv\begin{pmatrix} Y^{(4)}_7 \\Y^{(4)}_8 \\Y^{(4)}_9 \end{pmatrix}\,,\\
\nonumber
&& Y^{(5)}_\mathbf{\hat{2}}\equiv \begin{pmatrix} Y^{(5)}_1 \\Y^{(4)}_2  \end{pmatrix}, ~~
Y^{(5)}_\mathbf{\hat{3}}\equiv\begin{pmatrix} Y^{(5)}_3 \\Y^{(5)}_4 \\Y^{(5)}_5  \end{pmatrix}, ~~ Y^{(5)}_{\mathbf{\hat{3}'},I}\equiv\begin{pmatrix} Y^{(5)}_6 \\Y^{(5)}_7 \\Y^{(5)}_8 \end{pmatrix}, ~~ Y^{(5)}_{\mathbf{\hat{3}'},II}\equiv\begin{pmatrix} Y^{(5)}_9 \\Y^{(5)}_{10} \\Y^{(5)}_{11} \end{pmatrix}\,,\\
\nonumber
&& Y^{(6)}_\mathbf{1'}\equiv Y^{(6)}_1 , ~Y^{(6)}_\mathbf{1}\equiv Y^{(6)}_2 ,
~Y^{(6)}_\mathbf{2}\equiv\begin{pmatrix} Y^{(6)}_3 \\Y^{(6)}_4  \end{pmatrix},~ Y^{(6)}_{\mathbf{3},I}\equiv\begin{pmatrix} Y^{(6)}_5 \\Y^{(6)}_6 \\Y^{(6)}_7 \end{pmatrix}, ~Y^{(6)}_{\mathbf{3},II}\equiv\begin{pmatrix} Y^{(6)}_8 \\Y^{(6)}_9 \\Y^{(6)}_{10} \end{pmatrix},\\
&& Y^{(6)}_\mathbf{3'}\equiv\begin{pmatrix} Y^{(6)}_{11} \\Y^{(6)}_{12} \\Y^{(6)}_{13} \end{pmatrix}\,.
\end{eqnarray}

\begin{table}[t!]
\centering
\begin{tabular}{|c|c|}
\hline  \hline

Modular weight $k$ & Modular forms $Y^{(k)}_{\mathbf{r}}$ \\ \hline

$k=1$ & $Y^{(1)}_{\mathbf{\hat{3}'}}$\\  \hline

$k=2$ & $Y^{(2)}_{\mathbf{2}}, ~Y^{(2)}_{\mathbf{3}}$\\ \hline

$k=3$ & $Y^{(3)}_{\mathbf{\hat{1}'}},~ Y^{(3)}_{\mathbf{\hat{3}}},~Y^{(3)}_{\mathbf{\hat{3}'}}$\\ \hline

$k=4$ & $Y^{(4)}_{\mathbf{1}},~Y^{(4)}_{\mathbf{2}},~ Y^{(4)}_{\mathbf{3}}, ~Y^{(4)}_{\mathbf{3}'}$\\ \hline

$k=5$ & $Y^{(5)}_{\mathbf{\hat{2}}},~ Y^{(5)}_{\mathbf{\hat{3}}}, ~Y^{(5)}_{\mathbf{\hat{3}'},I},Y^{(5)}_{\mathbf{\hat{3}'},II}$\\ \hline

$k=6$ & $Y^{(6)}_{\mathbf{1'}},~Y^{(6)}_{\mathbf{1}},~Y^{(6)}_{\mathbf{2}},~Y^{(6)}_{\mathbf{3},I},~Y^{(6)}_{\mathbf{3},II},~Y^{(6)}_{\mathbf{3}'}$\\ \hline \hline
\end{tabular}
\caption{\label{Tab:Level4_MM}Summary of modular forms of level $N=4$ up to weight 6, the subscript $\mathbf{r}$ denote the transformation property under homogeneous finite modular group $S'_4$. }
\end{table}

\section{\label{sec:gCP-S4DC}Generalized CP consistent with $S'_4$ modular symmetry}

In order to consistently implement CP symmetry in the context of modular symmetry, the complex modulus $\tau$ should transform under the action CP as~\cite{Novichkov:2019sqv,Baur:2019kwi,Acharya:1995ag,Dent:2001cc,Giedt:2002ns}
\begin{equation}
\tau\stackrel{\mathcal{CP}}{\longmapsto}-\tau^{*}\,,
\end{equation}
up to modular transformations. A generic chiral superfield $\Phi(x)$ assigned to an irreducible representation $\mathbf{r}$ of the finite modular group $\Gamma'_N$ transforms under the action of $\Gamma'_N$ as
\begin{equation}
\Phi(x)\stackrel{\gamma}{\longmapsto} (c\tau+d)^{-k}\rho_{\mathbf{r}}(\gamma)\Phi(x),~~~\gamma=\begin{pmatrix}
a  & b \\
c  &  d
\end{pmatrix}\in\Gamma\,,
\end{equation}
where $-k$ is the modular weight of $\Phi$. We impose CP symmetry on the modular invariant theory. A generalized CP transformation acts on the chiral superfield $\Phi(x)$ as
\begin{equation}
\Phi(x)\stackrel{\mathcal{CP}}{\longmapsto} X_{\mathbf{r}}\overline{\Phi}(\mathcal{P}x)\,,
\end{equation}
where $\mathcal{P}x=(t, -\vec{x})$, and a bar denotes the hermitian conjugate superfield, $X_{\mathbf{r}}$ is not necessarily diagonal and it in general acts in a non-trivial way on the flavor space. As has been shown in~\cite{Novichkov:2019sqv}, constraints on the choice
of $X_{\mathbf{r}}$ arise from the requirement that the subsequent application of the CP transformation, the modular symmetry and the inverse CP transformation should be represented by another element of the
modular symmetry group, i.e.
\begin{equation}
\label{eq:cc}X_{\mathbf{r}}\rho^{*}_{\mathbf{r}}(\gamma)X^{-1}_{\mathbf{r}}=\rho_{\mathbf{r}}(u(\gamma))\,,
\end{equation}
where $u(\gamma)$ is an outer automorphism of the modular group,
\begin{equation}
\gamma=\begin{pmatrix}
a  &  b \\
c  &  d
\end{pmatrix}\longmapsto u(\gamma)=\begin{pmatrix}
a  &  -b \\
-c  &  d
\end{pmatrix}\,.
\end{equation}
Eq.~\eqref{eq:cc} is the so-called consistency condition which CP and modular symmetries have to obey in order to give a consistent definition of generalized CP transformations in setting with modular symmetry. It is notable that the consistency condition Eq.~\eqref{eq:cc} should be satisfied for all irreducible representations of the finite modular group $\Gamma'_N$. We see that the CP transformation $X_{\mathbf{r}}$ maps the modular group element $\gamma$ onto another element $u(\gamma)$ and the group structure of the modular symmetry is preserved, i.e. $u(\gamma_1\gamma_2)=u(\gamma_1)u(\gamma_2)$. Hence it is sufficient to  impose Eq.~\eqref{eq:cc} on the generators $S$ and $T$,
\begin{equation}
\label{eq:consc_gen}X_{\mathbf{r}}\rho^{*}_{\mathbf{r}}(S)X^{-1}_{\mathbf{r}}=\rho^{\dagger}_{\mathbf{r}}(S),\qquad X_{\mathbf{r}}\rho^{*}_{\mathbf{r}}(T)X^{-1}_{\mathbf{r}}=\rho^{\dagger}_{\mathbf{r}}(T)\,,
\end{equation}
where the identities $u(S)=S^{-1}$ and $u(T)=T^{-1}$ are used. The consistency condition in Eq.~\eqref{eq:cc} determines the CP transformation $X_{\mathbf{r}}$ up to an overall phase for a given irreducible representation $\mathbf{r}$. As regards the double covering group $S'_4$ with the basis given in table~\ref{tab:Rep_matrix}, solving the consistency conditions of Eq.~\eqref{eq:consc_gen}, we find that the generalized CP transformation $X_{\mathbf{r}}$ coincides with the representation matrix of $S$,
\begin{equation}
\label{eq:Xr_S4p}X_{\mathbf{r}}=\rho_{\mathbf{r}}(S)\,,
\end{equation}
which is a combination of the modular symmetry transformation $S$ and the  canonical CP transformation. Furthermore, we have checked that the modular forms $Y^{(k)}_{\mathbf{r}}(\tau)$ of section~\ref{sec:modularform_of_N=4} transforms in the same way as $\Phi(x)$ under CP:
\begin{equation}
\label{eq:MF-gCP}Y^{(k)}_{\mathbf{r}}(\tau)\stackrel{\mathcal{CP}}{\longmapsto} Y^{(k)}_{\mathbf{r}}(-\tau^{*})=X_{\mathbf{r}}[Y^{(k)}_{\mathbf{r}}(\tau)]^{*},~~\text{with}~~X_{\mathbf{r}}=\rho_{\mathbf{r}}(S)\,.
\end{equation}
Hence the above CP transformation $X_{\mathbf{r}}=\rho_{\mathbf{r}}(S)$ imposed on a modular invariant supersymmetric theory amounts to the canonical CP transformation. As shown in Appendix~\ref{sec:S4DC-app}, all the CG coefficients in our working basis are real, therefore the generalized CP symmetry would constrain all the couplings in the Lagrangian to be real.

\section{\label{sec:lepton-models}Lepton models based on $S'_4$ modular symmetry }

We work in the framework of the modular invariant supersymmetric theory~\cite{Ferrara:1989bc,Ferrara:1989qb,Feruglio:2017spp}. In the setting of $N=1$ global supersymmetry, the action can be generally written as
\begin{equation}
\mathcal{S}=\int d^4xd^2\theta d^2\bar{\theta} \, \mathcal{K}(\Phi_I,\bar{\Phi}_I,\tau,\bar{\tau}) + \left[\int d^4x d^2\theta \mathcal{W}(\Phi_I,\tau) + h.c.\right]\,,
\end{equation}
where $\mathcal{K}(\Phi_I,\bar{\Phi}_I,\tau,\bar{\tau})$ is K\"ahler potential, it is real gauge invariant function of the chiral superfields $\Phi$ and their hermitian conjugates $\bar{\Phi}$. $\mathcal{W}(\Phi_I,\tau)$ refers to the superpotential, and it is a holomorphic gauge invariant function of the chiral superfields $\Phi$. The whole action $\mathcal{S}$ should be modular invariant. The transformation properties of $\Phi_I$ are specified by its modular weight $-k_I$ and the representation $\mathbf{r}_I$ under $\Gamma'_N$,
\begin{equation}
\tau\rightarrow\gamma\tau=\frac{a\tau+b}{c\tau+d},~~~~~\Phi_I\rightarrow (c\tau+d)^{-k_I}\rho_{\mathbf{r}_I}(\gamma)\Phi_I\,.
\end{equation}
Following Ref.~\cite{Feruglio:2017spp}, we take the K\"ahler potential to be the minimal form,
\begin{equation}
\mathcal{K}(\Phi_I,\bar{\Phi}_I,\tau,\bar{\tau}) = -h\Lambda^2\log(-i\tau+i\bar{\tau})+\sum_I(-i\tau+i\bar{\tau})^{-k_I}|\Phi_I|^2\,,
\end{equation}
where $h$ is a positive constant. After the modulus $\tau$ gets a vacuum expectation, this K\"ahler potential gives the kinetic terms for the scalar components of the supermultiplet $\Phi_I$ and the modulus field $\tau$.
The K\"ahler potential is strongly constrained in some models based on string theory~\cite{Nilles:2020nnc,Nilles:2020kgo,Ohki:2020bpo}, and the above minimal K\"ahler potential as the leading order contribution could possibly be achieved. The superpotential $\mathcal{W}$ can be expanded into power series of supermultiplets $\Phi_I$
\begin{equation}
\mathcal{W}(\Phi_I,\tau)=\sum_n Y_{I_1...I_n}(\tau)\Phi_{I_1}...\Phi_{I_n}\,.
\end{equation}
Modular invariance requires the function $Y_{I_1...I_n}(\tau)$ should be a modular form of weight $k_Y$ of level $N$ and in the representation $\mathbf{r}_Y$ of $\Gamma'_N$:
\begin{equation}
Y(\tau)\to Y(\gamma\tau)=(c\tau+d)^{k_Y}\rho_{\mathbf{r}_Y}(\gamma)Y(\tau)\,,
\end{equation}
where $k_Y$ and $\mathbf{r}_Y$ should satisfy the conditions
\begin{equation}
k_Y=k_{1}+...+k_{n},~~~\rho_{\mathbf{r}_Y}\otimes \rho_{\mathbf{r}_{I_1}}\otimes\ldots\otimes\rho_{\mathbf{r}_{I_n}} \ni \mathbf{1}\,.
\end{equation}
In the present work, we shall study the modular symmetry group of level $N=4$, and a comprehensive analysis of lepton models with $S'_4$ modular symmetry is performed in the following. In the bottom-up approach of modular invariance~\cite{Feruglio:2017spp}, the representations and the weights of the matter fields are not subject to any constraint at all, and the number of modular invariant operators generally increase with the weights of the involved modular forms. The models are built aiming at minimizing the number of free parameters, consequently we will consider the weight 1, weight 2 and weight 3 modular forms for illustration in the following, and the cases with higher weight modular forms can be discussed in the same fashion.

\subsection{\label{subsec:charged_lepton}Charged lepton sector}
The left-handed lepton doublet fields are assigned to transform as triplet $\mathbf{3}$, $\mathbf{3'}$, $\mathbf{\hat{3}},\,\mathbf{\hat{3}'}$ of $S'_4$. There are multiple options for the assignments of the right-handed charged leptons. They can be assigned to three independent singlets, a triplet or the direct sum of a doublet and a singlet. In this section, we will focus on the first case, i.e., the right-handed charged leptons transform as singlets $\mathbf{1}$, $\mathbf{1'}$, $\mathbf{\hat{1}}$ or $\mathbf{\hat{1}'}$.  The other two cases and the corresponding charged lepton models are discussed in the Appendix~\ref{sec:lepton sector-app}. We follow the original paper~\cite{Feruglio:2017spp} and assume that the Higgs fields $H_{u,d}$ are invariant under $S'_4$, otherwise the modular forms would be involved in the Higgs potential and the dynamics of the electroweak symmetry breaking would be greatly complexified by the complex modulus $\tau$. The modular weights of $H_{u,d}$ can always be taken to zero through redefinition of the modular weights of matter fields. Thus the most general superpotential for the charged lepton masses can be written as :
\begin{equation}
\label{eq:We}
\mathcal{W}_e\,= \alpha (E^c_1 L f_{E_1}(Y))_\mathbf{1}H_d +\beta( E^c_2 L f_{E_2}(Y))_\mathbf{1}H_d +\gamma (E^c_3L f_{E_3}(Y))_\mathbf{1}H_d\,.
\end{equation}
The modular forms $f_{E_1}(Y)$, $f_{E_2}(Y)$ and $f_{E_3}(Y)$ should transform as three dimensional irreducible representations under $S'_4$, and their explicit forms depend on the weight and representation assignments for $L$ and $E^{c}_{1,2,3}$. In order to charged lepton mass matrix with rank less than three otherwise at least one charged lepton would be massless,  $f_{E_1}(Y)$, $f_{E_2}(Y)$ and $f_{E_3}(Y)$ must be different modular multiplets. For illustration, we consider modular forms of weight less than four, consequently $f_{E_1}(Y)$, $f_{E_2}(Y)$, $f_{E_3}(Y)$ can only be $Y^{(1)}_\mathbf{\hat{3}'}$, $Y^{(2)}_\mathbf{3}$, $Y^{(3)}_\mathbf{\hat{3}}$ and $Y^{(3)}_\mathbf{\hat{3}'}$. It is remarkable that the CG coefficients for the contraction \texttt{triplet}$\otimes$ \texttt{triplet}$\rightarrow$\texttt{singlet} are all the same in our basis. As a consequence, there are only four different structures of charged lepton mass matrix if the weights of the relevant modular forms are less than four.

\begin{itemize}[labelindent=-1.5em, leftmargin=1.6em]
\item[\textbf{(i)}]{$f_{E_1}(Y) = Y^{(1)}_\mathbf{\hat{3}'}$, $f_{E_2}(Y) = Y^{(2)}_\mathbf{3}$, $f_{E_3}(Y) = Y^{(3)}_\mathbf{\hat{3}}$}\\
In this case, there are four different representation assignments which give rise to the same charged lepton mass matrix:
\begin{equation}
\begin{cases}
\rho_L=\mathbf{3},~\rho_{E^{c}_1}=\mathbf{\hat{1}},~\rho_{E^{c}_2}=\mathbf{1},~\rho_{E^{c}_3}=\mathbf{\hat{1}'}\,, \\
\rho_L=\mathbf{3'},~\rho_{E^{c}_1}=\mathbf{\hat{1}'},~\rho_{E^{c}_2}=\mathbf{1'},~\rho_{E^{c}_3}=\mathbf{\hat{1}}\,, \\
\rho_L=\mathbf{\hat{3}},~\rho_{E^{c}_1}=\mathbf{1},~\rho_{E^{c}_2}=\mathbf{\hat{1}'},~\rho_{E^{c}_3}=\mathbf{1'}\,, \\
\rho_L=\mathbf{\hat{3}'},~\rho_{E^{c}_1}=\mathbf{1'},~\rho_{E^{c}_2}=\mathbf{\hat{1}},~\rho_{E^{c}_3}=\mathbf{1}\,.
\end{cases}
\end{equation}
The superpotential for the charged lepton masses are given by,
\begin{eqnarray}
\nonumber && \mathcal{W}_e = \alpha(E^c_1LY^{(1)}_{\mathbf{\hat{3}'}})_\mathbf{1}H_d+\beta (E^c_2L Y^{(2)}_{\mathbf{3}})_\mathbf{1}H_d+ \gamma(E^c_3LY^{(3)}_{\mathbf{\hat{3}}})_\mathbf{1}H_d \\
\nonumber&&~~~~ =\alpha E^c_1(L_1 Y_1+L_3Y_2+L_2Y_3)H_d+\beta E^c_2(L_1 Y^{(2)}_3+L_3Y^{(2)}_4+L_2Y^{(2)}_5)H_d\\
&&~~~~~~~ +\gamma E^c_3(L_1 Y^{(3)}_2+L_3Y^{(3)}_3+L_2Y^{(3)}_4)H_d\,.
\label{eq:WeI}
\end{eqnarray}
The condition of modular weight cancellation requires
\begin{equation}
k_{E_1}=k_{E_2}-1=k_{E_3}-2=1-k_L\,.
\end{equation}
\item[\textbf{(ii)}]{$f_{E_1}(Y) = Y^{(1)}_\mathbf{\hat{3}'}$, $f_{E_2}(Y) = Y^{(2)}_\mathbf{3}$, $f_{E_3}(Y) = Y^{(3)}_\mathbf{\hat{3}'}$ }\\
There are also four representation assignments for the lepton fields,
\begin{equation}
\begin{cases}
\rho_L=\mathbf{3},~\rho_{E^{c}_1}=\mathbf{\hat{1}},~\rho_{E^{c}_2}=\mathbf{1},~\rho_{E^{c}_3}=\mathbf{\hat{1}}\,, \\
\rho_L=\mathbf{3'},~\rho_{E^{c}_1}=\mathbf{\hat{1}'},~\rho_{E^{c}_2}=\mathbf{1'},~\rho_{E^{c}_3}=\mathbf{\hat{1}'}\,, \\
\rho_L=\mathbf{\hat{3}},~\rho_{E^{c}_1}=\mathbf{1},~\rho_{E^{c}_2}=\mathbf{\hat{1}'},~\rho_{E^{c}_3}=\mathbf{1}\,, \\
\rho_L=\mathbf{\hat{3}'},~\rho_{E^{c}_1}=\mathbf{1'},~\rho_{E^{c}_2}=\mathbf{\hat{1}},~\rho_{E^{c}_3}=\mathbf{1'}\,.
\end{cases}
\end{equation}
The superpotential for the charged lepton masses takes the following form,
\begin{eqnarray}
\nonumber && \mathcal{W}_e = \alpha(E^c_1LY^{(1)}_{\mathbf{\hat{3}'}})_\mathbf{1}H_d+\beta (E^c_2L Y^{(2)}_{\mathbf{3}})_\mathbf{1}H_d+ \gamma(E^c_3LY^{(3)}_{\mathbf{\hat{3}'}})_\mathbf{1}H_d \\
\nonumber&&~~~~ =\alpha E^c_1(L_1 Y_1+L_3Y_2+L_2Y_3)H_d+\beta E^c_2(L_1 Y^{(2)}_3+L_3Y^{(2)}_4+L_2Y^{(2)}_5)H_d\\
&&~~~~~~ +\gamma E^c_3(L_1 Y^{(3)}_5+L_3Y^{(3)}_6+L_2Y^{(3)}_7)H_d\,.
\label{eq:WeII}
\end{eqnarray}
Modular invariance imposes the following constraints on modular weights,
\begin{equation}
k_{E_1}=k_{E_2}-1=k_{E_3}-2=1-k_L\,.
\end{equation}
\item[\textbf{(iii)}]{$f_{E_1}(Y) = Y^{(1)}_\mathbf{\hat{3}'}$, $f_{E_2}(Y) = Y^{(3)}_\mathbf{\hat{3}'}$, $f_{E_3}(Y) = Y^{(3)}_\mathbf{\hat{3}}$}\\
Similar to previous cases, the lepton fields can be assigned to
\begin{equation}
\begin{cases}
\rho_L=\mathbf{3},~\rho_{E^{c}_1}=\mathbf{\hat{1}},~\rho_{E^{c}_2}=\mathbf{\hat{1}},~\rho_{E^{c}_3}=\mathbf{\hat{1}'}\,, \\
\rho_L=\mathbf{3'},~\rho_{E^{c}_1}=\mathbf{\hat{1}'},~\rho_{E^{c}_2}=\mathbf{\hat{1}'},~\rho_{E^{c}_3}=\mathbf{\hat{1}}\,, \\
\rho_L=\mathbf{\hat{3}},~\rho_{E^{c}_1}=\mathbf{1},~\rho_{E^{c}_2}=\mathbf{1},~\rho_{E^{c}_3}=\mathbf{1'}\,, \\
\rho_L=\mathbf{\hat{3}'},~\rho_{E^{c}_1}=\mathbf{1'},~\rho_{E^{c}_2}=\mathbf{1'},~\rho_{E^{c}_3}=\mathbf{1}\,.
\end{cases}
\end{equation}
The superpotential for the charged lepton masses is of the form,
\begin{eqnarray}
\nonumber && \mathcal{W}_e = \alpha(E^c_1LY^{(1)}_{\mathbf{\hat{3}'}})_\mathbf{1}H_d+\beta (E^c_2L Y^{(3)}_{\mathbf{\hat{3}'}})_\mathbf{1}H_d+ \gamma(E^c_3LY^{(3)}_{\mathbf{\hat{3}}})_\mathbf{1}H_d \\
\nonumber&&~~~~ =\alpha E^c_1(L_1 Y_1+L_3Y_2+L_2Y_3)H_d+\beta E^c_2(L_1 Y^{(3)}_5+L_3Y^{(3)}_6+L_2Y^{(3)}_7)H_d\\
&&~~~~~~~~ +\gamma E^c_3(L_1 Y^{(3)}_2+L_3Y^{(3)}_3+L_2Y^{(3)}_4)H_d\,,
\label{eq:WeIII}
\end{eqnarray}
with the modular weights
\begin{equation}
k_{E_1}=k_{E_2}-2=k_{E_3}-2=1-k_L\,.
\end{equation}
\item[\textbf{(iv)}]{$f_{E_1}(Y) = Y^{(3)}_\mathbf{\hat{3}'}$, $f_{E_2}(Y) = Y^{(2)}_\mathbf{3}$, $f_{E_3}(Y) = Y^{(3)}_\mathbf{\hat{3}}$}\\
Likewise, we have four different representation assignments which gives the same superpotential $\mathcal{W}_e$ as well as the same charged lepton mass matrix,
\begin{equation}
\begin{cases}
\rho_L=\mathbf{3},~\rho_{E^{c}_1}=\mathbf{\hat{1}},~\rho_{E^{c}_2}=\mathbf{1},~\rho_{E^{c}_3}=\mathbf{\hat{1}'}\,, \\
\rho_L=\mathbf{3'},~\rho_{E^{c}_1}=\mathbf{\hat{1}'},~\rho_{E^{c}_2}=\mathbf{1'},~\rho_{E^{c}_3}=\mathbf{\hat{1}}\,, \\
\rho_L=\mathbf{\hat{3}},~\rho_{E^{c}_1}=\mathbf{1},~\rho_{E^{c}_2}=\mathbf{\hat{1}'},~\rho_{E^{c}_3}=\mathbf{1'}\,, \\
\rho_L=\mathbf{\hat{3}'},~\rho_{E^{c}_1}=\mathbf{1'},~\rho_{E^{c}_2}=\mathbf{\hat{1}},~\rho_{E^{c}_3}=\mathbf{1}\,.
\end{cases}
\end{equation}
The superpotential for the charged lepton masses reads as,
\begin{eqnarray}
\nonumber && \mathcal{W}_e = \alpha(E^c_1LY^{(3)}_{\mathbf{\hat{3}'}})_\mathbf{1}H_d+\beta (E^c_2L Y^{(2)}_{\mathbf{3}})_\mathbf{1}H_d+ \gamma(E^c_3LY^{(3)}_{\mathbf{\hat{3}}})_\mathbf{1}H_d \\
\nonumber&&~~~~ =\alpha E^c_1(L_1 Y^{(3)}_5+L_3Y^{(3)}_6+L_2Y^{(3)}_7)H_d+\beta E^c_2(L_1 Y^{(2)}_3+L_3Y^{(2)}_4+L_2Y^{(2)}_5)H_d\\
&&~~~~~~~ +\gamma E^c_3(L_1 Y^{(3)}_2+L_3Y^{(3)}_3+L_2Y^{(3)}_4)H_d\,.
\label{eq:WeIV}
\end{eqnarray}
The modular weights $k_L$ and $k_{E_1, E_2, E_3}$ satisfy the constraints
\begin{equation}
k_{E_1}-1=k_{E_2}=k_{E_3}-1=2-k_L\,.
\end{equation}
\end{itemize}
It is straightforward to read out the predicted charged lepton mass matrix for each case discussed above, and results are summarized in table~\ref{tab:charged lepton}. We can exchange the assignments for the right-handed charged lepton fields $E^{c}_{1,2,3}$, accordingly the rows of the charged lepton mass matrix would be permutated. However, the hermitian combination $M^{\dagger}_eM_e$ are left invariant such that the predictions for charged lepton mass and the unitary rotation $U_e$ are unchanged, where $U_e$ diagonalize the charged lepton mass matrix via $U^{\dagger}_eM^{\dagger}_eM_eU_e=\text{diag}(m^2_e, m^2_{\mu}, m^2_{\tau})$.

\begin{table}[th!]
\renewcommand{\tabcolsep}{0.58mm}
\centering
\resizebox{0.95\textwidth}{!}{
\begin{tabular}{|c|c|c|c|} \hline\hline
\multirow{2}{*}{\texttt{Cases}} ~&~ \texttt{rep assignments} ~&~ \texttt{weights} &\multirow{2}{*}{\texttt{Charged lepton mass matrix}}\\
 ~&~ $(\rho_L,\rho_{E^c_1},\rho_{E^c_2},\rho_{E^c_3})$ ~&~ $k_L+k_{E^c_{1,2,3}}$ &  \\ \hline
 & & & \\[-0.15in]
$C_1$ & $\begin{cases}
(\mathbf{3},\mathbf{\hat{1}},\mathbf{1},\mathbf{\hat{1}'}) \\
(\mathbf{3'},\mathbf{\hat{1}'},\mathbf{1'},\mathbf{\hat{1}}) \\
(\mathbf{\hat{3}},\mathbf{1},\mathbf{\hat{1}'},\mathbf{1'}) \\
(\mathbf{\hat{3}'},\mathbf{1'},\mathbf{\hat{1}},\mathbf{1})\end{cases}$
& $(1,2,3)$ &~~ $ M_e =\begin{pmatrix}
 \alpha\,Y_1 ~&~ \alpha\,Y_3 ~&~ \alpha\,Y_2 \\
\beta Y^{(2)}_3 ~&~ \beta Y^{(2)}_5  ~&~ \beta Y^{(2)}_4 \\
 \gamma Y^{(3)}_2~&~ \gamma Y^{(3)}_4~&~\gamma Y^{(3)}_3\\
 \end{pmatrix} v_d $~~ \\ \hline
 & & & \\[-0.15in]
 $C_2$ & $\begin{cases}
(\mathbf{3},\mathbf{\hat{1}},\mathbf{1},\mathbf{\hat{1}})\\
(\mathbf{3'},\mathbf{\hat{1}'},\mathbf{1'},\mathbf{\hat{1}'}) \\
(\mathbf{\hat{3}},\mathbf{1},\mathbf{\hat{1}'},\mathbf{1}) \\
(\mathbf{\hat{3}'},\mathbf{1'},\mathbf{\hat{1}},\mathbf{1'})\end{cases}$
& $(1,2,3)$ &~~ $ M_e =\begin{pmatrix}
 \alpha\,Y_1 ~&~ \alpha\,Y_3 ~&~ \alpha\,Y_2 \\
\beta Y^{(2)}_3 ~&~ \beta Y^{(2)}_5  ~&~ \beta Y^{(2)}_4 \\
 \gamma Y^{(3)}_5~&~ \gamma Y^{(3)}_7~&~\gamma Y^{(3)}_6\\
 \end{pmatrix} v_d $~~ \\ \hline
 & & & \\[-0.15in]
$C_3$ & $\begin{cases}
(\mathbf{3},\mathbf{\hat{1}},\mathbf{\hat{1}},\mathbf{\hat{1}'}) \\
(\mathbf{3'},\mathbf{\hat{1}'},\mathbf{\hat{1}'},\mathbf{\hat{1}}) \\
(\mathbf{\hat{3}},\mathbf{1},\mathbf{1},\mathbf{1'}) \\
(\mathbf{\hat{3}'},\mathbf{1'},\mathbf{1'},\mathbf{1})\end{cases}$
& $(1,3,3)$ &~~ $ M_e =\begin{pmatrix}
 \alpha\,Y_1 ~&~ \alpha\,Y_3 ~&~ \alpha\,Y_2 \\
\beta Y^{(3)}_5 ~&~ \beta Y^{(3)}_7  ~&~ \beta Y^{(3)}_6 \\
 \gamma Y^{(3)}_2~&~ \gamma Y^{(3)}_4~&~\gamma Y^{(3)}_3\\
 \end{pmatrix} v_d $~~ \\ \hline
 & & & \\[-0.15in]
$C_4$ & $\begin{cases}
(\mathbf{3},\mathbf{\hat{1}},\mathbf{1},\mathbf{\hat{1}'}) \\
(\mathbf{3'},\mathbf{\hat{1}'},\mathbf{1'},\mathbf{\hat{1}}) \\
(\mathbf{\hat{3}},\mathbf{1},\mathbf{\hat{1}'},\mathbf{1'}) \\
(\mathbf{\hat{3}'},\mathbf{1'},\mathbf{\hat{1}},\mathbf{1})\end{cases}$
& $(3,2,3)$ &~~ $ M_e =\begin{pmatrix}
 \alpha\,Y^{(3)}_5 ~&~ \alpha\,Y^{(3)}_7 ~&~ \alpha\,Y^{(3)}_6 \\
\beta Y^{(2)}_3 ~&~ \beta Y^{(2)}_5  ~&~ \beta Y^{(2)}_4 \\
 \gamma Y^{(3)}_2~&~ \gamma Y^{(3)}_4~&~\gamma Y^{(3)}_3\\
 \end{pmatrix} v_d $~~ \\ \hline\hline
\end{tabular} }
\caption{\label{tab:charged lepton} The modular $S'_4$ models in charged lepton sector for different weight and representation assignments, where the charged lepton mass matrix $M_e$ is given in the convention $E^c\,M_e\,L$ with $ v_d =\langle H^0_d\rangle$.  }
\end{table}

\subsection{\label{subsec:neutrino}Neutrino sector}
In neutrino sector, we assume that neutrinos are Majorana particles,
and we consider two scenarios that the neutrino masses are described
by the effective Weinberg operator or arise from the type-I seesaw
mechanism. The left-handed lepton doublets would be assigned to
transform as triplet under $S'_4$. Guided by the principle of minimality and simplicity, we shall consider  modular multiplets with weight less than four similar to the charged lepton sector. For the cases involving higher weight modular forms, more modular invariant operators accompanied by free coupling constants would be allowed, and the predictive power of the models would be reduced.

\subsubsection{Weinberg operator}

From the $S'_4$ Kronecker products $\mathbf{3} \otimes \mathbf{3} = \mathbf{3'} \otimes \mathbf{3'} = \mathbf{1} \oplus \mathbf{2} \oplus \mathbf{3} \oplus \mathbf{3'}$,\, $\mathbf{\hat{3}} \otimes \mathbf{\hat{3}} = \mathbf{\hat{3}'} \otimes \mathbf{\hat{3}'} = \mathbf{1'} \oplus \mathbf{2} \oplus \mathbf{3} \oplus \mathbf{3'}$, we know that the operator $LLH_uH_u$ can not couple with odd weight modular forms such as $Y^{(1)}_\mathbf{\hat{3}'}$, $Y^{(3)}_{\mathbf{\hat{1}'}}$, $Y^{(3)}_{\mathbf{\hat{3}}}$, $Y^{(3)}_{\mathbf{\hat{3}'}}$ to form a $S'_4$ singlet. At the lowest order, the weight $2$ modular multiplets $Y^{(2)}_\mathbf{2}$, $Y^{(2)}_\mathbf{3}$ enter into the Weinberg operator, and the superpotential for neutrino masses are as follows.
\begin{itemize}
\item{$\rho_L=\mathbf{3}$ or $\mathbf{3'}$}
\begin{eqnarray}
\nonumber&& \mathcal{W}_\nu =  \frac{g_1}{\Lambda}((L L)_\mathbf{2}Y^{(2)}_{\mathbf{2}})_\mathbf{1}H_u H_u +\frac{g_2}{\Lambda}((L L)_\mathbf{3}Y^{(2)}_\mathbf{3})_\mathbf{1}H_u H_u \\
&&~~~~ =\big[g_1(2L_1L_2+L^2_3)Y^{(2)}_1+g_1(2L_1L_3+L^2_2)Y^{(2)}_2\big]\frac{H^2_u}{\Lambda}\,.
\label{eq:Wnu_w1}
\end{eqnarray}
The modular weight $k_L$ should be equal to 1, i.e. $k_L=1$. From the CG coefficients of $\mathbf{3}\otimes\mathbf{3}\rightarrow\mathbf{3}$ and $\mathbf{3'}\otimes\mathbf{3'}\rightarrow\mathbf{3}$, we know that the contraction $(L L)_\mathbf{3}$ is a antisymmetric combination of lepton fields $L$, while Lorentz invariance requires that the Majorana mass term $((L L)_\mathbf{3}Y^{(2)}_\mathbf{3})_{\mathbf{1}}H_u H_u$ should be symmetric with respect to $L$. As a result, the term proportional to $g_2$ is vanishing, and the corresponding neutrino mass matrix $M_\nu$ read as
\begin{equation}
\label{eq:Mnu_1}
M_\nu =g_1 \begin{pmatrix}
 0 ~&~ Y^{(2)}_1 ~&~ Y^{(2)}_2 \\
 Y^{(2)}_1 ~&~ Y^{(2)}_2  ~&~ 0 \\
 Y^{(2)}_2~&~ 0~&~ Y^{(2)}_1\\
 \end{pmatrix} \dfrac{v^2_u}{\Lambda}\,.
\end{equation}
where $ v_u =\langle H^0_u\rangle$
\item{$\rho_L=\mathbf{\hat{3}}$ or $\mathbf{\hat{3}'}$}
\begin{eqnarray}
\nonumber&& \mathcal{W}_\nu =  \frac{g_1}{\Lambda}((L L)_\mathbf{2}Y^{(2)}_\mathbf{2})_\mathbf{1}H_u H_u+\frac{g_2}{\Lambda}((L L)_\mathbf{3}Y^{(2)}_\mathbf{3})_\mathbf{1}H_u H_u\,\\
\nonumber&&~~~~ =\big[g_1(2L_1L_2+L^2_3)Y^{(2)}_1-g_1(2L_1L_3+L^2_2)Y^{(2)}_2 + g_2(2L^2_1-2L_2L_3)Y^{(2)}_3 \\
&&~~~~~~~~ +g_2(2L^2_2-2L_1L_3)Y^{(2)}_4+g_2(2L^2_3-2L_1L_2)Y^{(2)}_5\big]\frac{H^2_u}{\Lambda}\,,
\label{eq:Wnu_w2}
\end{eqnarray}
with the weight $k_L=1$. The light neutrino mass matrix $M_\nu$ is of the form
\begin{equation}
\label{eq:Mnu_2}
M_\nu =\begin{pmatrix}
 2g_2Y^{(2)}_3 ~& g_1Y^{(2)}_1-g_2Y^{(2)}_5 ~& -g_1Y^{(2)}_2-g_2Y^{(2)}_4 \\
 g_1Y^{(2)}_1-g_2Y^{(2)}_5~ & -g_1Y^{(2)}_2+2g_2Y^{(2)}_4 ~& -g_2Y^{(2)}_3 \\
 -g_1Y^{(2)}_2-g_2Y^{(2)}_4 ~& -g_2Y^{(2)}_3 ~& g_1Y^{(2)}_1+2g_2Y^{(2)}_5\\
 \end{pmatrix} \dfrac{v^2_u}{\Lambda}\,.
\end{equation}
\end{itemize}

\subsubsection{Type-I seesaw mechanism}

Three generations of right-handed neutrinos are introduced in the present work and they are assumed to transforms as triplet under the $S'_4$. Then the most general superpotential in neutrino sector can be written as
\begin{equation}
\mathcal{W}_\nu = g\left(N^cLH_uf_D(Y)\right)_\mathbf{1}
+\Lambda\left(N^cN^cf_M(Y)\right)_\mathbf{1}\,,
\label{eq:WnuII}
\end{equation}
where $f_N(Y)$ and $f_M(Y)$ are modular multiplets. Similar to the case of Weinberg operator, from the Kronecker products of two triplets, we know that $f_M(Y)$ can be $\tau-$independent constant\footnote{There are no non-trivial modular forms of weight zero.} or weight 2 modular form.
\begin{itemize}
\item{$f_M(Y) = 1$}\\
In this case, the right-handed neutrinos can transform as $\mathbf{3}$ or $\mathbf{3'}$ under $S'_4$ (i.e., $\rho_{N^c}=\mathbf{3}$ or $\mathbf{3'}$) and their modular weight should be vanishing with $k_{N^c}=0$. The heavy neutrino mass term is
\begin{eqnarray}
\label{eq:WN_1}
\mathcal{W}_N=\Lambda(N^c N^c)_\mathbf{1}=\Lambda(N^c_1N^c_1+N^c_2N^c_3+N^c_3N^c_2)\,,
\end{eqnarray}
which leads to the following heavy neutrino mass matrix,
\begin{equation}
\label{eq:MN_1}
M_N = \begin{pmatrix}
1 ~&~ 0 ~&~ 0 \\
0 ~&~ 0 ~&~ 1  \\
0 ~&~ 1 ~&~ 0
\end{pmatrix} \Lambda\,.
\end{equation}
\item{$f_M(Y) = Y^{(2)}_\mathbf{2},\,Y^{(2)}_\mathbf{3}$}\\
If the right-handed neutrinos are assigned to transform as unhatted triplet $\rho_{N^c}=\mathbf{3}$ or $\mathbf{3'}$ with $k_{N^c}=1$, we have
\begin{eqnarray}
\label{eq:WN_2}
\nonumber&& \mathcal{W}_N=\Lambda((N^c N^c)_\mathbf{2}Y^{(2)}_\mathbf{2})_\mathbf{1} +\Lambda'((N^c N^c)_\mathbf{3}Y^{(2)}_\mathbf{3})_\mathbf{1}\\
&&~~~~ = \Lambda\big[(2N^c_1N^c_2+N^c_3N^c_3)Y^{(2)}_1+(2N^c_1N^c_3+N^c_2N^c_2)Y^{(2)}_2\big]\,.
\end{eqnarray}
Notice that the term $[(N^c N^c)_\mathbf{3}Y^{(2)}_\mathbf{3}]_\mathbf{1}$ is vanishing because the contractions for both $\mathbf{3}\otimes\mathbf{3}\rightarrow\mathbf{3}$ and $\mathbf{3'}\otimes\mathbf{3'}\rightarrow\mathbf{3}$ are antisymmetric combinations. The corresponding heavy Majorana mass matrix $M_N$ can be easily read out as
\begin{equation}
\label{eq:MN_2}
M_N = \begin{pmatrix}
0 ~&~ Y^{(2)}_1 ~&~ Y^{(2)}_2 \\
Y^{(2)}_1 ~&~ Y^{(2)}_2 ~&~ 0  \\
Y^{(2)}_2 ~&~ 0 ~&~ Y^{(2)}_1
\end{pmatrix} \Lambda\,.
\end{equation}
On the other hand, we can also assign the right-handed neutrinos to hatted triplets $\rho_{N^c}=\mathbf{\hat{3}}$ or $\mathbf{\hat{3}'}$ with $k_{N^c}=1$. Then the superpotential $\mathcal{W}_{N}$ is
\begin{eqnarray}
\label{eq:WN_3}
\nonumber \mathcal{W}_N&=&\Lambda[(N^c N^c)_\mathbf{2}Y^{(2)}_\mathbf{2}]_\mathbf{1} +\Lambda'[(N^c N^c)_\mathbf{3}Y^{(2)}_\mathbf{3}]_\mathbf{1}\\
\nonumber &=& \Lambda\big[(2N^c_1N^c_2+N^c_3N^c_3)Y^{(2)}_1-(2N^c_1N^c_3+N^c_2N^c_2)Y^{(2)}_2] + \Lambda'[(2N^c_1N^c_1-2N^c_2N^c_3)Y^{(2)}_3 \\
&&~~ +(2N^c_2N^c_2-2N^c_1N^c_3)Y^{(2)}_4+(2N^c_3N^c_3-2N^c_1N^c_2)Y^{(2)}_5\big]\,,
\end{eqnarray}
which gives rise to
\begin{equation}
\label{eq:MN_3}
M_N=\begin{pmatrix}
 2\Lambda' Y^{(2)}_3 ~ & \Lambda Y^{(2)}_1-\Lambda' Y^{(2)}_5 ~ & -\Lambda Y^{(2)}_2-\Lambda' Y^{(2)}_4 \\
 \Lambda Y^{(2)}_1-\Lambda' Y^{(2)}_5 ~ & -\Lambda Y^{(2)}_2+2\Lambda' Y^{(2)}_4 ~ & -\Lambda' Y^{(2)}_3 \\
 -\Lambda Y^{(2)}_2-\Lambda' Y^{(2)}_4 ~ & -\Lambda' Y^{(2)}_3 ~ & \Lambda Y^{(2)}_1+2\Lambda' Y^{(2)}_5 \\
\end{pmatrix}\,.
\end{equation}
\end{itemize}
Now we proceed to discuss the neutrino Yukawa interaction term $g\left(N^cLH_uf_D\left(Y\right)\right)_\mathbf{1}$. The modular form $f_{D}(Y)$ is fixed by the assignments for $L$ and $N^c$, it can  $1$, $Y^{(1)}_\mathbf{\hat{3}'}$, $Y^{(2)}_\mathbf{2}$, $Y^{(2)}_\mathbf{3}$, $Y^{(3)}_\mathbf{\hat{1}'}$, $Y^{(3)}_\mathbf{\hat{3}}$ and $Y^{(3)}_\mathbf{\hat{3}'}$ up to weight 3. We shall report the predictions for Dirac neutrino mass matrix for each possible cases.

\begin{itemize}[labelindent=-1.5em, leftmargin=1.6em]
\item[\textbf{(i)}]{$f_D(Y) = 1$}\\
In this case, left-handed lepton doublet $L$ and right-handed neutrinos $N$ contract to a singlet, hence their assignments can be $(\rho_{N^c},\,\rho_L)=(\mathbf{3},\,\mathbf{3})$ or $(\mathbf{3'},\,\mathbf{3'})$ or $(\mathbf{\hat{3}},\,\mathbf{\hat{3}'})$ or $(\mathbf{\hat{3}'},\,\mathbf{\hat{3}})$. The Dirac neutrino mass term is
\begin{eqnarray}
\label{eq:WD_1}
\mathcal{W}_D= g(N^cL)_\mathbf{1}H_u =g(L_1N^c_1 + L_2N^c_3+ L_3N^c_2)H_u \,,
\end{eqnarray}
with $k_{N^c}+k_L=0$. Consequently the Dirac neutrino mass matrix read as
\begin{equation}
\label{eq:MD_1}
M_D =g\begin{pmatrix}
 1 ~& 0 ~& 0 \\
 0 ~& 0 ~& 1 \\
 0 ~& 1 ~& 0
 \end{pmatrix}v_u\,.
\end{equation}
\item[\textbf{(ii)}]{$f_D(Y) = Y^{(1)}_\mathbf{\hat{3}'}$}\\
There are eight possible assignments for $\rho_L$ and $\rho_{N^c}$, and they can divided into two categories. In the case of $(\rho_{N^c},\,\rho_L)=(\mathbf{3},\,\mathbf{\hat{3}})$, $(\mathbf{3'},\,\mathbf{\hat{3}'})$ or $(\mathbf{\hat{3}},\,\mathbf{3})$ , $(\mathbf{\hat{3}'},\,\mathbf{3'})$, we have
\begin{eqnarray}
\label{eq:WD_2}
\nonumber&& \mathcal{W}_D= g\Big((N^cL)_\mathbf{\hat{3}}Y^{(1)}_\mathbf{\hat{3}'}\Big)_\mathbf{1}H_u \\
&&~~~~ = g[(L_2N^c_3 - L_3N^c_2)Y_1+(L_3N^c_1 - L_1N^c_3)Y_2 +(L_1N^c_2 - L_2N^c_1)Y_3 ]H_u \,,
\end{eqnarray}
with the modular weights $k_{N^c}+k_L=1$. We can read out the Dirac neutrino mass matrix is
\begin{equation}
\label{eq:MD_2}
M_D =g\begin{pmatrix}
 0 ~& -Y_3 ~& Y_2 \\
 Y_3 ~& 0 ~& -Y_1\\
 -Y_2 ~& Y_1 ~& 0 \\
 \end{pmatrix}v_u\,.
\end{equation}
For the second type of assignments $(\rho_{N^c},\,\rho_L)=(\mathbf{3},\,\mathbf{\hat{3}'})$, $(\mathbf{3'},\,\mathbf{\hat{3}})$, $(\mathbf{\hat{3}'},\,\mathbf{3})$ or $(\mathbf{\hat{3}},\,\mathbf{3'})$ with $k_{N^c}+k_L=1$, we find
\begin{eqnarray}
\label{eq:WD_3}
\nonumber \mathcal{W}_D&=& g\Big((N^cL)_\mathbf{\hat{3}}Y^{(1)}_\mathbf{\hat{3}'}\Big)_\mathbf{1}H_u \\
\nonumber &=&g[(2L_1N^c_1-L_2N^c_3 - L_3N^c_2)Y_1+(2L_2N^c_2-L_1N^c_3 - L_3N^c_1)Y_2 \\
&& +(2L_3N^c_3-L_1N^c_2 - L_2N^c_1)Y_3 ]H_u \,,
\end{eqnarray}
which leads to
\begin{equation}
\label{eq:MD_3}
M_D =g\begin{pmatrix}
 2Y_1 & -Y_3 & -Y_2 \\
 -Y_3 & 2Y_2 & -Y_1\\
 -Y_2 & -Y_1 & 2Y_3
 \end{pmatrix}v_u \,.
\end{equation}

\item[\textbf{(iii)}]{$f_D(Y) = Y^{(2)}_\mathbf{2},\,Y^{(2)}_\mathbf{3}$}\\
The modular weights of $L$ and $N^c$ should compensate that of $f_D(Y)$, they satisfy the condition $k_{N^c}+k_L=2$. For the assignments $(\rho_{N^c},\,\rho_L)=(\mathbf{3},\,\mathbf{3})$ , $(\mathbf{3'},\,\mathbf{3'})$ , $(\mathbf{\hat{3}},\,\mathbf{\hat{3}'})$ or $(\mathbf{\hat{3}'},\,\mathbf{\hat{3}})$, we have
\begin{eqnarray}
\label{eq:WD_4}
\nonumber \mathcal{W}_D&=& \Big(g_1(N^cL)_\mathbf{2}Y^{(2)}_\mathbf{2}\Big)_\mathbf{1}H_u+g_2\Big((N^cL)_\mathbf{3}Y^{(2)}_\mathbf{3}\Big)_\mathbf{1}H_u \\
\nonumber &=&g_1[(L_2N^c_1 + L_1N^c_2+ L_3N^c_3)Y^{(2)}_1+(L_3N^c_1 + L_1N^c_3 + L_2N^c_2)Y^{(2)}_2]H_u \\
&&+ g_2[(L_2N^c_3-L_3N^c_2)Y^{(2)}_3+(L_3N^c_1-L_1N^c_3)Y^{(2)}_4+(L_1N^c_2- L_2N^c_1)Y^{(2)}_5]H_u\,.
\end{eqnarray}
Accordingly the Dirac neutrino mass matrix is of the following form
\begin{equation}
\label{eq:MD_4}
M_D =\begin{pmatrix}
 0 ~& g_1Y^{(2)}_1-g_2Y^{(2)}_5 ~& g_1Y^{(2)}_2+ g_2Y^{(2)}_4 \\
 g_1Y^{(2)}_1+g_2Y^{(2)}_5 ~& g_1Y^{(2)}_2 ~& -g_2Y^{(2)}_3 \\
 g_1Y^{(2)}_2-g_2Y^{(2)}_4 ~& g_2Y^{(2)}_3 ~& g_1Y^{(2)}_1 \\
 \end{pmatrix}v_u\,.
\end{equation}
We can also assign $N^c$ and $L$ to the $S'_4$ triplets  $(\rho_{N^c},\,\rho_L)=(\mathbf{\hat{3}},\, \mathbf{\hat{3}})$, $(\mathbf{\hat{3}'},\,\mathbf{\hat{3}'})$ , $(\mathbf{3},\,\mathbf{3'})$ or $(\mathbf{3'},\,\mathbf{3})$, and thus
\begin{eqnarray}
\label{eq:WD_5}
\nonumber \mathcal{W}_D&=& \Big(g_1(N^cL)_\mathbf{2}Y^{(2)}_\mathbf{2}\Big)_\mathbf{1}H_u +g_2\Big((N^cL)_\mathbf{3}Y^{(2)}_\mathbf{3}\Big)_\mathbf{1}H_u \\
\nonumber &=&g_1[(L_2N^c_1 + L_1N^c_2+ L_3N^c_3)Y^{(2)}_1-(L_3N^c_1 + L_1N^c_3 + L_2N^c_2)Y^{(2)}_2] H_u\\
\nonumber&&+g_2[(2L_1N^c_1-L_2N^c_3-L_3N^c_2)Y^{(2)}_3+(2L_2N^c_2-L_3N^c_1-L_1N^c_3)Y^{(2)}_4 \\
&&+(2L_3N^c_3-L_1N^c_2- L_2N^c_1)Y^{(2)}_5]H_u \,.
\end{eqnarray}
The Dirac neutrino mass matrix read as
\begin{equation}
\label{eq:MD_5}
M_D =\begin{pmatrix}
 2g_2Y^{(2)}_3 ~& g_1Y^{(2)}_1-g_2Y^{(2)}_5
 ~& -g_1Y^{(2)}_2-g_2Y^{(2)}_4 \\
 g_1Y^{(2)}_1-g_2Y^{(2)}_5 ~& -g_1Y^{(2)}_2+2g_2Y^{(2)}_4 ~& -g_2Y^{(2)}_3 \\
 -g_1Y^{(2)}_2-g_2Y^{(2)}_4 ~& -g_2Y^{(2)}_3 ~& g_1Y^{(2)}_1+2g_2Y^{(2)}_5 \\
 \end{pmatrix}v_u\,.
\end{equation}

\item[\textbf{(iv)}]{$f_D(Y) = Y^{(3)}_\mathbf{\hat{1}'},\, Y^{(3)}_\mathbf{\hat{3}},\,Y^{(3)}_\mathbf{\hat{3}'}$}\\
The weight cancellation requires $k_{L}$ and $k_{N^c}$ fulfill the condition $k_{N^c}+k_L=3$. Invariance of the neutrino Yukawa coupling under $S'_4$ entails $N^c$ and $L$ should contract to $\mathbf{\hat{1}}$, $\mathbf{\hat{3}}$ and $\mathbf{\hat{3}'}$. Therefore $N^c$ and $L$ can be assigned to $(\rho_{N^c},\,\rho_L)=(\mathbf{3},\,\mathbf{\hat{3}})$, $(\mathbf{3'},\,\mathbf{\hat{3}'})$, $(\mathbf{\hat{3}},\,\mathbf{3})$ or $(\mathbf{\hat{3}'},\,\mathbf{3'})$, then the superpotential $\mathcal{W}_D$ is of the form
\begin{eqnarray}
\label{eq:WD_6}
\nonumber \mathcal{W}_D&=& g_1\Big((N^cL)_\mathbf{\hat{3}}Y^{(3)}_\mathbf{\hat{3}'}\Big)_\mathbf{1} H_u + g_2\Big((N^cL)_\mathbf{\hat{3}'}Y^{(3)}_\mathbf{\hat{3}}\Big)_\mathbf{1} H_u+g_3\Big((N^cL)_\mathbf{\hat{1}}Y^{(3)}_\mathbf{\hat{1}'} \Big)_\mathbf{1} H_u \\
\nonumber &=&\pm g_1[(L_2N^c_3 - L_3N^c_2)Y^{(3)}_5+(L_3N^c_1 - L_1N^c_3)Y^{(3)}_6 +(L_1N^c_2 - L_2N^c_1)Y^{(3)}_7 ]H_u\\
\nonumber&&+g_2[(2L_1N^c_1-L_2N^c_3-L_3N^c_2)Y^{(3)}_2+(2L_2N^c_2-L_1N^c_3-L_3N^c_1)Y^{(3)}_3 \\
&& +(2L_3N^c_3-L_1N^c_2-L_2N^c_1)Y^{(3)}_4]H_u+g_3[L_1N^c_1+L_2N^c_3+L_3N^c_2]H_u\,,
\end{eqnarray}
which gives rise to
\begin{eqnarray}
\label{eq:MD_6}
 M_D=\begin{pmatrix}
 2g_2Y^{(3)}_2 + g_3Y^{(3)}_1 ~& -g_1Y^{(3)}_7-g_2Y^{(3)}_4 ~& g_1Y^{(3)}_6-g_2Y^{(3)}_3 \\
 g_1Y^{(3)}_7-g_2Y^{(3)}_4 ~& 2g_2Y^{(3)}_3 ~& -g_1Y^{(3)}_5-g_2Y^{(3)}_2 + g_3Y^{(3)}_1 \\
 -g_1Y^{(3)}_6-g_2Y^{(3)}_3 ~& g_1Y^{(3)}_5-g_2Y^{(3)}_2 + g_3Y^{(3)}_1 ~& 2g_2Y^{(3)}_4\\
 \end{pmatrix}v_u\,.
\end{eqnarray}
We can also assign $N^c$ and $L$ to transform as $(\rho_{N^c},\,\rho_L)=(\mathbf{3},\,\mathbf{\hat{3}'})$, $(\mathbf{3'},\,\mathbf{\hat{3}})$, $(\mathbf{\hat{3}'},\,\mathbf{3})$ or $(\mathbf{\hat{3}},\,\mathbf{3'})$, then we have
\begin{eqnarray}
\label{eq:WD_7}
\nonumber \mathcal{W}_D&=& g_1\Big((N^cL)_\mathbf{\hat{3}}Y^{(3)}_\mathbf{\hat{3}'}\Big)_\mathbf{1}H_u+g_2\Big((N^cL)_\mathbf{\hat{3}'}Y^{(3)}_\mathbf{\hat{3}}\Big)_\mathbf{1}H_u \\
\nonumber&=&g_1[(2L_1N^c_1-L_2N^c_3-L_3N^c_2)Y^{(3)}_5+(2L_2N^c_2-L_1N^c_3-L_3N^c_1)Y^{(3)}_6 \\
\nonumber&&+(2L_3N^c_3-L_1N^c_2-L_2N^c_1)Y^{(3)}_7]H_u + g_2[(L_2N^c_3 - L_3N^c_2)Y^{(3)}_2 \\
&&+(L_3N^c_1 - L_1N^c_3)Y^{(3)}_3 +(L_1N^c_2 - L_2N^c_1)Y^{(3)}_4 ]H_u \,.
\end{eqnarray}
The Dirac neutrino mass matrix is determined to be
\begin{equation}
\label{eq:MD_7}
M_D =\begin{pmatrix}
 2g_1Y^{(3)}_5 & -g_1Y^{(3)}_7-g_2Y^{(3)}_4 & -g_1Y^{(3)}_6+g_2Y^{(3)}_3 \\
 -g_1Y^{(3)}_7+g_2Y^{(3)}_4 & 2g_1Y^{(3)}_6 & -g_1Y^{(3)}_5-g_2Y^{(3)}_2 \\
 -g_1Y^{(3)}_6-g_2Y^{(3)}_3 & -g_1Y^{(3)}_5+g_2Y^{(3)}_2 & 2g_1Y^{(3)}_7\\
 \end{pmatrix}v_u\,.
\end{equation}
\end{itemize}
For all the above type-I seesaw models, the effective light neutrino mass matrix is given by the seesaw formula,
\begin{equation}
\label{eq:see-saw}
M_\nu= -M_D^TM_N^{-1}M_D\,.
\end{equation}
We are interested in the models with less free parameters, and we list the possible neutrino models in table~\ref{tab:neutrino} for which the resulting light neutrino mass matrices contain less than four free parameters excluding the modulus $\tau$.

\begin{table}[t!]
\centering
\resizebox{1.0\textwidth}{!}{
\begin{tabular}{|c||c|c|c|c|c|c|c|c|c|c|c|c|} \hline\hline
\texttt{Cases} & $W_1$ & $W_2$ & $S_1$ & $S_2$ & $S_3$ & $S_4$ \\ \hline
\texttt{Irrep}$(\rho_{N^c},\,\rho_L)$ & $\begin{cases}(-,\mathbf{3})\\(-,\mathbf{3'}) \end{cases}$ & $\begin{cases}(-,\mathbf{\hat{3}})\\(-,\mathbf{\hat{3}'}) \end{cases}$ & \multicolumn{3}{c|}{$\begin{cases}(\mathbf{3},\, \mathbf{3})\\(\mathbf{3'},\,\mathbf{3'}) \end{cases}$ }& $\begin{cases}(\mathbf{\hat{3}},\, \mathbf{\hat{3}'})\\(\mathbf{\hat{3}'},\,\mathbf{\hat{3}}) \end{cases}$    \\ \hline
\texttt{weight}$(k_{N^c},k_L)$ &$(-,1)$ & $(-,1)$ & $(1, -1)$ & $(0, 2)$ & $(1,1)$ & $(1,-1)$ \\ \hline
\texttt{Neutrino} &  \multirow{2}{*}{Eq.~\eqref{eq:Mnu_1}}  & \multirow{2}{*}{Eq.~\eqref{eq:Mnu_2}}   &  Eq.~\eqref{eq:MN_2} & Eq.~\eqref{eq:MN_1} & Eq.~\eqref{eq:MN_2} & Eq.~\eqref{eq:MN_3}
 \\

\texttt{mass matrices} & & & Eq.~\eqref{eq:MD_1} & Eq.~\eqref{eq:MD_4} & Eq.~\eqref{eq:MD_4} & Eq.~\eqref{eq:MD_1}  \\  \hline \hline

\texttt{Cases} & $S_5$ & $S_6$ & $S_7$ & $S_8$ & $S_9$ & $S_{10}$ \\ \hline
\texttt{Irrep}$(\rho_{N^c},\,\rho_L)$ & \multicolumn{2}{c|}{$\begin{cases}(\mathbf{3},\,\mathbf{3'})\\(\mathbf{3'},\,\mathbf{3}) \end{cases}$} & \multicolumn{4}{c|}{$\begin{cases}(\mathbf{3},\, \mathbf{\hat{3}})\\(\mathbf{3'},\,\mathbf{\hat{3}'}) \end{cases}$ } \\ \hline
\texttt{weight}$(k_{N^c},k_L)$ &$(0,2)$ & $(1,1)$ & $(0,1)$ & $(1,0)$ & $(0,3)$ & $(1,2)$ \\ \hline
\texttt{Neutrino}  & Eq.~\eqref{eq:MN_1} & Eq.~\eqref{eq:MN_2}& Eq.~\eqref{eq:MN_1} & Eq.~\eqref{eq:MN_2} & Eq.~\eqref{eq:MN_1} & Eq.~\eqref{eq:MN_2} \\
\texttt{mass matrices} & Eq.~\eqref{eq:MD_5} & Eq.~\eqref{eq:MD_5} & Eq.~\eqref{eq:MD_2} & Eq.~\eqref{eq:MD_2} & Eq.~\eqref{eq:MD_6} & Eq.~\eqref{eq:MD_6}\\  \hline \hline

\texttt{Cases} & $S_{11}$ & $S_{12}$ & $S_{13}$ & $S_{14}$ & $S_{15}$ & $S_{16}$ \\ \hline
\texttt{Irrep}$(\rho_{N^c},\,\rho_L)$ & $\begin{cases}(\mathbf{\hat{3}},\,\mathbf{3})\\(\mathbf{\hat{3}'},\,\mathbf{3'}) \end{cases}$ & \multicolumn{4}{c|}{$\begin{cases}(\mathbf{3},\, \mathbf{\hat{3}'})\\(\mathbf{3'},\,\mathbf{\hat{3}}) \end{cases}$ } & $\begin{cases}(\mathbf{\hat{3}'},\,\mathbf{3})\\(\mathbf{\hat{3}},\,\mathbf{3'}) \end{cases}$ \\ \hline
\texttt{weight}$(k_{N^c},k_L)$ &$(1,0)$ & $(0,1)$ & $(1,0)$ & $(0,3)$ & $(1,2)$ & $(1,0)$ \\ \hline
\texttt{Neutrino} &  Eq.~\eqref{eq:MN_3} & Eq.~\eqref{eq:MN_1}& Eq.~\eqref{eq:MN_2} & Eq.~\eqref{eq:MN_1} & Eq.~\eqref{eq:MN_2} & Eq.~\eqref{eq:MN_3} \\
\texttt{mass matrices} &
Eq.~\eqref{eq:MD_2} & Eq.~\eqref{eq:MD_3} & Eq.~\eqref{eq:MD_3} & Eq.~\eqref{eq:MD_7} & Eq.~\eqref{eq:MD_7} & Eq.~\eqref{eq:MD_3}  \\  \hline \hline
\end{tabular} }
\caption{\label{tab:neutrino} Summary of neutrino models with less than four free parameters excluding $\tau$, $W_{1,2}$ and $S_{i}$ ($i=1,\ldots,16$) denote the models in which neutrino masses arise from Weinberg operator and type-I seesaw mechanism respectively. }
\end{table}

\subsection{\label{subsec:numerical-analysis}Numerical results}

In short, the charged lepton can take four possible forms shown in table~\ref{tab:charged lepton} if only modular forms of weight less than 4 are considered, and there are eighteen neutrino models with parameters less than 4, as summarized in table~\ref{tab:neutrino}. Combining charged lepton sector with neutrino sector, we obtain totally $4\times18=72$ lepton models which are denoted as $C_i$-$W_{1}$, $C_i$-$W_{2}$ and $C_i$-$S_{j}$ with the indices $i=1,\ldots,4$, $j=1,\ldots,16$. We see that for the four cases $C_{1,2,3,4}$ the charged lepton mass matrix $M_e$ depends on three parameters $\alpha$, $\beta$ and $\gamma$ which can be made real by redefining the phases of the right-handed charged leptons $E^c_{1,2,3}$. The three parameters $\alpha$, $\beta$ and $\gamma$ are in one-to-one correspondence with the charged lepton masses. The electron, muon, tau masses can be reproduced by adjusting the parameters $\alpha$, $\beta$ and $\gamma$. We confront each model with the neutrino oscillation data and charged lepton masses, we perform a conventional $\chi^2$ analysis to optimize the model parameters and determine how well each model can be compatible with the observations. The overall mass scale $\alpha v_d$ in the charged lepton mass matrix and $g^2v^2_u/\Lambda$ in the neutrino mass matrix can be fixed by requiring the electron mass and the mass splitting $\Delta m^2_{21}$ are reproduced. Since the overall factor of the mass matrix doesn't affect the predictions for mass ratios, mixing angles and CP violating phases, we construct the $\chi^2$ function using the lepton mixing angles $\theta_{12}$, $\theta_{13}$, $\theta_{23}$ and the mass ratios $m_e/m_{\mu}$, $m_{\mu}/m_{\tau}$, $\Delta m^2_{21}/\Delta m^2_{31}$. The neutrino oscillation parameters are taken from the latest global fit results of NuFIT v4.1 including the atmospheric neutrino data from Super-Kamiokande~\cite{Esteban:2018azc}. Since the current data somewhat prefer normal ordering (NO) over inverted neutrino (IO) mass ordering, we shall focus on NO neutrino masses in the numerical analysis. The best fit values and $1\sigma$ ranges of the three lepton mixing angles CP violating phase $\delta_{CP}$ and the neutrino mass squared differences are as follows
\begin{equation}
\label{eq:lepton-observables}\begin{aligned}
&\sin^2\theta_{12}=0.310^{+0.013}_{-0.012}, ~~~ \sin^2\theta_{13}=0.02237^{+0.00066}_{-0.00065},~~~\sin^2\theta_{23}=0.563^{+0.018}_{-0.024},\\
&\delta^l_{CP}/\pi=1.2278^{+0.2167}_{-0.1556},~~~\frac{\Delta m^2_{21}}{10^{-5}\text{eV}^2}=7.39^{+0.21}_{-0.20},~~~\frac{\Delta m^2_{31}}{10^{-3}\text{eV}^2}=2.528^{+0.029}_{-0.031}\,.
\end{aligned}
\end{equation}
The ratios of charged lepton masses taken from ~\cite{Ross:2007az},
\begin{equation}
m_e/m_\mu =0.0048\pm 0.0002, \quad m_\mu/m_\tau=0.0565 \pm 0.0045
\end{equation}
The leptonic Dirac CP phase $\delta^{l}_{CP}$ is not measured precisely at present and the indication of a preferred value of $\delta^{l}_{CP}$ from global data analyses is rather weak, we don't include the information of $\delta^{l}_{CP}$ in the $\chi^2$ function.

It is an open question to dynamically determine the VEV of the complex modulus $\tau$. It has been conjectured that the VEV of the complex modulus is pure imaginary or along the border of the fundamental domain in the modular invariant $N=1$ supergravity theories~\cite{Cvetic:1991qm}. It has been shown that the complex modulus could possibly be stabilized at some $Z_2$ fixed points in string compactifications~\cite{Kobayashi:2020uaj}. Following the original work~\cite{Feruglio:2017spp,Criado:2018thu}, we will not address the vacuum selection mechanism here and consequently we will not attempt to build the most general supersymmetric and modular invariant scalar potential for $\tau$ in a more fundamental theory. The VEV of $\tau$ will be treated as a free parameter, to be varied to maximize the agreement with data.

The absolute values of all coupling constants are scanned in the region $[0, 10^4]$ and the phases are freely varied in the range $[0, 2\pi]$, and the modulus $\tau$ is restricted in the right-hand part of the fundamental domain $\mathcal{D}$ with $0\leq\texttt{Re}(\tau)\leq0.5$, the reason of why not scanning the complete fundamental domain is explained below. We numerically minimize the $\chi^2$ function by using the minimization algorithms incorporated in the package MINUIT developed by CERN to determine the optimum values of the input parameters. We find that 15 models can give very good fit to the data for certain values of input parameters. We display the best fit values of the input parameter for which the $\chi^2$ function reach a global minimum $\chi^2_{\text{min}}$ in table~\ref{tab:goodmodel_para_obs}, and we also give the predictions for lepton mixing parameters and neutrino masses at the best fitting point in table~\ref{tab:goodmodel_para_obs}. We see that the charged lepton mass hierarchies require hierarchical values of the parameters $\alpha$, $\beta$ and $\gamma$, this can be naturally realized by the weighton mechanism~\cite{King:2020qaj}. For all the 15 phenomenologically viable models, the light neutrino mass matrix $M_{\nu}$ depends on a single complex parameter $g_2/g_1$ and the complex modulus $\tau$ besides the overall scale $g^2_1v^2_u/\Lambda$. Hence the three lepton mixing angles, Dirac and Majorana CP phases and three light neutrino masses are completely determined by five real parameters $|g_2/g_1|$, $\arg{(g_2/g_1)}$, $\texttt{Re}(\tau)$, $\texttt{Im}(\tau)$, and $g^2_1v_u^2/\Lambda$ whose values can be fixed by the precisely measured lepton mixing angles and neutrino mass squared splittings shown in Eq.~\eqref{eq:lepton-observables}. The number of free parameters is four less than the that of observables, therefore these models are quite predictive. It is remarkable that all these 15 models can predict the unknown values of absolute neutrino masses, the Dirac and Majorana CP violation phases, and the effective neutrino masses in neutrinoless double beta decay. These predictions could be tested in future more sensitive experiments. Note that the models $C_i$-$S_9$ and $C_i$-$S_{10}$ contain more free parameters,   consequently we don't not show the numerical results of these model here. As can be seen from table~\ref{tab:goodmodel_para_obs},
the three lepton mixing angles $\theta_{12}$, $\theta_{13}$, $\theta_{23}$ and the neutrino mass squared difference $\Delta m^2_{21}$, $\Delta m^2_{31}$ fall in the $1\sigma$ experimental range for these 15 viable models except the models $C_1$-$S_3$ and $C_4$-$S_3$ where $\sin^2\theta_{23}=0.4949$ is outside the $1\sigma$ region but still in the $3\sigma$ region. Although predictions for the lepton mixing angles are quite similar, the predictions for CP violation phases $\delta^{l}_{CP}$, $\alpha_{21}$ and $\alpha_{31}$ as well as light neutrino masses $m_{1, 2, 3}$ and $|m_{ee}|$ are different. The future long baseline neutrino experiments DUNE~\cite{Acciarri:2016crz,Acciarri:2015uup,Strait:2016mof,Acciarri:2016ooe} and T2HK~\cite{Abe:2018uyc}, if running in both neutrino and anti-neutrino modes, will significantly improve the precision on $\theta_{23}$ and $\delta^{l}_{CP}$. Hence future neutrino oscillation facilities have the potential to discriminate among the above possible cases, or rule out some of them  completely. It seems extremely difficult or impossible to directly measure the two Majorana CP violating phases from any feasible measurements of the lepton number violating processes. However, the Majorana phases play an important role in the neutrinoless double beta, and most of our predictions for the effective neutrino mass $|m_{ee}|$ is within the reach of future neutrinoless double decay experiments, as discussed below. Moreover, the predictions for neutrino masses in our models could be tested at cosmological experiments such as Planck which can constrain the sum of light neutrino masses.

Furthermore, we notice that the modular forms have the property $Y^{(k)}_{\mathbf{r}}(-\tau^{*})=\rho_{\mathbf{r}}(S)[Y^{(k)}_{\mathbf{r}}(\tau)]^{*}$ as shown in Eq.~\eqref{eq:MF-gCP}. Therefore if we make the replacement $\tau\rightarrow-\tau^{*}$, $g_{1,2}\rightarrow g^{*}_{1,2}$ and perform the $S$ transformation on both lepton and right-handed neutrino fields, the charged lepton and neutrino mass matrices would become their complex conjugate. Hence under such transformation, lepton masses and mixing angles are unchanged while the signs of all CP violating phases are
flipped. As a consequence, the complex modulus $\tau$ is limited in the right-hand part of the fundamental domain $\mathcal{D}$ with $0\leq\texttt{Re}(\tau)\leq0.5$ when we scan over the parameter space in numerical minimisation. The predictions of the mixing parameters in the left-hand part of $\mathcal{D}$ with $-0.5\leq\texttt{Re}(\tau)\leq0$ can be easily obtained by reversing the overall signs of the Dirac and Majorana CP phases. Hence all the numerical results given in table~\ref{tab:goodmodel_para_obs} should understand to come in pair with opposite CP violating phases.

\begin{figure}[t!]
\centering
\includegraphics[width=0.70\linewidth]{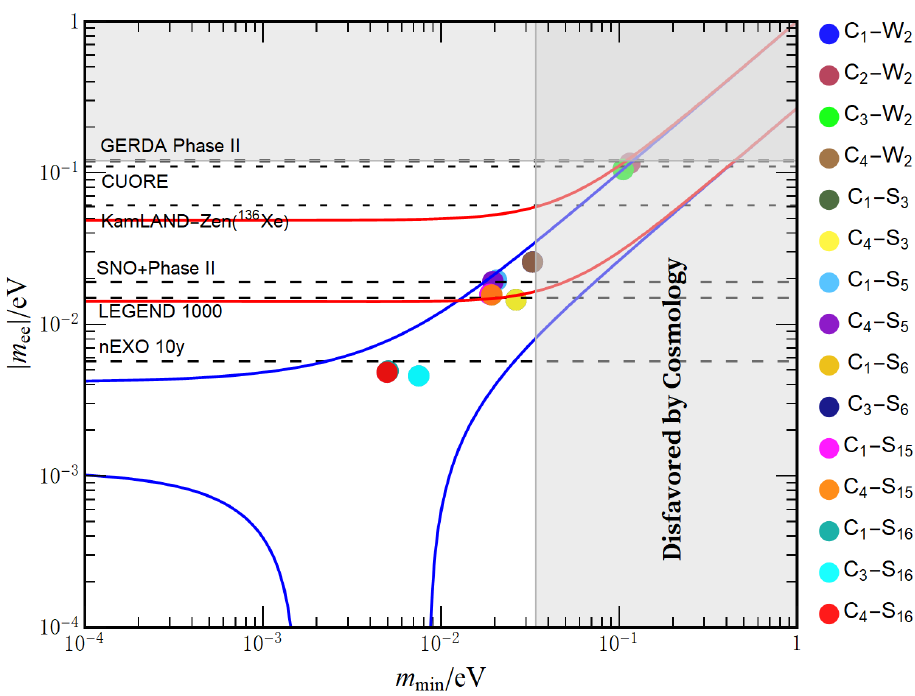}
\caption{The predictions for lightest neutrino mass $m_1$ and the effective Majorana mass $|m_{ee}|$ for the 15 phenomenologically viable models at the best fit points shown in table~\ref{tab:goodmodel_para_obs}. The blue (red) lines denote the most general allowed regions for NO (IO) where the neutrino oscillation parameters are freely varied in their $3\sigma$ regions~\cite{Esteban:2018azc}. The vertical grey exclusion band denotes the bound on the lightest neutrino mass coming from the cosmological data $\Sigma_{i}m_{i}<0.120\,\text{eV}$ at $95\%$ confidence level obtained by the Planck collaboration~\cite{Aghanim:2018eyx}. The values of $|m_{ee}|$ in the models $C_1$-$S_6$ and $C_3$-$S_6$ are too tiny to be visible.  }
\label{fig:m1mee}
\end{figure}

It is known that the neutrino mass spectrum tends to be nearly degenerate in modular invariant models based on inhomogeneous finite modular group. As can be seen from table~\ref{tab:goodmodel_para_obs}, a remarkable feature of these modular $S'_4$ models is that the neutrino masses are hierarchical except the models $C_2$-$W_2$ and $C_3$-$W_2$. From the predictions for neutrino masses, mixing angles and CP violating phases in table~\ref{tab:goodmodel_para_obs}, we can pin down the effective neutrino mass $|m_{ee}|$ relevant to neutrinoless double beta decay. We displayed the lightest neutrino mass and $|m_{ee}|$ of each viable model in figure~\ref{fig:m1mee}, where the experimental bound of KamLAND-Zen~\cite{KamLAND-Zen:2016pfg} and the expected sensitivities of future experiments~\cite{Alduino:2017ehq,Albert:2017owj,Agostini:2018tnm,Andringa:2015tza,Abgrall:2017syy,Albert:2017hjq} are indicated by the horizontal lines. For the models $C_1$-$S_6$ and $C_3$-$S_6$, the effective Majorana mass are $|m_{ee}|\simeq 2.864\times10^{-6}$ meV and $|m_{ee}|=8.620\times10^{-7}$ meV respectively with the lightest neutrino mass $m_1\simeq 6.953$ meV.  Hence the corresponding points are not visible in the figure. The future neutrinoless double beta decay experiments are designed at the tonne scale, and the sensitivity is expected to be improved by about two orders of magnitude over current experiments. Thus we expect that some of our predictions could be tested in future neutrinoless double beta decay experiments, as indicated in the figure.

If we require the theory to be invariant under both $S'_4$ modular symmetry and the generalized CP symmetry, all the couplings would be restricted to be real in our working basis, as shown in section~\ref{sec:gCP-S4DC}. Thus the number of free parameters in a model would be reduced further. For the 15 viable models listed in table~\ref{tab:goodmodel_para_obs}, the generalized CP symmetry enforces both coupling constants $g_1$ and $g_2$ to be real such that the phase $\arg{(g_2/g_1)}$ is equal to zero or $\pi$. As a consequence, the minimal CP-invariant models with $S'_4$ modular symmetry are characterized by only 7 free parameters:
$\beta/\alpha$, $\gamma/\alpha$, $\alpha v_d$, $g_2/g_1$, $g^2_1v_u^2/\Lambda$, $\texttt{Re}(\tau)$ and $\texttt{Im}(\tau)$, the predictive power of the models are enhanced. The former three parameters $\beta/\alpha$, $\gamma/\alpha$ and $\alpha v_d$ in the charged lepton mass matrix are still fixed by the charged lepton masses $m_{e, \mu, \tau}$. The remaining four parameters $g_2/g_1$, $g^2_1v_u^2/\Lambda$, $\texttt{Re}(\tau)$ and $\texttt{Im}(\tau)$ describe the entire neutrino sector including the three neutrino masses $m_{1,2,3}$, three neutrino mixing angles $\theta_{12}$, $\theta_{13}$, $\theta_{23}$,
the Dirac CP violation phase $\delta^{l}_{CP}$ and the Majorana CP phases $\alpha_{21}$ and $\alpha_{31}$. In particular, the complex modulus $\tau$ and modular forms would be sources of all CP violation phases. We find only seven out of the fifteen models are compatible with data, the numerical results are shown in table~\ref{tab:goodmodel_gCP}.

\begin{table}[ht!]
\centering
\renewcommand{\arraystretch}{1.1}
\begin{tabular}{c}
\resizebox{1.0\textwidth}{!}{
\begin{tabular}{|c|c|c|c|c|c|c|c|c|c|c|c|}  \hline\hline
\multirow{2}{*}{ Models} & \multicolumn{8}{c|}{ Best fit values of the input parameters for NO} &\multirow{2}{*}{$\chi^2_{\mathrm{min}}$}  \\ \cline{2-9}
& $\texttt{Re}\langle \tau \rangle$ &$\texttt{Im}\langle \tau \rangle$  &$\beta/\alpha$ &$\gamma/\alpha$  &$|g_2/g_1|$  &$\arg{(g_2/g_1)}/\pi$ & $\alpha v_d$/MeV  & $\dfrac{g^2_1v_u^2}{\Lambda}/$meV & \\ \hline
 $C_1$-$W_2$ & 0.3656   & 1.1638  & 670.6170 & 13.7484   & 0.7008 & 0.0261  & 0.1995 & 2.7462 & 2.277\\ \hline
 $C_2$-$W_2$ & 0.4600  & 0.8911  & 34.6012 & 203.4790   &0.8046   & 1.8275  &  0.1700 & 5.2737 &0.005  \\ \hline
 $C_3$-$W_2$  & 0.4519  &0.8957 & 196.2490 & 57.8262  &0.7826  &0.1813 &  0.1705 & 4.9698 & $ 7.570\times 10^{-5}$ \\ \hline
 $C_4$-$W_2$  & 0.3658 & 1.1639 & 4536.5100& 92.9859  &0.7007  &0.0261   &0.0295 & 2.7461 & 2.261\\ \hline
 $C_1$-$S_3$ &  0.3860 & 1.3025  & 717.9890 & 13.6791 & 0.5157  &1.8065 &0.2030  & 4.5728 & 8.211\\ \hline
 $C_4$-$S_3$ & 0.3860  & 1.3025 & 6074.9500  & 115.7380 & 0.5157  & 0.8065 &0.0240  & 4.5727 & 8.211\\ \hline
 $C_1$-$S_5$& 0.1470  & 0.9994  &0.0001 & 0.0031 & 0.3405 & 0.3505 & 653.0910 & 0.2332 & $7.780\times 10^{-6}$\\ \hline
 $C_4$-$S_5$& 0.0582 &1.0131 & 9584.4200 & 253.7910  & 0.3889 &  0.3067 & 0.0123 & 0.2139 & $2.269\times 10^{-5}$\\ \hline
 $C_1$-$S_6$ &0.1764  &0.9915 &0.0016 &0.8839 & 1.5584 &1.8651 &38.3255 & 0.0846 & 1.581 \\ \hline
 $C_3$-$S_6$ &0.1763  &0.9914 &0.0003 &0.8838 & 1.5576 &1.8646 & 38.3203 & 0.0847 & 1.577 \\ \hline
 $C_1$-$S_{15}$ &0.4792 &1.1710  &44.6105 &215.7250  &1.8280 &1.3567 &0.1934 & 0.0357 & $1.665\times 10^{-5}$ \\ \hline
 $C_4$-$S_{15}$ &0.4881 &1.1629 &244.8360 &1192.4900 &1.7919 &0.6360 &0.0349 & 0.0363 &$1.011\times 10^{-5}$ \\ \hline
 $C_1$-$S_{16}$ &0.2609 &1.1527 &608.7890 &13.2779 &0.2412 &0.0940  &0.2178 &7.3242 & 4.323\\ \hline
 $C_3$-$S_{16}$ &0.3065 &1.0220 &229.7120 &15.8054 &0.2119 &0.0672 &0.1860 &8.0831 & 4.323 \\ \hline
 $C_4$-$S_{16}$ &0.2673 &1.1501 &5679.8400 &123.5270 &0.2342 &1.9063 &0.0233 &7.2261 & 4.309\\ \hline \hline
\end{tabular}}\\
  \resizebox{1.0\textwidth}{!}{
  \begin{tabular}{|c|c|c|c|c|c|c|c|c|c|c|c|}
\multirow{2}{*}{ Models} & \multicolumn{10}{c|}{Predictions for mixing parameters and neutrino masses at best fitting point}  \\ \cline{2-11}
& $\sin^2\theta_{12}$ &$\sin^2\theta_{13}$ &$\sin^2\theta_{23}$&$\delta^l_{CP}/\pi$ &$\alpha_{21}/\pi$  &$\alpha_{31}/\pi$ & $m_1$/meV & $m_2$/meV & $m_3$/meV & $|m_{ee}|$/meV  \\ \hline
 $C_1$-$W_2$ & 0.3100   & 0.02184 & 0.5326 & 1.3285  & 1.5086 & 0.5246  & 32.4905 & 33.6085 & 59.8177 & 25.7353 \\ \hline
 $C_2$-$W_2$ & 0.3100  & 0.02237 & 0.5643 & 0.4160  &1.9869   & 0.9920  & 115.0660 & 115.3870 & 125.5740 & 115.0220  \\ \hline
 $C_3$-$W_2$  & 0.3100  &0.02237 & 0.5628 & 0.4325  &0.0046 &1.0061 & 104.7660 & 105.1180 & 116.2070 & 104.8860 \\ \hline
 $C_4$-$W_2$  & 0.3100 & 0.02184 & 0.5327& 1.3285   &1.5085 &0.5247   & 32.4781 & 33.5965 & 59.8100 & 25.7233\\ \hline
 $C_1$-$S_3$ & 0.3136 & 0.02254  & 0.4949& 0.9971  & 1.3424 & 1.0547 & 26.3479 & 27.7148 & 56.7644 & 14.5905\\ \hline
 $C_4$-$S_3$ & 0.3136  & 0.02254 & 0.4949  & 0.9971 & 1.3424  &1.0547 & 26.3474 & 27.7144 & 56.7638 & 14.5901\\ \hline
 $C_1$-$S_5$& 0.3100  & 0.02237  &0.5630 & 0.4389   & 0.0776 & 0.7294 & 19.3629 & 21.1854 & 53.8788 & 20.3940\\ \hline
 $C_4$-$S_5$& 0.3100 &0.02237 & 0.5630 & 0.8519  & 0.0920 & 0.3068 & 19.5386 & 21.3461 & 53.9455  & 19.2309 \\ \hline
 $C_1$-$S_6$ &0.3220  &0.02227 &0.5435 &1.0014   & 1.0004  &1.0015 & 6.9527  & 11.0562  & 50.6110 &$2.8645\times10^{-6}$  \\ \hline
 $C_3$-$S_6$ & 0.3221  & 0.02226 &0.5435 &1.0012   & 1.0004 &1.0012 & 6.9534 & 11.0560 & 50.6117 &$8.6208\times10^{-7}$ \\ \hline
 $C_1$-$S_{15}$ &0.3100 &0.02237  &0.5629 &1.1729 &1.6460 &1.5562 & 18.8024 & 20.6744 & 53.6802 & 15.8409 \\ \hline
 $C_4$-$S_{15}$ &0.3100 &0.02237 &0.5630 &0.5832 &0.3523 &0.3831 &19.1636 & 21.0035 & 53.8077 & 15.6386  \\ \hline
 $C_1$-$S_{16}$ &0.3008 &0.02147 &0.5630&1.6759 &1.5782 &1.7340 &5.0645 & 9.9774 & 49.5504 & 4.9267\\ \hline
 $C_3$-$S_{16}$ & 0.3008 &0.02147 &0.5630 &0.6638&0.7622 &1.5373 &7.4979 & 11.4070 & 49.8573 & 4.5791 \\ \hline
 $C_4$-$S_{16}$ &0.3008 &0.02147 &0.5630 &0.2980 &1.4096 &0.2421 &4.9822 & 9.9359 & 49.5433 & 4.8485 \\ \hline\hline
\end{tabular}}
\end{tabular}
\caption{\label{tab:goodmodel_para_obs}The best fit values of the input parameters at the minimum of the $\chi^2$ under the assumption of NO neutrino masses. We give the predictions for neutrino mixing angles $\theta_{12}$, $\theta_{13}$, and Dirac CP violating phase $\delta^l_{CP}$ as well as Majorana CP violating phases $\alpha_{21}$, $\alpha_{31}$, and the light neutrino masses $m_{1,2,3}$ and the effective mass $|m_{ee}|$ in neutrinoless double decay. Notice in the CP dual point $\tau\rightarrow-\tau^{*}$, $g_{1,2}\rightarrow g^{*}_{1,2}$, the signs of Dirac and Majorana CP phases are reversed while the predictions for lepton mixing angles and neutrino masses are unchanged. }
\end{table}

\begin{table}[ht!]
\centering
\renewcommand{\arraystretch}{1.1}
\begin{adjustbox}{max width=\textwidth}
  \begin{tabular}{|c|c|c|c|c|c|c|c|c|c|c|}  \hline \hline
Models with gCP & $C_1$-$S_5$ & $C_4$-$S_5$ &$C_1$-$S_6$ & $C_3$-$S_6$ & $C_1$-$S_{16}$ & $C_3$-$S_{16}$ & $C_4$-$S_{16}$ \\ \hline
$\texttt{Re}\langle \tau \rangle$  & 0.1997 & 0.2118   & 0.1745 & 0.1745  & 0.3028   & 0.3166 & 0.3028 \\
$\texttt{Im}\langle \tau \rangle$  & 0.9969 & 0.9709 & 0.9846   & 0.9847 & 1.1351   & 1.0086    & 1.1351 \\
$\beta/\alpha$                     & 0.0001 & 435.0270 & 0.0017 & 0.0003 & 629.3540 & 227.1940 & 4958.2700  \\
$\gamma/\alpha$   & 0.0031 & 1696.6900 & 0.8776 & 0.8776 & 13.4699 & 16.5263  & 106.1150 \\
$g_2/g_1$   & $-0.0066$ & 0.6520 & 1.6936  & 1.6939 & 0.1779   & 0.1899  & 0.1780 \\
$\alpha v_d$/MeV  & 654.2130 & 0.0199 & 38.1991   & 38.1998 & 0.2079 & 0.1843   & 0.0264 \\
$\dfrac{g^2_1v_u^2}{\Lambda}$/meV  & 0.3712 & 0.1600  & 0.0718 & 0.0718 & 6.5644  & 7.8506 & 6.5646\\ \hline
$\sin^2\theta_{12}$   & 0.3105 & 0.3145 & 0.3234  & 0.3234 & 0.3008  & 0.3008  & 0.3008 \\
$\sin^2\theta_{13}$ & 0.02239 & 0.02289 & 0.02230 & 0.02230 & 0.02147 & 0.02147 & 0.02147 \\
$\sin^2\theta_{23}$  & 0.5057 & 0.4491 & 0.5460 & 0.5461 & 0.5630 & 05630 & 0.5630 \\
$\delta_{CP}/\pi$ & 0.5405 & 1.7891 & 1.0000 & 1.0001 & 0 & 0.8932 & 0.0001\\
$\alpha_{21}/\pi$ & 0.0857 & 0.8369 & 1.0000 & 1.0000 & 1.4223 & 0.7645 & 1.4222\\
$\alpha_{31}/\pi$ & 1.0569 & 1.7147 & 1.0000 & 1.0001 & 0 & 1.7866 & 0 \\ \hline
$m_e/m_{\mu}$     & 0.0048 & 0.0048 & 0.0048  & 0.0048 & 0.0048  & 0.0048   & 0.0048  \\
$m_{\mu}/m_{\tau}$& 0.0564 & 0.0561 & 0.0565 & 0.0565 & 0.0565  & 0.0565 & 0.0565  \\ \hline
$m_1$/meV & 14.6582 & 37.3832 & 7.0066 & 7.0063 & 4.4561 & 7.4556 & 4.4564 \\
$m_2$/meV & 16.9930 & 38.3589 & 11.0902 & 11.0900 & 9.6828 & 11.3792 & 9.6830\\
$m_3$/meV & 52.3800 & 62.7243 & 50.6287 & 50.6287 & 49.4911 & 49.8506 & 49.4918\\
$\sum_im_i$/meV & 84.0312 & 138.4660 & 68.7255 & 68.7250 & 63.6300 & 68.6854 & 63.6312\\
$|m_{ee}|$/meV & $16.0735$ & $17.2902$ & $2.8886\times10^{-6}$ & $8.6181\times 10^{-7}$ & $4.4001$ & $4.3325$ & $4.3998$ \\ \hline
$\chi^2_{\mathrm{min}}$ & 5.694 & 23.279 & 1.605 & 1.601 & 4.328 & 4.331 & 4.328\\ \hline\hline
  \end{tabular}
\end{adjustbox}
\caption{\label{tab:goodmodel_gCP}
The best fit values of the input parameters after imposing generalized CP. We give the predictions for lepton mixing parameters and neutrino masses at the best fit points. Notice in the CP dual point $\tau\rightarrow-\tau^{*}$, the signs of Dirac and Majorana CP phases are reversed while the predictions for lepton mixing angles and neutrino masses are unchanged. }
\end{table}

\begin{figure}[ht]
\centering
\includegraphics[width=6.5in]{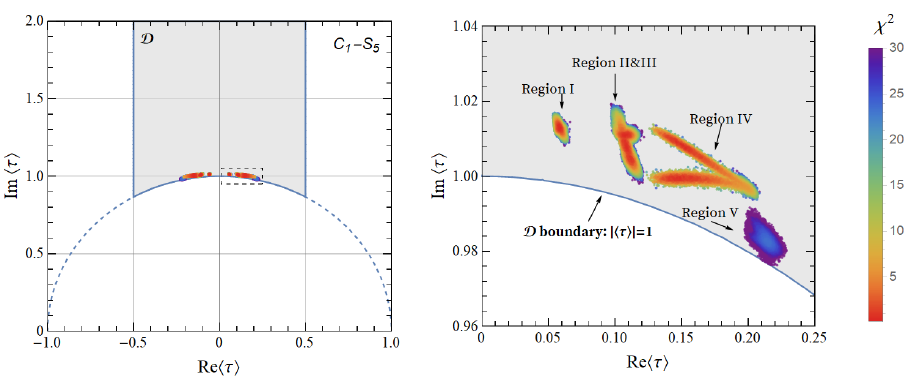}
\caption{The regions of the complex modulus $\langle \tau \rangle$ compatible with experimental data in the fundamental domain $\mathcal{D}$ for the model $C_1$-$S_5$ without gCP. There are five disconnected parameter regions, and two of them (region II and region III) coincide on the $\langle \tau \rangle$ plane as can be seen in the right panel. The values of $\chi^2$ are represented by different colors, as shown in the color bar. Here we focus on the right-hand part of $\mathcal{D}$ with $0\leq\texttt{Re}(\tau)\leq0.5$. The predictions for mixing angles are unchanged and the signs of the CP violating phases are reversed in the CP dual regions $\tau\rightarrow\tau^{*}$, $g_{i}\rightarrow g^{*}_i$. }
\label{fig:C1S5}
\end{figure}

\begin{figure}[ht]
\centering
\includegraphics[width=6.5in]{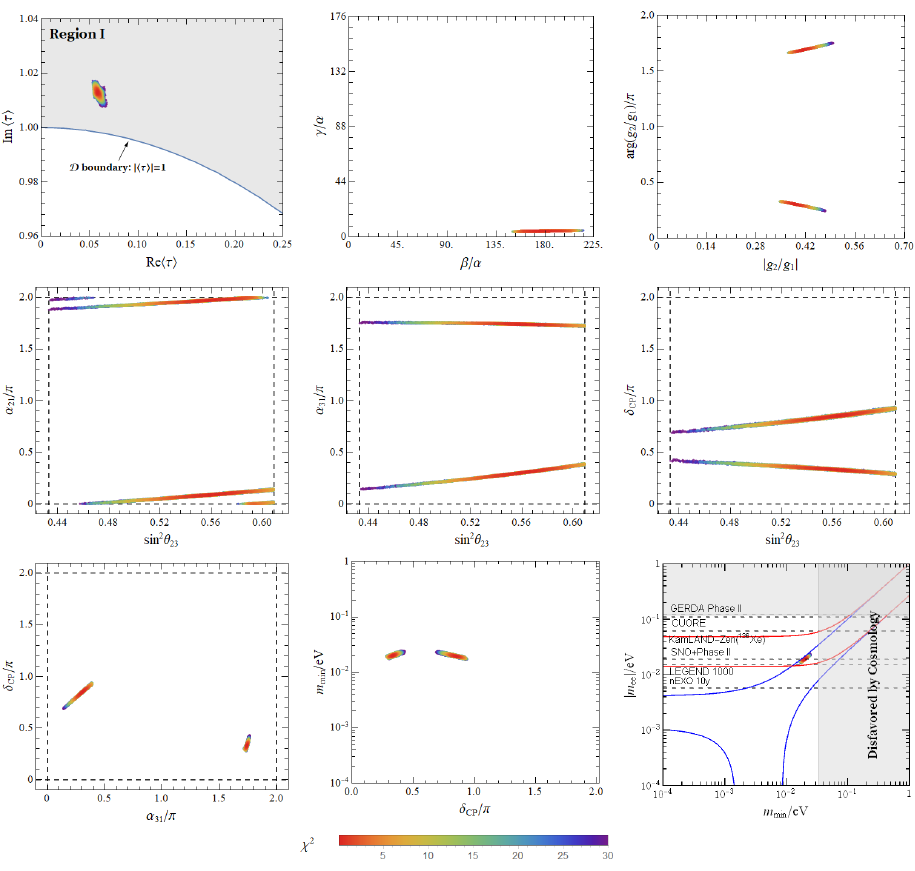}
\caption{The predictions for the correlations among the input free parameters, neutrino mixing angles, CP violating phases and neutrino masses in the region I of model $C_1$-$S_5$ without gCP symmetry.}
\label{fig:C1S5_I}
\end{figure}

\begin{figure}[ht]
\centering
\includegraphics[width=6.5in]{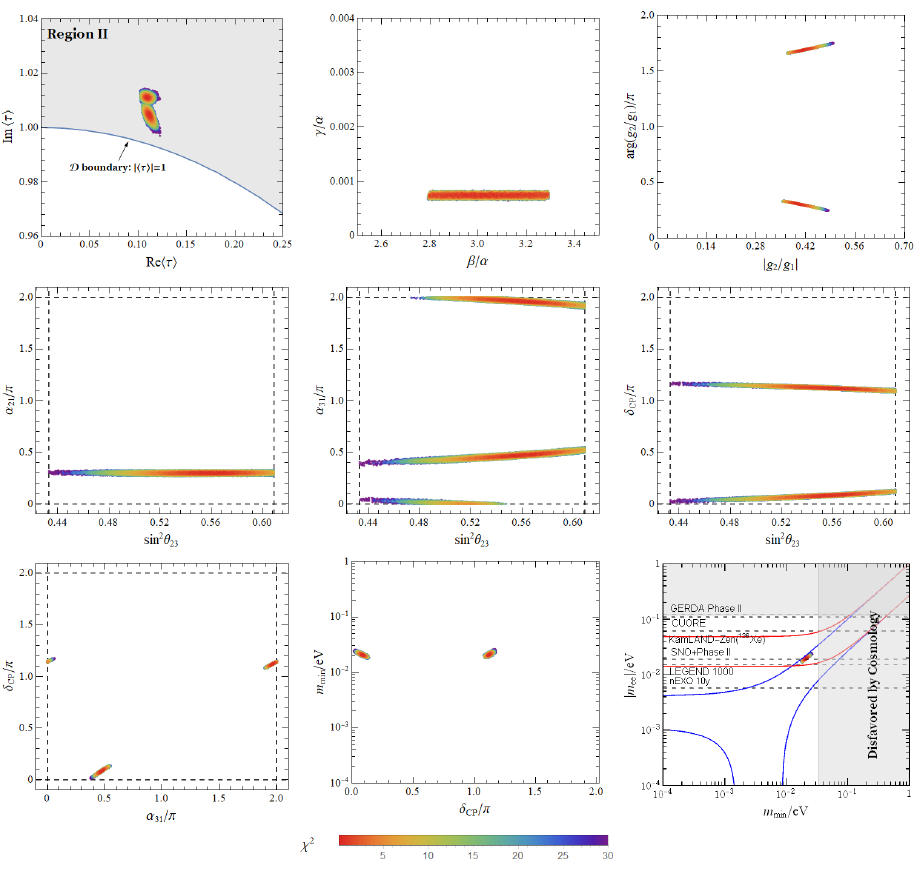}
\caption{The predictions for the correlations among the input free parameters, neutrino mixing angles, CP violating phases and neutrino masses in the region II of model $C_1$-$S_5$ without gCP symmetry.}
\label{fig:C1S5_II}
\end{figure}

\begin{figure}[ht]
\centering
\includegraphics[width=6.5in]{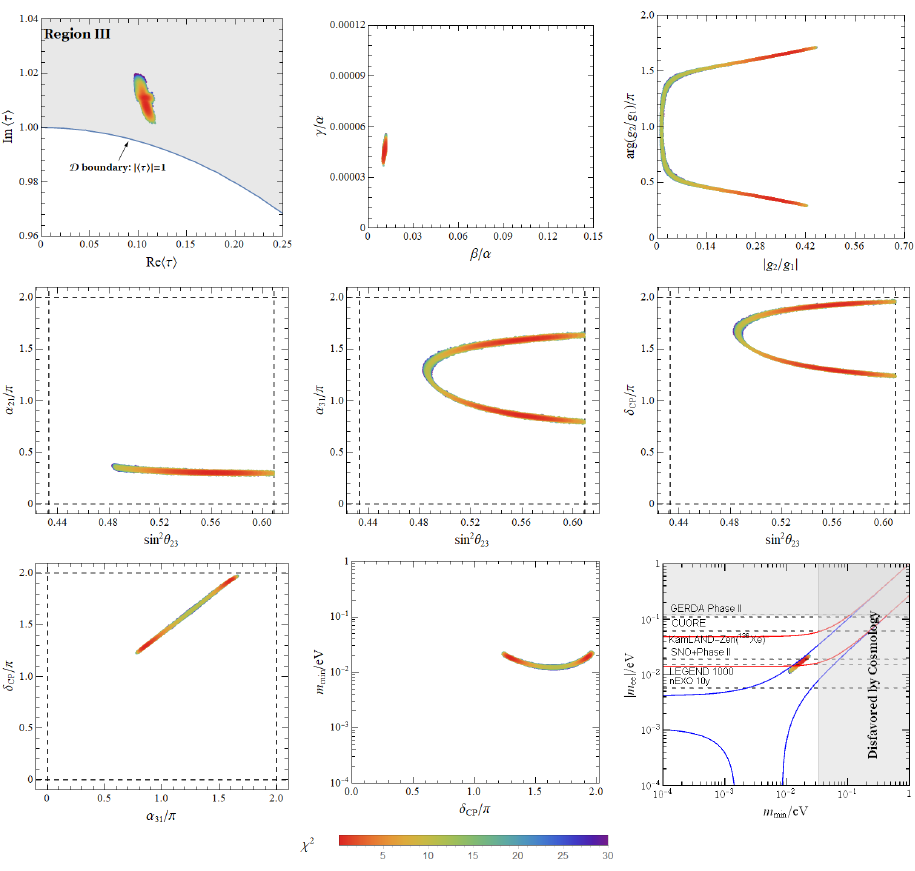}
\caption{The predictions for the correlations among the input free parameters, neutrino mixing angles, CP violation phases and neutrino masses in the region III of model $C_1$-$S_5$ without gCP symmetry.}
\label{fig:C1S5_III}
\end{figure}

\begin{figure}[ht]
\centering
\includegraphics[width=6.5in]{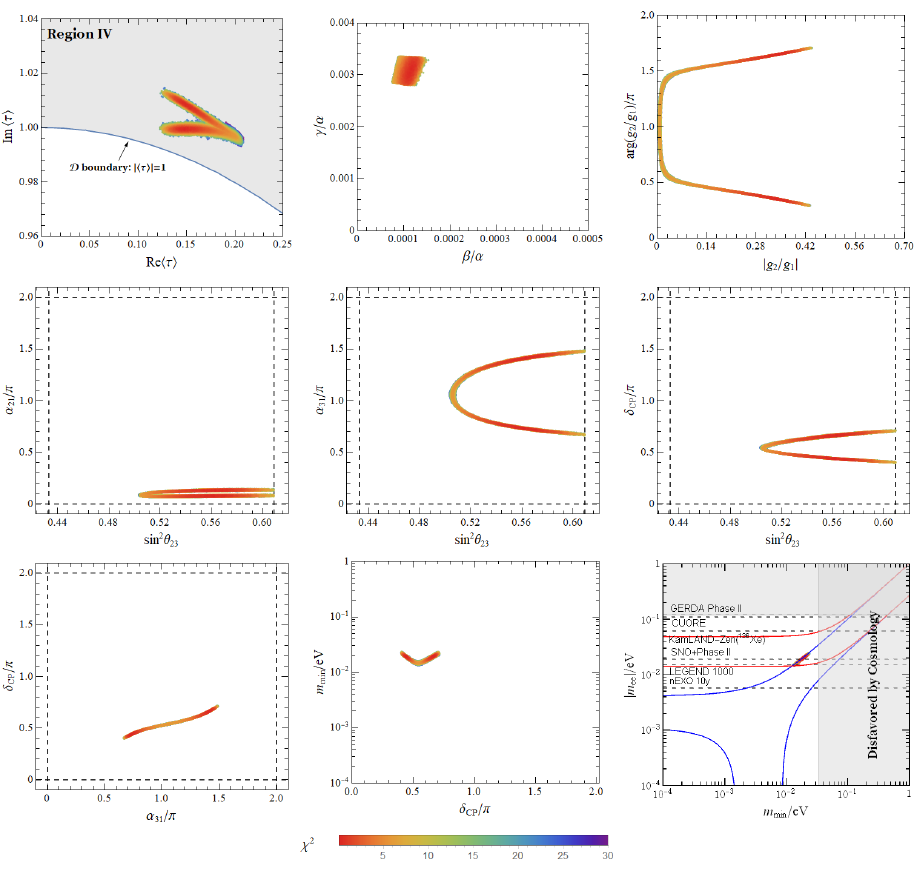}
\caption{The predictions for the correlations among the input free parameters, neutrino mixing angles, CP violating phases and neutrino masses in the region IV of model $C_1$-$S_5$ without gCP symmetry.}
\label{fig:C1S5_IV}
\end{figure}

\begin{figure}[ht]
\centering
\includegraphics[width=6.5in]{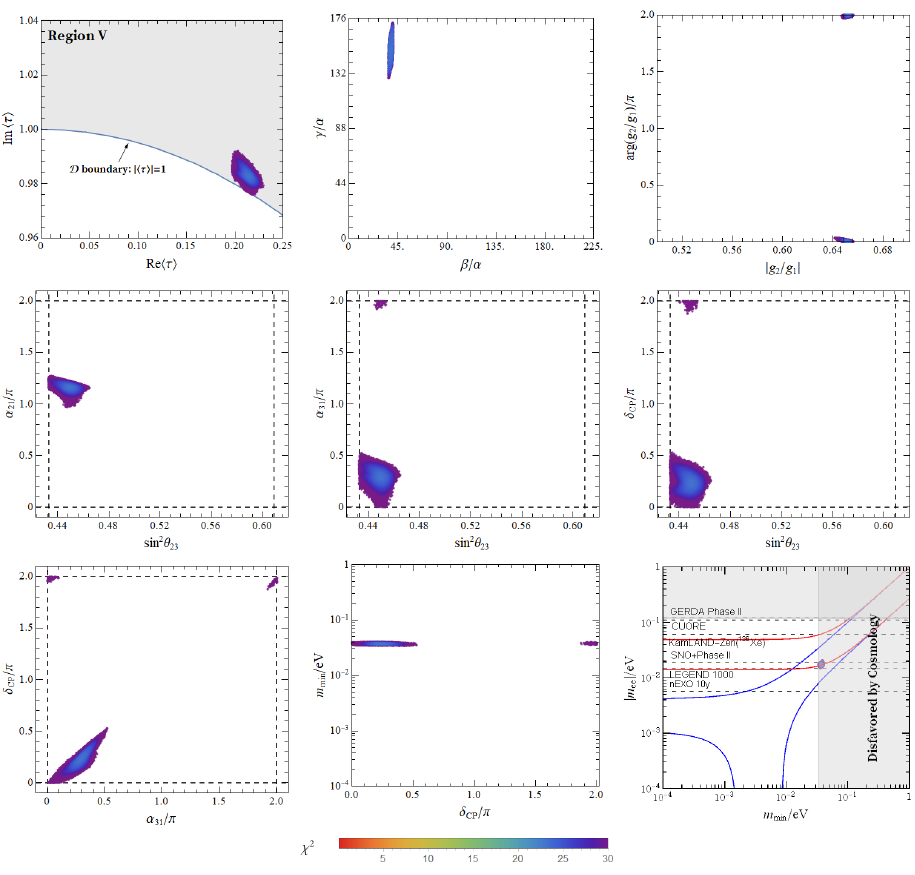}
\caption{The predictions for the correlations among the input free parameters, neutrino mixing angles, CP violating phases and neutrino masses in the region V of model $C_1$-$S_5$ without gCP symmetry.}
\label{fig:C1S5_V}
\end{figure}

\begin{figure}[ht]
\centering
\includegraphics[width=6.5in]{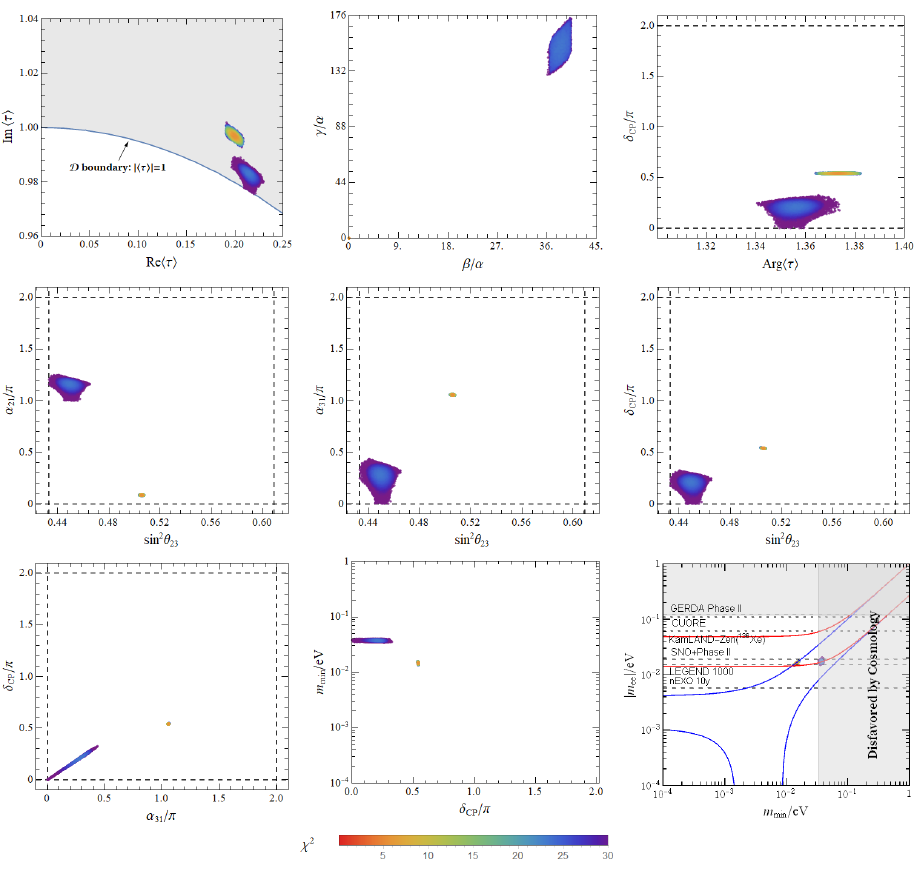}
\caption{The predictions for the correlations among the input free parameters, neutrino mixing angles, CP violating phases and neutrino masses in the model $C_1$-$S_5$ with gCP symmetry.}
\label{fig:C1S5_gCP}
\end{figure}

The modular symmetry models are quite predictive, the mixing parameters and neutrino masses are generally correlated with each other since the number of free parameter is generally less than the number of observables. As an example, we take the model $C_1$-$S_{5}$ for illustration, and we use the popular tool \texttt{MultiNest}~\cite{Feroz:2007kg,Feroz:2008xx} to scan the parameter space fully and efficiently. We require all the three lepton mixing angles $\theta_{12}$, $\theta_{13}$, $\theta_{23}$ and the mass ratios $\Delta m^2_{21}/\Delta m^2_{31}$, $m_{\mu}/m_e$, $m_{\tau}/m_e$ lie in the $3\sigma$ allowed regions, the observed values of the neutrino mass squared differences and charged lepton masses can be reproduced by adjusting the overall mass scales $\alpha v_d$ and $g^2_1v^2_u/\Lambda$. The experimentally allowed values of the complex modulus $\tau$ are displayed in figure~\ref{fig:C1S5}, and they are all located near the boundary $|\tau|=1$ of $\mathcal{D}$. There are five independent and disconnected regions compatible with experiment data in the parameter space. Notice that although region II and region III coincide on the $\langle \tau \rangle$ plane, the allowed regions in the $\beta/\alpha-\gamma/\alpha$ plane and $|g_2/g_1|-\text{arg}(g_2/g_1)$ plane are different. The correlations between the input parameters, neutrino mixing parameters and neutrino masses are shown in figure~\ref{fig:C1S5_I}, figure~\ref{fig:C1S5_II}, figure~\ref{fig:C1S5_III}, figure~\ref{fig:C1S5_IV} and figure~\ref{fig:C1S5_V}. We see that observables are really strongly correlated and show different patterns in each region. After gCP symmetry is imposed, only two independent and disconnected regions together with their CP dual regions in $\mathcal{D}$ are compatible with experimental data, and all observables are predicted to vary in quite small regions, as can be seen from figure~\ref{fig:C1S5_gCP}. Hence gCP make the predictive power of modular invariant models increase considerably.

We have focused on NO neutrino masses in the above numerical analysis. However, IO mass spectrum still isn't excluded although it
is slightly disfavoured by the current data~\cite{Esteban:2018azc}. Analogous to the NO cases, we can numerically scan the parameter space of each model, search for the minimum of the $\chi^2$ function built with the leptonic data of IO case, and eventually find out the models compatible with experimental data of IO. In the following, we take the model $C_2$-$S_9$ as an example, the best fit values of the input parameters are found to be
\begin{equation}
\begin{gathered}
\tau=-0.0131+1.0418i\,,\quad \beta/\alpha=3.0561\,, \quad \gamma/\alpha=0.0003\,,\\
 g_2/g_1=0.5459-0.1804i\,,\quad  g_1^2v_u^2/\Lambda=0.0221\text{ meV}\,,\quad \alpha v_d = 39.5755 \text{ MeV}\,,
\end{gathered}
\end{equation}
which gives rise to the following predictions for lepton masses and mixing parameters,
\begin{equation}
\label{eq:mixpara}
\begin{gathered}
m_e/m_{\mu}=0.0048\,,\quad
m_{\mu}/m_{\tau}=0.0566\,, \\
\sin^2\theta_{12}=0.3100\,,\quad
\sin^2\theta_{13}=0.02258\,,\quad
\sin^2\theta_{23}=0.504,\\
\delta_{CP}/\pi=1.5281\,,\quad
\alpha_{21}/\pi=0.0266\,,\quad \alpha_{31}/\pi=0.9935\,,\\
m_1=51.0533\text{ meV},~
m_2=51.7720\text{ meV},~
m_3=13.0471\text{ meV},~ |m_{ee}|=50.3679\text{ meV}\,.
\end{gathered}
\end{equation}
It is remarkable that the lepton mixing angles $\theta_{12}$, $\theta_{13}$ and neutrino mass squared differences fall in the $1\sigma$ ranges~\cite{Esteban:2018azc}, and $\theta_{23}$ is slightly below its $1\sigma$ lower limit. The Dirac CP phase is very closed to $3\pi/2$, and the effective neutrino mass $|m_{ee}|$ is within the reach of future neutrinoless double decay experiments. The sum of neutrino masses $\sum_i m_i\simeq115.8724 \text{ meV}$ is compatible with the latest Planck bound on neutrino mass sum $\sum_{i}m_{i}<0.12 \text{ eV}-0.60\text{ eV}$ at $95\%$ confidence level~\cite{Aghanim:2018eyx}.

\section{\label{sec:quark-models}Quark models based on $S'_4$ modular symmetry}

In this section, we will exploit the $S'_4$ modular symmetry to understand the quark mass hierarchies and the observed pattern of hierarchial quark mixing angles and CP violating phase encoded in the CKM matrix. We aim to construct viable quark mass models with a minimal amount of free parameters. The quark fields can be assigned to triplet of $S'_4$, the direct product of a doublet and a singlet, or the direct sum of three singlets. Similar to what we have done in the charged lepton sector, we can classify the structures of the quark mass matrix for each assignment. For instance, analogous to the charged lepton sector, we could assign the three generations of quark doublets $Q$ to a triplet of $S'_4$, the right-handed up type quarks $u^{c}$, $c^c$, $t^c$ and down type quarks $d^c$, $s^c$, $b^c$ transform as singlets of $S'_4$, then the up type quark mass matrix $M_u$ and down type quark mass matrix $M_d$ can only take the four possible forms shown in table~\ref{tab:charged lepton} if modular forms of weight less than four are used. Consequently $M_u$ and $M_d$ would depend on three coupling constants $\alpha_u$, $\beta_u$, $\gamma_u$ and $\alpha_d$, $\beta_d$, $\gamma_d$ respectively which can be taken real by field redefinitions. We can tune the values of $\alpha_u$, $\beta_u$, $\gamma_u$ to match the up type quark masses $m_{u,c,t}$, and the down type quark masses $m_{d,s,b}$ can be reproduced  by adjusting the parameters $\alpha_d$, $\beta_d$ and $\gamma_d$. Hence the CKM quark mixing matrix is completely determined by the modulus $\tau$, we find it is impossible to reproduce the three hierarchical quark mixing angles and CP phase by varying a single complex parameter $\tau$. We have constructed tens of thousands quark models by using Wolfram Mathematica for different possible weight and representation assignments of quark fields.

The modular symmetry is broken by the vacuum expectation value of the modulus $\tau$ at high energy scale. We assume that modular invariance breaking scale is around the grand unified theory (GUT) scale $2\times10^{16}$ GeV. From the up type and down type quark mass matrices $M_u$ and $M_{d}$ we can calculate the quark masses, mixing angles and CP violation phase in
terms of the input parameters of the model. In order to find the point in parameter space which optimizes the agreement between predictions and data, we generalize the numerical analysis strategy of section~\ref{subsec:numerical-analysis} to the quark sector, and we search the minimum of the $\chi^2$ contributions from quark mass ratios and CKM parameters. For the calculation of $\chi^2_{\text{min}}$, we use the values of quark masses and the CKM parameters calculated at the GUT scale from a minimal supersymmetry (SUSY) breaking scenario, with SUSY breaking scale $M_{\text{SUSY}}=1$ TeV and $\tan\beta=7.5$, $\bar{\eta}_b=0.09375$~\cite{Antusch:2013jca},
\begin{align}
\nonumber
&m_u/m_c=(1.9286\pm 0.6017)\times 10^{-3}, ~\quad~ m_c/m_t=(2.7247 \pm 0.1200)\times 10^{-3},\\
\nonumber
&m_d/m_s=(5.0528 \pm 0.6192)\times 10^{-2}, ~\quad~  m_s/m_b=(1.7684 \pm 0.0975)\times 10^{-2}\,,\\
\nonumber
&m_t=89.5335 ~\text{GeV}\,,\quad m_b=0.9336 ~\text{GeV}\,, \quad \delta^q_{CP}=69.213^{\circ}\pm 3.115^{\circ}\,,\\
&\theta^q_{12}=0.22736\pm 0.00073,~~\theta^q_{13}=0.00338 \pm 0.00012,~\theta^q_{23}=0.03888 \pm 0.00062\,.
\label{eq:exp-data-quark}
\end{align}
where $\bar{\eta}_b$ denotes the contribution from SUSY threshold corrections which mainly affects the bottom quark Yukawa coupling. After examining tens of thousands quark models constructed by using Wolfram Mathematica, we succeeded in finding some models which can accommodate the experimental data of quark masses and CKM matrix. In the following, we present eight benchmark models with small number of free parameters. The transformation properties of the quark fields under $S'_4$ and their modular weights are summarized in table~\ref{tab:quark_mod}.

\begin{table}[t!]
\centering
\begin{tabular}{|c||c|c|c|c|c|c|c|c|c|c|c} \hline\hline
\multicolumn{2}{|c|}{} & $u^c$ & $c^c$ & $t^c$ & $d^c$ & $s^c$ & $b^c$ & $Q_1$ & $Q_2$ & $Q_3$  \\ \hline

\multirow{2}{*}{\texttt{ Model I}} & $S'_4$ &  \multicolumn{2}{c|}{$\mathbf{2}$} & $\mathbf{1}$ &   \multicolumn{2}{c|}{$\mathbf{\hat{2}}$} & $\mathbf{1'}$ & \multicolumn{2}{c|}{$\mathbf{2}$} & $\mathbf{1}$ \\ \cline{2-11}
  & $k_I$  &\multicolumn{2}{c|}{$6-k_{Q_3}$}   & $6-k_{Q_3}$  & \multicolumn{2}{c|}{$7-k_{Q_3}$}   & $6-k_{Q_3}$ & \multicolumn{2}{c|}{$k_{Q_3}-2$}   & $k_{Q_3}$ \\ \hline

\multirow{2}{*}{\texttt{ Model II}} & $S'_4$ &  \multicolumn{2}{c|}{$\mathbf{2}$} & $\mathbf{1'}$ &   \multicolumn{2}{c|}{$\mathbf{\hat{2}}$} & $\mathbf{1}$ & \multicolumn{2}{c|}{$\mathbf{2}$} & $\mathbf{1}$ \\ \cline{2-11}
  & $k_I$  &\multicolumn{2}{c|}{$6-k_{Q_3}$}   & $6-k_{Q_3}$  & \multicolumn{2}{c|}{$7-k_{Q_3}$}   & $6-k_{Q_3}$ & \multicolumn{2}{c|}{$k_{Q_3}-2$}   & $k_{Q_3}$ \\ \hline

\multirow{2}{*}{\texttt{ Model III}} & $S'_4$ &  \multicolumn{2}{c|}{$\mathbf{\hat{2}}$} & $\mathbf{1'}$ &   \multicolumn{2}{c|}{$\mathbf{2}$} & $\mathbf{1}$ & \multicolumn{2}{c|}{$\mathbf{2}$} & $\mathbf{1}$ \\ \cline{2-11}
  & $k_I$  &\multicolumn{2}{c|}{$5-k_{Q_3}$}   & $6-k_{Q_3}$  & \multicolumn{2}{c|}{$6-k_{Q_3}$}   & $6-k_{Q_3}$ & \multicolumn{2}{c|}{$k_{Q_3}-2$}   & $k_{Q_3}$ \\ \hline

\multirow{2}{*}{\texttt{ Model IV}} & $S'_4$ &  \multicolumn{2}{c|}{$\mathbf{2}$} & $\mathbf{\hat{1}'}$ &   \multicolumn{2}{c|}{$\mathbf{\hat{2}}$} & $\mathbf{1}$ & \multicolumn{3}{c|}{$\mathbf{3'}$} \\ \cline{2-11}
  & $k_I$  &\multicolumn{2}{c|}{$2-k_{Q}$}   & $5-k_{Q}$  & \multicolumn{2}{c|}{$5-k_{Q}$}   & $4-k_{Q}$ & \multicolumn{3}{c|}{$k_{Q}$} \\ \hline

\multirow{2}{*}{\texttt{ Model V}} & $S'_4$ &  \multicolumn{2}{c|}{$\mathbf{\hat{2}}$} & $\mathbf{1'}$ &   \multicolumn{2}{c|}{$\mathbf{2}$} & $\mathbf{\hat{1}}$ & \multicolumn{3}{c|}{$\mathbf{3'}$} \\ \cline{2-11}
  & $k_I$  &\multicolumn{2}{c|}{$3-k_{Q}$}   & $6-k_{Q}$  & \multicolumn{2}{c|}{$4-k_{Q}$}   & $5-k_{Q}$ & \multicolumn{3}{c|}{$k_{Q}$} \\ \hline

\multirow{2}{*}{\texttt{ Model VI}} & $S'_4$ & $\mathbf{\hat{1}}$ & $\mathbf{1}$~ &~ $\mathbf{\hat{1}'}$ & $\mathbf{\hat{1}}$ & $\mathbf{\hat{1}'}$ & $\mathbf{\hat{1}}$ & \multicolumn{3}{c|}{$\mathbf{3}$}   \\ \cline{2-11}
 & $k_I$ & $1-k_Q$  & $2-k_Q$  & $5-k_Q$ & $1-k_Q$ & $5-k_Q$ & $5-k_Q$ & \multicolumn{3}{c|}{$k_Q$} \\ \hline

\multirow{2}{*}{\texttt{ Model VII}} &  $S'_4$ & $\mathbf{1}$ & $\mathbf{1}$  & $\mathbf{\hat{1}}$ & $\mathbf{\hat{1}}$ & $\mathbf{1}$ &  $\mathbf{1}$ & \multicolumn{3}{c|}{$\mathbf{3}$}   \\ \cline{2-11}

 & $k_I$ & $2-k_Q$  & $4-k_Q$ & $5-k_Q$ & $1-k_Q$ & $2-k_Q$ & $6-k_Q$ & \multicolumn{3}{c|}{$k_Q$} \\ \hline

\multirow{2}{*}{\texttt{ Model VIII}} &  $S'_4$ & $\mathbf{\hat{1}}$ & $\mathbf{1}$ & $\mathbf{\hat{1}}$ & $\mathbf{\hat{1}}$ & $\mathbf{1}$ & $\mathbf{1}$ & \multicolumn{3}{c|}{$\mathbf{3}$}   \\ \cline{2-11}
 & $k_I$ & $1-k_Q$ & $4-k_Q$ & $5-k_Q$ & $1-k_Q$ & $2-k_Q$ & $6-k_Q$ & \multicolumn{3}{c|}{$k_Q$} \\ \hline\hline
\end{tabular}
\caption{\label{tab:quark_mod} Transformation properties of the quark fields under the $S'_4$ modular symmetry and the modular weight assignments. The Higgs fields $H_{u,d}$ are invariant under $S'_4$ with vanishing modular weight. }
\end{table}

\begin{description}[labelindent=-0.8em, leftmargin=0.3em]
\item[\textcircled{1}]{\textbf{Model I with gCP: 9 free real parameters including $\texttt{Re}(\tau)$ and $\texttt{Im}(\tau)$ }}\\
In this model, left-handed quarks $Q$ and right-handed up quarks $u^c,~c^c,~t^c$ are assigned to a direct sum of doublet and singlet $\mathbf{2}\oplus \mathbf{1}$ of $S'_4$, and the right-handed down quarks $d^c$, $s^c$, $b^c$ are assigned to $\mathbf{\hat{2}} \oplus \mathbf{1'}$ of $S'_4$. For convenience, we use the subscript ``$D$'' to denote the doublet assignment, i.e. $Q_D\equiv (Q_1,~ Q_2)^T $, $u^c_D\equiv (u^c,~d^c)^T$, $d^c_D\equiv (d^c, ~s^c)^T$. The modular weights of the quark superfields are set to
\begin{equation}
\label{eq:wt1-QMI}k_{Q_3}=k_{Q_D}+2=6-k_{u^c_D}=6-k_{t^c}=7-k_{d^c_D}=6-k_{b^c}\,.
\end{equation}
Thus the modular invariant superpotentials for quark masses read as follows,
\begin{eqnarray}
\nonumber&&\hskip-0.2in \mathcal{W}_u= \alpha_{u1} (u^c_D Q_D)_\mathbf{1} Y^{(4)}_{\mathbf{1}} H_u+\alpha_{u2} (u^c_D Q_D Y^{(4)}_{\mathbf{2}})_\mathbf{1}H_u + \beta_t t^c(Q_D Y^{(4)}_{\mathbf{2}})_\mathbf{1} + \gamma_u Q_3 (u^c_D Y^{(6)}_{\mathbf{2}})_\mathbf{1}H_u\,, \\
&&\hskip-0.2in \mathcal{W}_d= \alpha_d (d^c_D Q_D Y^{(5)}_{\mathbf{\hat{2}}})_\mathbf{1}H_d + \beta_d b^c( Q_D Y^{(4)}_{\mathbf{2}})_\mathbf{1'}H_d + \gamma_{d} b^c Q_3 Y^{(6)}_{\mathbf{1'}} H_d \,.
\label{eq:Wq1}
\end{eqnarray}
From the CG coefficients of $S'_4$ group in Appendix~\ref{sec:S4DC-app}, we find the up and down quark mass matrices are given by
\begin{eqnarray}
\label{eq:Mq_1}
M_{u}=\begin{pmatrix}
 \alpha_{u2}  Y_2^{(4)} ~& \alpha_{u1}  Y_1^{(4)} ~& \gamma_u  Y_4^{(6)} \\
 \alpha_{u1}  Y_1^{(4)} ~& \alpha_{u2}  Y_3^{(4)} ~& \gamma_u  Y_3^{(6)} \\
 \beta_t  Y_3^{(4)} ~& \beta_t  Y_2^{(4)} ~& 0 \\
\end{pmatrix}v_{u}\,,\quad M_{d}=\begin{pmatrix}
 \alpha_d  Y_1^{(5)} ~& 0 ~& 0 \\
 0 ~& -\alpha_d  Y_2^{(5)} ~& 0 \\
 -\beta_d  Y_3^{(4)} ~& \beta_d  Y_2^{(4)} ~& \gamma_d  Y_1^{(6)} \\
\end{pmatrix}v_{d}\,,
\end{eqnarray}
We see that this model makes use of five real positive parameters $\alpha_{u1,d}$, $\beta_{t,d}$, $\gamma_{d}$ and two complex parameter $\alpha_{u2}, \gamma_u$ to describe quark masses and CKM matrix. If we impose gCP symmetry on this model, $\alpha_{u2}$ and $\gamma_u$ restricted to be real and they can be either positive or negative. A good agreement between data and
predictions is obtained for the following values of input parameters
\begin{equation}
\begin{gathered}
\langle\tau\rangle=-0.4385+0.9100i\,,\quad \alpha_{u2}/\alpha_{u1}=-1.8814\,,\quad \gamma_u/\alpha_{u1}=0.1846\,, \\
\beta_t/\alpha_{u1}=719.0101\,,\quad \beta_{d}/\alpha_d=23.3376\,,\quad \gamma_{d}/\alpha_d=0.0225\,,\\
\alpha_{u1} v_u=0.00080~\text{GeV},\quad \alpha_d v_d=0.00025~\text{GeV}\,.
\end{gathered}
\end{equation}
The quark mass ratios and mixing parameters are determined to be
\begin{equation}
\begin{gathered}
\theta^q_{12}=0.22732\,,\quad \theta^q_{13}=0.00338\,,\quad \theta^q_{23}=0.03880\,,\quad \delta^q_{CP}=68.0952^{\circ}\,,\\
m_u/m_c=0.001927\,,~ m_c/m_t=0.002726\,,~ m_d/m_s=0.060247\,,~ m_s/m_b=0.017679\,.
\end{gathered}
\end{equation}

\item[\textcircled{2}]{\textbf{Model II with gCP: 9 free real parameters including $\texttt{Re}(\tau)$ and $\texttt{Im}(\tau)$}}\\
In this model, the representation assignments for quark fields are the same as \textbf{Model I} except changing the assignments of $t^c$ and $b^c$ to $\mathbf{1'}$ and $\mathbf{1}$ respectively under $S'_4$.
The modular weights of quark fields still satisfy the condition of Eq.~\eqref{eq:wt1-QMI}. Consequently the superpotential in quark sector is given by
\begin{eqnarray}
\nonumber&&\hskip-0.2in \mathcal{W}_u= \alpha_{u1} (u^c_D Q_D)_\mathbf{1} Y^{(4)}_{\mathbf{1}} H_u+\alpha_{u2} (u^c_D Q_D Y^{(4)}_{\mathbf{2}})_\mathbf{1}H_u + \beta_t t^c(Q_D Y^{(4)}_{\mathbf{2}})_\mathbf{1'} + \gamma_u Q_3 (u^c_D Y^{(6)}_{\mathbf{2}})_\mathbf{1}H_u\,, \\
&&\hskip-0.2in \mathcal{W}_d= \alpha_d (d^c_D Q_D Y^{(5)}_{\mathbf{\hat{2}}})_\mathbf{1}H_d + \beta_d b^c( Q_D Y^{(4)}_{\mathbf{2}})_\mathbf{1}H_d + \gamma_{d} b^c Q_3 Y^{(6)}_{\mathbf{1}} H_d \,,
\label{eq:Wq2}
\end{eqnarray}
which give rise to the following up and down quark mass matrices
\begin{eqnarray}
\label{eq:Mq_2}
M_{u}=\begin{pmatrix}
 \alpha_{u2}  Y_2^{(4)} ~& \alpha_{u1}  Y_1^{(4)} ~& \gamma_u  Y_4^{(6)} \\
 \alpha_{u1}  Y_1^{(4)} ~& \alpha_{u2}  Y_3^{(4)} ~& \gamma_u  Y_3^{(6)} \\
 -\beta_t  Y_3^{(4)} ~& \beta_t  Y_2^{(4)} ~& 0
\end{pmatrix}v_{u}\,,\quad M_{d}=\begin{pmatrix}
 \alpha_d  Y_1^{(5)} ~& 0 ~& 0 \\
 0 ~& -\alpha_d  Y_2^{(5)} ~& 0 \\
 \beta_d  Y_3^{(4)} ~& \beta_d  Y_2^{(4)} ~& \gamma_d  Y_2^{(6)}
\end{pmatrix}v_{d}\,.
\end{eqnarray}
The parameters $\alpha_{u1,d}$, $\beta_{t,d}$, $\gamma_{d}$ can be taken real by field redefinition, $\alpha_{u1}$ and $\gamma_u$ are generically real numbers if we impose gCP symmetry on this model. We find that the agreement between predictions and data is optimized for the following values of input parameters
\begin{equation}
\begin{gathered}
\langle\tau\rangle=0.4894+0.9423i\,,\quad \alpha_{u2}/\alpha_{u1}=-2.1364\,,\quad \gamma_u/\alpha_{u1}=0.2163\,, \\
\beta_t/\alpha_{u1}=814.6742\,,\quad \beta_{d}/\alpha_d=22.7465\,,\quad \gamma_{d}/\alpha_d=0.0113\,,\\
\alpha_{u1} v_u=0.00075~\text{GeV},\quad \alpha_d v_d=0.00028~\text{GeV}\,.
\end{gathered}
\end{equation}
The quark mass ratios and mixing parameters are determined to be
\begin{equation}
\begin{gathered}
\theta^q_{12}=0.22731\,,\quad \theta^q_{13}=0.00338\,,\quad \theta^q_{23}=0.03876\,,\quad \delta^q_{CP}=67.9162^{\circ}\,,\\
m_u/m_c=0.001956\,,~ m_c/m_t=0.002724\,,~ m_d/m_s=0.060138\,,~ m_s/m_b=0.017783\,.
\end{gathered}
\end{equation}

\item[\textcircled{3}]{\textbf{Model III with gCP: 9 free real parameters including $\texttt{Re}(\tau)$ and $\texttt{Im}(\tau)$ }}\\
In comparison with \textbf{Model I}, the representation assignments for the right-handed up quarks and down quarks are interchanged. We choose the  modular weights of quark fields as
\begin{equation}
k_{Q_3}=k_{Q_D}+2=5-k_{u^c_D}=6-k_{t^c}=6-k_{d^c_D}=6-k_{b^c}\,.
\end{equation}
The superpotentials for quark masses takes the following form
\begin{eqnarray}
&&\hskip-0.2in \mathcal{W}_u= \alpha_u (d^c_D Q_D Y^{(3)}_{\mathbf{\hat{1}'}})_\mathbf{1}H_u + \beta_u t^c( Q_D Y^{(4)}_{\mathbf{2}})_\mathbf{1'}H_u + \gamma_{u} t^c Q_3 Y^{(6)}_{\mathbf{1'}} H_u \,,\\
\nonumber&&\hskip-0.2in \mathcal{W}_d= \alpha_{d1} (u^c_D Q_D)_\mathbf{1} Y^{(4)}_{\mathbf{1}} H_u+\alpha_{d2} (u^c_D Q_D Y^{(4)}_{\mathbf{2}})_\mathbf{1}H_d + \beta_d b^c(Q_D Y^{(4)}_{\mathbf{2}})_\mathbf{1} + \gamma_d Q_3 (u^c_D Y^{(6)}_{\mathbf{2}})_\mathbf{1}H_d\,.
\label{eq:Wq3}
\end{eqnarray}
Then we can straightforwardly read out the quark mass matrices,
\begin{eqnarray}
\label{eq:Mq_3}
\quad M_{u}=\begin{pmatrix}
 0 ~& \alpha_u Y_1^{(3)} ~& 0 \\
 \alpha_u Y_1^{(3)} ~& 0 ~& 0 \\
 -\beta_u  Y_3^{(4)} ~& \beta_u  Y_2^{(4)} ~& \gamma_u  Y_1^{(6)} \\
\end{pmatrix}v_{u}\,,~~~
M_{d}=\begin{pmatrix}
 \alpha_{d2}  Y_2^{(4)} ~& \alpha_{d1}  Y_1^{(4)} ~& \gamma_d  Y_4^{(6)} \\
 \alpha_{d1}  Y_1^{(4)} ~& \alpha_{d2}  Y_3^{(4)} ~& \gamma_d  Y_3^{(6)} \\
 \beta_d  Y_3^{(4)} ~& \beta_d  Y_2^{(4)} ~& 0 \\
\end{pmatrix}v_{d}\,,
\end{eqnarray}
where $\alpha_{u,d1}$, $\beta_{u,d}$, $\gamma_{u}$ are positive real parameters since their phases are unphysical, $\alpha_{d2}$ and $\gamma_d$ should be real because of the invariance of the model under gCP symmetry. The best fit values of these input parameters determined to be
\begin{equation}
\begin{gathered}
\langle\tau\rangle=0.4522+0.9262i\,,\quad \beta_{u}/\alpha_u=124.4933\,,\quad \gamma_u/\alpha_{u}=0.0475\,, \\
\alpha_{d2}/\alpha_{d1}=3.1475\,,\quad \gamma_{d}/\alpha_{d1}=-0.3056\,,\quad \beta_{d}/\alpha_{d1}=183.1602\,,\\
\alpha_{u} v_u=0.00476~\text{GeV},\quad \alpha_{d1} v_d=3.37231 \times 10^{-5}~\text{GeV}\,.
\end{gathered}
\end{equation}
Accordingly the quark mass ratios and mixing parameters at the best fit point are
\begin{equation}
\begin{gathered}
\theta^q_{12}=0.22736\,,\quad \theta^q_{13}=0.00333\,,\quad \theta^q_{23}=0.03888\,,\quad \delta^q_{CP}=69.2142^{\circ}\,,\\
m_u/m_c=0.003322\,,~ m_c/m_t=0.002725\,,~ m_d/m_s=0.050560\,,~ m_s/m_b=0.017689\,.
\end{gathered}
\end{equation}

\item[\textcircled{4}]{\textbf{Model IV with gCP: 9 free real parameters including $\texttt{Re}(\tau)$ and $\texttt{Im}(\tau)$ }} \\
The left-handed quarks $Q$ are embedded into a triplet $\mathbf{3'}$ of $S'_4$, the right-handed up and down type quark fields are assigned to $\mathbf{2} \oplus \mathbf{\hat{1}'}$ and $\mathbf{\hat{2}} \oplus \mathbf{1}$ respectively, as shown in table~\ref{tab:quark_mod}.
The modular weights of the quark fields satisfy the condition
\begin{equation}
k_{Q} = 2-k_{u_D^c} = 5-k_{t^c} = 5-k_{d_D^c} = 4-k_{b^c}\,.
\end{equation}
We can read out the superpotential for quark masses as follows,
\begin{eqnarray}
\nonumber&&\hskip-0.2in \mathcal{W}_u=\alpha_{u} (u^c_D QY_{\mathbf{3}}^{(2)})_\mathbf{1} H_u+\beta_{u} t^{c}(Q Y_{\mathbf{\hat{3}'},I}^{(5)})_\mathbf{\hat{1}} H_u + \gamma_{u}  t^{c}(Q Y_{\mathbf{\hat{3}'},II}^{(5)})_\mathbf{\hat{1}}H_u \,,\\
&&\hskip-0.2in \mathcal{W}_{d}=  \alpha_{d} (d^c_D Q Y_{\mathbf{\hat{3}}}^{(5)})_\mathbf{1}H_d + \beta_{d} (d^c_D Q Y_{\mathbf{\hat{3}'},I}^{(5)})_\mathbf{1}H_d + \gamma_{d}(d^{c}_{D} Q  Y_{\mathbf{\hat{3}'},II}^{(5)})_\mathbf{1}H_d +\delta_{d} b^c(Q Y_{\mathbf{3}'}^{(4)})_\mathbf{1}H_d\,,
\label{eq:Wq4}
\end{eqnarray}
which leads to the following quark mass matrices
\begin{eqnarray}
\label{eq:Mq_4}
\nonumber&& M_{u}=\begin{pmatrix}
 -\alpha_{u} Y_4^{(2)} ~& -\alpha_{u} Y_3^{(2)} ~& -\alpha_{u} Y_5^{(2)} \\
 \alpha_{u} Y_5^{(2)} ~& \alpha_{u} Y_4^{(2)} ~& \alpha_{u} Y_3^{(2)} \\
 \beta_{u} Y_6^{(5)}+\gamma_{u} Y_9^{(5)} ~& \beta_{u} Y_8^{(5)}+\gamma_{u} Y_{11}^{(5)} ~& \beta_{u} Y_7^{(5)}+\gamma_{u} Y_{10}^{(5)} \\
\end{pmatrix}v_{u}\,,\\[0.1in]
\nonumber
&& M_{d}=\begin{pmatrix}
 \alpha_{d} Y_{4}^{(5)}-\beta_{d} Y_{7}^{(5)}-\gamma_{d} Y_{10}^{(5)} & \alpha_{d} Y_{3}^{(5)}-\beta_{d} Y_{6}^{(5)}-\gamma_{d} Y_{9}^{(5)} & \alpha_{d} Y_{5}^{(5)}-\beta_{d} Y_{8}^{(5)}-\gamma_{d} Y_{11}^{(5)} \\
 \alpha_{d} Y_{5}^{(5)}+\beta_{d} Y_{8}^{(5)}+\gamma_{d} Y_{11}^{(5)} & \alpha_{d} Y_{4}^{(5)}+\beta_{d} Y_{7}^{(5)}+\gamma_{d} Y_{10}^{(5)} & \alpha_{d} Y_{3}^{(5)}+\beta_{d} Y_{6}^{(5)}+\gamma_{d} Y_{9}^{(5)} \\
 \delta_{d} Y_{7}^{(4)} & \delta_{d} Y_{9}^{(4)} & \delta_{d} Y_{8}^{(4)} \\
\end{pmatrix}v_{d}\,.\\
\end{eqnarray}
The parameters $\alpha_{u,d}$, $\beta_{u}$ and $\delta_d$ can be taken real and positive by field redefinitions, and the gCP symmetry constrains  $\gamma_{u,d}$ and $\beta_d$ to be real numbers. Note that $\gamma_{u,d}$ and $\beta_d$ would be complex without gCP. The predictions approach the experimental data best for the following values of input parameters
\begin{equation}
\begin{gathered}
\langle\tau\rangle=-0.3382+1.4779i\,,\quad \beta_u/\alpha_u=0.1960\,,\quad \gamma_u/\alpha_u=19.8731\,, \\
\delta_{d}/\alpha_d= 1.3238\,, \quad  \beta_d/\alpha_d=1.7610\,,\quad \gamma_{d}/\alpha_d=0.1085\,,\\
\alpha_u v_u=0.01653~\text{GeV},\quad \alpha_d v_d=0.00117~\text{GeV}\,.
\end{gathered}
\end{equation}
The quark mass ratios and mixing parameters are determined to be
\begin{equation}
\begin{gathered}
\theta^q_{12}=0.22747\,,\quad \theta^q_{13}=0.00338\,,\quad \theta^q_{23}=0.03849\,,\quad \delta^q_{CP}=70.1665^{\circ}\,,\\
m_u/m_c=0.001570\,,~ m_c/m_t=0.002736\,,~ m_d/m_s=0.050425\,,~ m_s/m_b=0.018259\,,
\end{gathered}
\end{equation}
which are compatible with experimental data at $1\sigma$ level.

\item[\textcircled{5}]{\textbf{Model V with gCP: 9 free real parameters including $\texttt{Re}(\tau)$ and $\texttt{Im}(\tau)$ }} \\
Similar to \texttt{Model IV}, the left-handed quarks $Q$ transform as  $\mathbf{3'}$ under $S'_4$, the right-handed up and down type quarks are assigned to the direct sum of doublet and singlet $\mathbf{\hat{2}} \oplus \mathbf{1'}$ and $\mathbf{2} \oplus \mathbf{\hat{1}}$ respectively. The modular weights satisfy the relations $k_{Q} = 3-k_{u_D^c} = 6-k_{t^c} = 4-k_{d_D^c} = 5-k_{b^c}$. Consequently the superpotential for quark masses is given by
\begin{eqnarray}
\nonumber&&\hskip-0.2in \mathcal{W}_{u}=\alpha_{u} (u^{c}_{D} Q Y_{\mathbf{\hat{3}}}^{(3)})_\mathbf{1}H_u+\beta_{u} (u^{c}_{D}Q Y_{\mathbf{\hat{3}'}}^{(3)})_\mathbf{1}H_u+\gamma_{u} t^{c}(Q Y_{\mathbf{3},I}^{(6)})_\mathbf{1'}H_u+\delta_{u} t^{c}(Q Y_{\mathbf{3},II}^{(6)})_\mathbf{1'}H_u \,,\\
&&\hskip-0.2in \mathcal{W}_{d}=\alpha_{d} ( d^{c}_{D}Q Y_{\mathbf{3}}^{(4)} )_\mathbf{1}H_d+\beta_{d}( d^{c}_{D} Q  Y_{\mathbf{3}'}^{(4)} )_\mathbf{1}H_d + \gamma_{d}  b^{c}(Q  Y_{\mathbf{\hat{3}}}^{(5)})_\mathbf{\hat{1}'} H_d\,,
\label{eq:Wq5}
\end{eqnarray}
where $\alpha_{u,d}$ and $\gamma_{u,d}$ can be taken real and positive without loss of generality, the couplings $\beta_{u,d}$ and $\delta_u$ are real since gCP symmetry is imposed on the model. Subsequently we can read out the up and down type quark mass matrices
\begin{eqnarray}
\label{eq:Mq_5}
\nonumber&&M_{u}=\begin{pmatrix}
 \alpha_{u} Y_{3}^{(3)}-\beta_{u} Y_{6}^{(3)} ~& \alpha_{u} Y_{2}^{(3)}-\beta_{u} Y_{5}^{(3)} ~& \alpha_{u} Y_{4}^{(3)}-\beta_{u} Y_{7}^{(3)} \\
 \beta_{u} Y_{7}^{(3)}+\alpha_{u} Y_{4}^{(3)} ~& \beta_{u} Y_{6}^{(3)}+\alpha_{u} Y_{3}^{(3)} ~& \beta_{u} Y_{5}^{(3)}+\alpha_{u} Y_{2}^{(3)} \\
 \gamma_{u} Y_{5}^{(6)}+\delta_{u} Y_{8}^{(6)} & \gamma_{u} Y_{7}^{(6)}+\delta_{u} Y_{10}^{(6)} & \gamma_{u} Y_{6}^{(6)}+\delta_{u} Y_{9}^{(6)}
\end{pmatrix}v_{u}\,,\\[0.1in]
&& M_{d}=\begin{pmatrix}
 \beta_{d} Y_{8}^{(4)}-\alpha_{d} Y_{5}^{(4)} ~& \beta_{d} Y_{7}^{(4)}-\alpha_{d} Y_{4}^{(4)} ~& \beta_{d} Y_{9}^{(4)}-\alpha_{d} Y_{6}^{(4)} \\
 \beta_{d} Y_{9}^{(4)}+\alpha_{d} Y_{6}^{(4)} ~& \beta_{d} Y_{8}^{(4)}+\alpha_{d} Y_{5}^{(4)} ~& \beta_{d} Y_{7}^{(4)}+\alpha_{d} Y_{4}^{(4)} \\
 \gamma_{d} Y_{3}^{(5)} ~& \gamma_{d} Y_{5}^{(5)} ~& \gamma_{d} Y_{4}^{(5)}
\end{pmatrix}v_{d}\,.
\end{eqnarray}
The $\chi^2$ analysis gives the best fit values of input parameters and the predictions for quark masses and mixing parameters at the best fit point as follow
\begin{eqnarray}
\nonumber&&\hskip-0.4in \langle\tau\rangle=-0.4382+2.0445i\,,\quad \gamma_u/\alpha_u=103.4796\,,\quad \beta_u/\alpha_u=-0.0105\,, \\
\nonumber&&\hskip-0.4in \delta_{u}/\alpha_u= -51.8872\,, \quad  \gamma_d/\alpha_d=362.6550\,,\quad \beta_{d}/\alpha_d=-189.0835\,,\\
\nonumber&&\hskip-0.4in \alpha_u v_u=0.00450~\text{GeV},\quad \alpha_d v_d=5.61219\times 10^{-6}~\text{GeV}\,,\\
\nonumber&&\hskip-0.4in \theta^q_{12}=0.22736\,,\quad \theta^q_{13}=0.00338\,,\quad \theta^q_{23}=0.03888\,,\quad \delta^q_{CP}=69.1973^{\circ}\,,\\
&&\hskip-0.4in m_u/m_c=0.001928\,,~ m_c/m_t=0.002724\,,~ m_d/m_s=0.050546\,, ~ m_s/m_b=0.017684\,.
\end{eqnarray}
It is notable that the experimental data are accommodated within $1\sigma$.

\item[\textcircled{6}]{\textbf{Model VI without gCP: 10 free real parameters including $\texttt{Re}(\tau)$ and $\texttt{Im}(\tau)$}} \\
In this model, we assume the left-handed quarks $Q$ transform as triplet $\mathbf{3}$ of $S'_4$, the right-handed up quark fields $u^c$, $c^c$, $t^c$ transform as $\mathbf{\hat{1}}$, $\mathbf{1}$, $\mathbf{\hat{1}'}$, the right-handed down quark fields $d^c$, $s^c$, $b^c$ are assigned to $\mathbf{\hat{1}},\,\mathbf{\hat{1}'},\,\mathbf{\hat{1}}$. We take the weights to fulfill the relations,
\begin{equation}
k_{Q} = 1-k_{u^c} = 2-k_{c^c} = 5-k_{t^c} = 1-k_{d^c} = 5-k_{s^c} = 5-k_{b^c}\,.
\end{equation}
Thus the modular invariant superpotentials for quark masses read as follows,
\begin{eqnarray}
\nonumber&&\hskip-0.2in \mathcal{W}_u = \alpha_u u^c(QY^{(1)}_{\mathbf{\hat{3}'}})_\mathbf{\hat{1}'}H_u + \beta_u c^c(QY^{(2)}_{\mathbf{3}})_\mathbf{1}H_u + \gamma_u t^c(QY^{(5)}_{\mathbf{\hat{3}}})_\mathbf{\hat{1}}H_u \,,\\
&&\hskip-0.2in \mathcal{W}_d= \alpha_d d^c(QY^{(1)}_{\mathbf{\hat{3}'}})_\mathbf{\hat{1}'}H_d + \beta_d s^c(QY^{(5)}_{\mathbf{\hat{3}}})_\mathbf{\hat{1}}H_d + \gamma_{d1} b^c(QY^{(5)}_{\mathbf{\hat{3}'},I})_\mathbf{\hat{1}'}H_d + \gamma_{d2} b^c(QY^{(5)}_{\mathbf{\hat{3}'},II})_\mathbf{\hat{1}'}H_d \,.
\label{eq:Wq6}
\end{eqnarray}
The phases of $\alpha_u$, $\beta_u$, $\gamma_u$, $\alpha_d$, $\beta_d$ and $\gamma_{d1}$ can be absorbed into the quark fields while the phase of $\gamma_{d2}$ can not be removed by field redefinition. Using the CG coefficients of $S'_4$ group in Appendix~\ref{sec:S4DC-app}, we find the up and down quark mass matrices are given by
\begin{eqnarray}
\label{eq:Mq_6}
\nonumber&& M_u =\begin{pmatrix}
 \alpha_u Y^{(1)}_1 ~& \alpha_u Y^{(1)}_3 ~& \alpha_u Y^{(1)}_2 \\
 \beta_u Y^{(2)}_3 ~& \beta_u Y^{(2)}_5 ~& \beta_u Y^{(2)}_4 \\
 \gamma_u Y^{(5)}_3~ & \gamma_u Y^{(5)}_5 ~& \gamma_u Y^{(5)}_4\\
 \end{pmatrix}v_u\,, \\[0.1in]
&& M_d =\begin{pmatrix}
 \alpha_d Y^{(1)}_1 ~& \alpha_d Y^{(1)}_3 ~& \alpha_d Y^{(1)}_2 \\
 \beta_d Y^{(5)}_3 ~& \beta_d Y^{(5)}_5 ~& \beta_d Y^{(5)}_4 \\
 \gamma_{d1} Y^{(5)}_6 + \gamma_{d2}Y^{(5)}_9 ~& \gamma_{d1} Y^{(5)}_8+\gamma_{d2}Y^{(5)}_{11}  ~& \gamma_{d1} Y^{(5)}_7 + \gamma_{d2}Y^{(5)}_{10}\\
 \end{pmatrix}v_d\,,
\end{eqnarray}
We see that this model make uses of six real parameters $\alpha_{u,d}$, $\beta_{u,d}$, $\gamma_{u,d1}$ and one complex parameter $\gamma_{d2}$ to describe quark masses and CKM matrix. A good agreement between data and
predictions is obtained for the following values of input parameters
\begin{equation}
\begin{gathered}
\langle\tau\rangle=-0.4999+0.8958i\,,\quad \beta_u/\alpha_u=62.2142\,,\quad \gamma_u/\alpha_u=0.00104\,, \\
\beta_d/\alpha_d=0.7378\,,\quad \gamma_{d1}/\alpha_d=1.4946\,,\quad \gamma_{d2}/\alpha_d=-0.1958-0.2762i\,,\\
\alpha_u v_u=0.07989~\text{GeV},\quad \alpha_d v_d=0.00091~\text{GeV}\,.
\end{gathered}
\end{equation}
The quark mass ratios and mixing parameters are determined to be
\begin{equation}
\begin{gathered}
\theta^q_{12}=0.22731\,,\quad \theta^q_{13}=0.00298\,,\quad \theta^q_{23}=0.04873\,,\quad \delta^q_{CP}=67.1962^{\circ}\,,\\
m_u/m_c=0.00204\,,\quad m_c/m_t=0.00268\,,\quad m_d/m_s=0.05182\,,\quad m_s/m_b=0.01309\,,
\end{gathered}
\end{equation}
which are compatible with the experimental data in Eq.~\eqref{eq:exp-data-quark} except that $\theta^q_{23}$ is somewhat larger. Notice that the top and bottom quark masses can be reproduced by adjusting the parameters $\alpha_u$ and $\alpha_d$.

\item[\textcircled{7}]{\textbf{Model VII with gCP: 10 free real parameters including $\texttt{Re}(\tau)$ and $\texttt{Im}(\tau)$}} \\
The left-handed quarks $Q$ are assigned to triplet $\mathbf{3}$ of $S'_4$, $u^c$, $c^c$ and $t^c$ transform as $\mathbf{1}$, $\mathbf{1}$ and $\mathbf{\hat{1}}$ respectively under $S'_4$, down type quarks $d^c,\,s^c,\,b^c$ transform as $\mathbf{\hat{1}},\,\mathbf{1},\,\mathbf{1}$ respectively. Note that $u^c$ and $c^c$ are distinguished by their different modular weight, and similarly for $s^c$ and $b^c$. We choose of the weights of quark fields to fulfill $k_{Q} = 2-k_{u^c} = 4-k_{c^c} = 5-k_{t^c} = 1-k_{d^c} = 2-k_{s^c} = 6-k_{b^c}$. The superpotential of the quark sector is given by,
\begin{eqnarray}
\nonumber&& \mathcal{W}_u = \alpha_u u^c(QY^{(2)}_{\mathbf{3}})_\mathbf{\hat{1}}H_u + \beta_u c^c(QY^{(4)}_{\mathbf{3}})_\mathbf{1}H_u + \gamma_{u1} t^c(QY^{(5)}_{\mathbf{\hat{3}'},I})_\mathbf{\hat{1}'}H_u + \gamma_{u2} t^c(QY^{(5)}_{\mathbf{\hat{3}'},II})_\mathbf{\hat{1}'}H_u \,,\\
&& \mathcal{W}_d= \alpha_d d^c(QY^{(1)}_{\mathbf{\hat{3}'}})_\mathbf{\hat{1}'}H_d + \beta_d s^c(QY^{(2)}_{\mathbf{3}})_\mathbf{1}H_d + \gamma_{d1} b^c(QY^{(6)}_{\mathbf{3},I})_\mathbf{1}H_d + \gamma_{d2} b^c(QY^{(6)}_{\mathbf{3},II})_\mathbf{1}H_d \,,
\label{eq:Wq7}
\end{eqnarray}
which lead to the quark mass matrices,
\begin{eqnarray}
\label{eq:Mq_7}
\nonumber&& M_u =\begin{pmatrix}
 \alpha_u Y^{(2)}_3 & \alpha_u Y^{(2)}_5 & \alpha_u Y^{(2)}_4 \\
 \beta_u Y^{(4)}_4 & \beta_u Y^{(4)}_6 & \beta_u Y^{(4)}_5 \\
 \gamma_{u1} Y^{(5)}_6 + \gamma_{u2} Y^{(5)}_9  ~& \gamma_{u1} Y^{(5)}_8 +\gamma_{u2} Y^{(5)}_{11}  ~& \gamma_{u1} Y^{(5)}_7 + \gamma_{u2} Y^{(5)}_{10} \\
 \end{pmatrix}v_u\,, \\[0.1in]
&& M_d =\begin{pmatrix}
 \alpha_d Y^{(1)}_1 & \alpha_d Y^{(1)}_3 & \alpha_d Y^{(1)}_2 \\
 \beta_d Y^{(2)}_3 & \beta_d Y^{(2)}_5 & \beta_d Y^{(2)}_4 \\
 \gamma_{d1} Y^{(6)}_5 + \gamma_{d2}Y^{(6)}_8 ~& \gamma_{d1} Y^{(6)}_7+\gamma_{d2}Y^{(6)}_{10}  ~& \gamma_{d1} Y^{(6)}_6 + \gamma_{d2}Y^{(6)}_{9}\\
 \end{pmatrix}v_d\,,
\end{eqnarray}
The parameters $\alpha_{u,d}$, $\beta_{u,d}$ and $\gamma_{u1,d1}$ can be made real and positive by field redefinition while $\gamma_{u2}$ and $\gamma_{d2}$ are complex. If we impose CP as symmetry on model, all couplings are constrained to be real, and $\gamma_{u2}$ and $\gamma_{d2}$ are either positive or negative. The best fit values of input parameters and the predictions for quark mass ratios and  CKM mixing parameters are
\begin{eqnarray}
\nonumber&&\langle\tau\rangle=-0.4362+1.8184i\,,~~ \beta_u/\alpha_u=9104.8600\,,~~ \gamma_{u1}/\alpha_u=19.7442\,,~~ \gamma_{u2}/\alpha_u=-19.9232\,,\\
\nonumber&&\beta_d/\alpha_d=0.0244\,,\quad \gamma_{d1}/\alpha_d=0.2479\,,\quad \gamma_{d2}/\alpha_d=-0.0021\,,\\
\nonumber&&\alpha_u v_u=0.00009~\text{GeV},\quad \alpha_d v_d=0.00672~\text{GeV}\,, \\
\nonumber&&\theta^q_{12}=0.22735\,,\quad \theta^q_{13}=0.00337\,,\quad \theta^q_{23}=0.03922\,,\quad \delta^q_{CP}=68.9352^{\circ}\,,\\
&&m_u/m_c=0.001811\,,~~ m_c/m_t=0.002714\,,~~ m_d/m_s=0.050529\,,~~ m_s/m_b=0.017686\,.
\end{eqnarray}
It is remarkable that all observables are within the $1\sigma$ experimental ranges.

\item[\textcircled{8}]{\textbf{Model VIII with gCP: 10 free real parameters including $\texttt{Re}(\tau)$ and $\texttt{Im}(\tau)$ }} \\
This model is different from \texttt{Model VII} in the assignment of $u^c$ which is assigned to $\mathbf{\hat{1}}$ with weight $k_{u^c}= 1-k_{Q}$.
Thus the superpotential in the quark sector reads as,
\begin{eqnarray}
\nonumber&& \mathcal{W}_u = \alpha_u u^c(QY^{(1)}_{\mathbf{\hat{3}'}})_\mathbf{\hat{1}}H_u + \beta_u c^c(QY^{(4)}_{\mathbf{3}})_\mathbf{1}H_u + \gamma_{u1} t^c(QY^{(5)}_{\mathbf{\hat{3}'},I})_\mathbf{\hat{1}'}H_u+ \gamma_{u2} t^c(QY^{(5)}_{\mathbf{\hat{3}'},II})_\mathbf{\hat{1}'}H_u \,,\\
&& \mathcal{W}_d= \alpha_d d^c(QY^{(1)}_{\mathbf{\hat{3}'}})_\mathbf{\hat{1}'}H_d + \beta_d s^c(QY^{(2)}_{\mathbf{3}})_\mathbf{1}H_d +\gamma_{d1} b^c(QY^{(6)}_{\mathbf{3},I})_\mathbf{1}H_d + \gamma_{d2} b^c(QY^{(6)}_{\mathbf{3},II})_\mathbf{1}H_d \,,
\label{eq:Wq8}
\end{eqnarray}
where all the couplings are enforced to be real by the generalized CP symmetry. The resulting quark mass matrices are different from
those of Eq.~\eqref{eq:Mq_2} in the first row of $M_u$,
\begin{eqnarray}
\label{eq:Mq_8}
\nonumber && M_u =\begin{pmatrix}
 \alpha_u Y^{(1)}_1 ~& \alpha_u Y^{(1)}_3 ~& \alpha_u Y^{(1)}_2 \\
 \beta_u Y^{(4)}_4 ~& \beta_u Y^{(4)}_6 ~& \beta_u Y^{(4)}_5 \\
 \gamma_{u1} Y^{(5)}_6 + \gamma_{u2} Y^{(5)}_9  ~& \gamma_{u1} Y^{(5)}_8 +\gamma_{u2} Y^{(5)}_{11}  ~& \gamma_{u1} Y^{(5)}_7 + \gamma_{u2} Y^{(5)}_{10} \\
 \end{pmatrix}v_u\,, \\[0.1in]
&& M_d =\begin{pmatrix}
 \alpha_d Y^{(1)}_1 ~& \alpha_d Y^{(1)}_3 ~& \alpha_d Y^{(1)}_2 \\
 \beta_d Y^{(2)}_3 ~& \beta_d Y^{(2)}_5 ~& \beta_d Y^{(2)}_4 \\
 \gamma_{d1} Y^{(6)}_5 + \gamma_{d2}Y^{(6)}_8 ~& \gamma_{d1} Y^{(6)}_7+\gamma_{d2}Y^{(6)}_{10}  ~& \gamma_{d1} Y^{(6)}_6 + \gamma_{d2}Y^{(6)}_{9}\\
 \end{pmatrix}v_d\,.
\end{eqnarray}
The numerical minimisation of the $\chi^2$ function gives the best fit point of the model,
\begin{eqnarray}
\nonumber&&\langle\tau\rangle=0.0617+1.5127i\,,~~ \beta_u/\alpha_u=16.5002\,,~~ \gamma_{u1}/\alpha_u=3919.0300\,,~~ \gamma_{u2}/\alpha_u=-1945.1100\,,\\
\nonumber&&\beta_d/\alpha_d=91.1983\,,\quad \gamma_{d1}/\alpha_d=0.3027\,,\quad \gamma_{d2}/\alpha_d=-1.9970\,,\\
\nonumber&&\alpha_u v_u=0.00019~\text{GeV},\quad \alpha_d v_d=0.00035~\text{GeV}\,, \\
\nonumber&&\theta^q_{12}=0.22737\,,\quad \theta^q_{13}=0.00338\,,\quad \theta^q_{23}=0.03889\,,\quad \delta^q_{CP}=69.2564^{\circ}\,,\\
&&m_u/m_c=0.001965\,,~~ m_c/m_t=0.002729\,,~~ m_d/m_s=0.050363\,,~~ m_s/m_b=0.017679\,,
\end{eqnarray}
which are compatible with experimental data at $1\sigma$ level as well.

\end{description}

\section{\label{sec:quark-lepton-unification} A complete model for quark and lepton }

As shown in sections~\ref{sec:lepton-models} and~\ref{sec:quark-models}, the charged lepton masses and the neutrino oscillation data can be explained very well in the $S'_4$ modular symmetry models, and the $S'_4$ modular symmetry can also help to understand the quark mass hierarchies and CKM mixing matrix. In this section, we shall investigate the quark-lepton unified models which can explain the experimental data of quarks and leptons simultaneously for certain common value of modulus $\tau$. Such kind of models at level $N=3$ has been studied~\cite{Lu:2019vgm,Okada:2018yrn,Okada:2019uoy,Okada:2020rjb} in the literature. It is generally not an easy task to construct a quark-lepton unified model. After trying many possibilities, we find out a realistic quark-lepton unified model with 15 real free parameters. Note that 16 real input parameters~\cite{Lu:2019vgm} or more~\cite{Okada:2018yrn,Okada:2019uoy,Okada:2020rjb} are required to describe the masses and mixing patterns of quark and lepton in the models based on level $N=3$ modular group. Consequently our model is the most predictive modular quark-lepton unification model as far as we know at present.

In the lepton sector, we assign the three generations of left-handed lepton doublets $L$ and right-handed neutrinos $N^c$ to two triplets $\mathbf{3}$ of $S'_4$, while the right-handed charged leptons $E_1^c$, $E_2^c$ and $E_3^c$ transform as $\mathbf{1}$, $\mathbf{1}$ and $\mathbf{\hat{1}'}$ respectively. We choose the modular weights of lepton fields as $k_{N^c}=0$, $k_L=2$, $k_{E_1^c}=2$, $k_{E_2^c}=0$ and $k_{E_3^c}=1$. Then the modular invariant superpotential of the lepton sector is given by
\begin{eqnarray}
\nonumber && \mathcal{W}_{e}=\alpha_e (E^{c}_{1}L Y_{\mathbf{3}}^{(4)} )_{\mathbf{1}}H_{d}+\beta_e (E^{c}_{2}L Y_{\mathbf{3}}^{(2)} )_{\mathbf{1}}H_{d} +\gamma_e (E^{c}_{3}L  Y_{\mathbf{\hat{3}}}^{(3)} )_{\mathbf{1}}H_{d}\,,\\
&& \mathcal{W}_{\nu}=g_1( N^{c} LY_{\mathbf{2}}^{(2)})_{\mathbf{1}}H_{u}+g_2(N^{c} L Y_{\mathbf{3}}^{(2)})_{\mathbf{1}}H_{u}+\Lambda (N^{c}N^{c})_{\mathbf{1}}\,,
\label{eq:unification}
\end{eqnarray}
where all couplings $\alpha_{e}$, $\beta_{e}$, $\gamma_{e}$, $g_1$, $g_2$ and $\Lambda$ are real because of the generalized CP invariance. We see that the charged lepton sector is different from $C_4$ in the values of $f_{E_1}(Y)$ and the neutrino sector is exactly the model $S_2$. Then we can read out the lepton mass matrices as follows,
\begin{eqnarray}
\label{eq:Ml}
\nonumber&& M_{e}=\begin{pmatrix}
\alpha_e  Y_{4}^{(4)} ~&~ \alpha_e  Y_{6}^{(4)} ~&~  \alpha_e  Y_{5}^{(4)}\\
\beta_e  Y_{3}^{(2)} ~&~ \beta_e  Y_{5}^{(2)} ~&~ \beta_e Y_{4}^{(2)} \\
 \gamma_e  Y_{2}^{(3)} ~&~ \gamma_e  Y_{4}^{(3)} ~&~ \gamma_e  Y_{3}^{(3)}
\end{pmatrix}v_{d}\,,~\quad~ M_{N}=\begin{pmatrix}
 1 ~&~ 0 ~&~ 0 \\
 0 ~&~ 0 ~&~ 1 \\
 0 ~&~ 1 ~&~ 0
\end{pmatrix}\Lambda\,, \\
&&  M_{D}=\begin{pmatrix}
0 ~&~ g_1 Y_{1}^{(2)}-g_2 Y_{5}^{(2)} ~&~ g_1 Y_{2}^{(2)}+g_2 Y_{4}^{(2)}\\
g_1 Y_{1}^{(2)}+g_2 Y_{5}^{(2)} ~&~ g_1 Y_{2}^{(2)} ~&~ -g_2 Y_{3}^{(2)} \\
g_1 Y_{2}^{(2)}-g_2 Y_{4}^{(2)} ~&~ g_2 Y_{3}^{(2)} ~&~ g_1 Y_{1}^{(2)}
\end{pmatrix}v_{u}\,.
\end{eqnarray}
In the quark sector, the left-handed quarks $Q$ are assigned to triplet $\mathbf{3}$ of $S'_4$, $u^c$, $c^c$ and $t^c$ transform as $\mathbf{1}$, $\mathbf{1}$ and $\mathbf{\hat{1}'}$ respectively under $S'_4$, down type quarks $d^c,\,s^c,\,b^c$ transform as $\mathbf{1}',\,\mathbf{\hat{1}},\,\mathbf{\hat{1}'}$ respectively. Note that $u^c$ and $c^c$ are distinguished by their different modular weights. We choose the weights of quark fields to fulfill the conditions $k_{Q} = 4-k_{u^c} = 6-k_{c^c} = 3-k_{t^c} = 4-k_{d^c} = 5-k_{s^c} = 5-k_{b^c}$. The superpotential of the quark sector is given by,
\begin{eqnarray}
\nonumber&& \mathcal{W}_{u}=\alpha_{u}(u^{c}Q Y_{\mathbf{3}}^{(4)} )_{\mathbf{1}}H_{u}+\beta_{u}(c^{c}QY_{\mathbf{3},I}^{(6)} )_{\mathbf{1}}H_{u}+\gamma_{u}( c^{c}QY_{\mathbf{3},II}^{(6)} )_{\mathbf{1}}H_{u}+\delta_{u}(t^{c}QY_{\mathbf{\hat{3}}}^{(3)})_{\mathbf{1}}H_{u}\,,\\
&& \mathcal{W}_{d}=\alpha_{d}(d^{c}Q Y_{\mathbf{3}'}^{(4)} )_{\mathbf{1}}H_{d}+\beta_{d}(s^{c} QY_{\mathbf{\hat{3}'},I}^{(5)} )_{\mathbf{1}}H_{d}+\gamma_{d}(s^{c} QY_{\mathbf{\hat{3}'},II}^{(5)})_{\mathbf{1}}H_{d}+\delta_{d}(b^{c}Q Y_{\mathbf{\hat{3}}}^{(5)})_{\mathbf{1}}H_{d} \,,
\label{eq:WqII}
\end{eqnarray}
which lead to the quark mass matrices,
\begin{eqnarray}
\label{eq:MqII}
\nonumber&& M_{u}=\begin{pmatrix}
 \alpha_{u} Y_{4}^{(4)} ~&~ \alpha_{u} Y_{6}^{(4)} ~&~ \alpha_{u} Y_{5}^{(4)} \\
 \beta_{u} Y_{5}^{(6)}+\gamma_{u} Y_{8}^{(6)} ~&~ \beta_{u} Y_{7}^{(6)}+\gamma_{u} Y_{10}^{(6)} ~&~ \beta_{u} Y_{6}^{(6)}+\gamma_{u} Y_{9}^{(6)} \\
 \delta_{u} Y_{2}^{(3)} ~&~ \delta_{u} Y_{4}^{(3)} ~&~ \delta_{u} Y_{3}^{(3)} \\
\end{pmatrix}v_{u}\,, \\[0.1in]
&& M_{d}=\begin{pmatrix}
\alpha_{d} Y_{7}^{(4)} ~&~ \alpha_{d} Y_{9}^{(4)} ~&~ \alpha_{d} Y_{8}^{(4)} \\
\beta_{d} Y_{6}^{(5)}+\gamma_{d} Y_{9}^{(5)} ~&~ \beta_{d} Y_{8}^{(5)}+\gamma_{d} Y_{11}^{(5)} ~&~ \beta_{d} Y_{7}^{(5)}+\gamma_{d} Y_{10}^{(5)} \\
\delta_{d} Y_{3}^{(5)} ~&~ \delta_{d} Y_{5}^{(5)} ~&~ \delta_{d} Y_{4}^{(5)}
\end{pmatrix}v_{d}\,.
\end{eqnarray}
The parameters $\alpha_{u,d}$, $\beta_{u,d}$ and $\delta_{u,d}$ can be made real and positive by field redefinition while $\gamma_{u}$ and $\gamma_{d}$ are complex. If we impose CP as symmetry on this model, all couplings are constrained to be real, and $\gamma_{u}$ and $\gamma_{d}$ are either positive or negative.

It is notable that the model has less free parameters  than the number of observable quantities including quark and lepton masses and mixing parameters. We perform a comprehensive numerical scan over the parameter space, we find that good agreement with experimental data can be achieved for the following value of $\tau$ common to quark and lepton sectors,
\begin{equation}
\langle\tau\rangle=-0.2123+1.5201i\,,
\end{equation}
which is mainly determined by the quark masses and CKM mixing parameters. Given this value of $\tau$, the charged lepton masses can be reproduced by adjusting $\alpha_e$, $\beta_{e}$ and $\gamma_{e}$, only two real parameters $g_1^2v_u^2/\Lambda$ and $g_2/g_1$ describe the nine neutrino observables: three neutrino masses, three neutrino mixing angles and three CP violating phases. The best fit values of the free parameters of both lepton and quark sectors are found to be
\begin{eqnarray}
\nonumber&&\hskip-0.3in \beta_u/\alpha_u=325.6502\,,~~ \gamma_{u}/\alpha_u=2427.3101\,,~~\delta_{u}/\alpha_u=219.3019\,,\\
\nonumber&&\hskip-0.3in  \alpha_u v_u= 2.7758\times 10^{-5}~\text{GeV}\,,~~\beta_d/\alpha_d=466.6990\,,~~\gamma_{d}/\alpha_d=-234.0473\,,\\
\nonumber&&\hskip-0.3in  \delta_{d}/\alpha_d=2.3388\,,~~\alpha_d v_d= 1.72111\times 10^{-5}~\text{GeV}\,,~~\beta_{e}/\alpha_e=0.0187\,,\\
&& \hskip-0.3in  \gamma_{e}/\alpha_e=0.1466\,,~~ g_2/g_1=0.6834\,,~\alpha_e v_d=16.8880\text{MeV}\,,~g_1^2v^2_u/\Lambda= 0.3043 ~\text{meV}\,.
\end{eqnarray}
The masses and mixing parameters of quarks and leptons are predicted to be
\begin{eqnarray}
\nonumber&&\theta^q_{12}=0.22752\,,~~ \theta^q_{13}=0.003379\,,~~ \theta^q_{23}=0.038886\,,~~ \delta^q_{CP}=75.9958^{\circ}\,,\\
\nonumber&&m_u/m_c=0.001929\,,~~ m_c/m_t=0.002725\,,~~ m_d/m_s=0.050345\,,~~ m_s/m_b=0.017726\,,\\
\nonumber&&\sin^2\theta^l_{12}=0.34981\,,~~ \sin^2\theta^l_{13}=0.02193\,,~~ \sin^2\theta^l_{23}=0.56393\,,\\
\nonumber&&\delta^l_{CP}=266.1824^{\circ}\,,~~ \alpha_{21}=1.1482\pi\,,~~ \alpha_{31}=0.1522\pi\,,\\
\nonumber&&m_1=3.5269 ~\text{meV}\,,~~ m_2=9.2919 ~\text{meV}\,,~~ m_3=50.2404 ~\text{meV}\,,\\
\label{eq:num-quark-lepton-unif}&&\sum_im_i=63.0592 ~\text{meV}\,,~~ |m_{ee}|=2.5480~\text{meV}\,.
\end{eqnarray}
We see that the solar mixing angle $\theta^l_{12}$ and $\delta^q_{CP}$ are within the $3\sigma$ experimental region, $\delta^l_{CP}$ are within the $2\sigma$ experimental region, and all other observables fall in the $1\sigma$ ranges. The sum of neutrino masses is determined to be 63.0592 meV, this is compatible with the latest bound $\sum_im_i<120$ meV at $95\%$ confidence level from Planck~\cite{Aghanim:2018eyx}.

Moreover, we find another quite similar model which can be obtained from the above model by taking the down quark modular weight $k_{d^{c}}=6-k_Q$. Thus the superpotential of the down quark sector reads as
\begin{eqnarray}
&& \mathcal{W}_{d}=\alpha_{d}(d^{c}Q Y_{\mathbf{3}'}^{(6)} )_{\mathbf{1}}H_{d}+\beta_{d}(s^{c} QY_{\mathbf{\hat{3}'},I}^{(5)} )_{\mathbf{1}}H_{d}+\gamma_{d}(s^{c} QY_{\mathbf{\hat{3}'},II}^{(5)})_{\mathbf{1}}H_{d}+\delta_{d}(b^{c}Q Y_{\mathbf{\hat{3}}}^{(5)})_{\mathbf{1}}H_{d} \,.
\label{eq:WqIII}
\end{eqnarray}
All the other superpotentials for lepton and quark masses are not changed. After numerically scanning overall the parameter space, we find the numerical results are quite similar to those of the above model given in Eq.~\eqref{eq:num-quark-lepton-unif}, consequently we will not show the corresponding numbers here.

\section{\label{sec:conclusion}Conclusion}

The modular invariance is a promising framework to describe the masses and mixing of both quarks and leptons~\cite{Feruglio:2017spp}. The homogeneous finite modular group $\Gamma'_N$ provides new opportunity for understanding the flavor structure of quarks and leptons based on modular invariance. $\Gamma'_2$ is identical to $\Gamma_2\cong S_3$, $\Gamma'_N$ is the double covering of the inhomogeneous finite modular group $\Gamma_N$ for $N>2$, and $\Gamma_N$ is isomorphic to the quotient of $\Gamma'_N$ over its center $\{1, R\}$, i.e., $\Gamma_{N}\cong\Gamma'_N/\{1, R\}$. It is notable that texture zeros of fermion mass matrices can be naturally obtained from $\Gamma'_N$, and $\Gamma'_3\cong T'$ has been studied in~\cite{Liu:2019khw,Ding:2019xna}. In the present work, we have considered the modular group $\Gamma'_4\equiv S'_4$ in the setup of modular invariance approach.

The weight 1 modular forms of level 4 are constructed in terms of the Dedekind eta function, and they can be arranged into a triplet  $Y^{(1)}_{\mathbf{\hat{3}'}}(\tau)=(Y_1(\tau), Y_2(\tau), Y_3(\tau))^{T}$ which transforms as $\mathbf{\hat{3}'}$ of $S'_4$. The higher weight modular forms up to weight 6 are built from the tensor products of $Y^{(1)}_{\mathbf{\hat{3}'}}(\tau)$, and they are homogeneous polynomials of $Y_{1,2,3}$. The odd weight modular forms can be decomposed into the hatted representations  $\mathbf{\hat{1}}$, $\mathbf{\hat{1}}^{\prime}$, $\mathbf{\hat{2}}$, $\mathbf{\hat{3}}$ and $\mathbf{\hat{3}}^{\prime}$ of $S'_4$ while the even weight modular forms can be organized into the other representations $\mathbf{1}$, $\mathbf{1}^{\prime}$, $\mathbf{2}$, $\mathbf{3}$ and $\mathbf{3}^{\prime}$ in common with $S_4$. The results are summarized in table~\ref{Tab:Level4_MM}. Solving the consistency condition, we find the generalized CP transformation corresponding to $\tau\rightarrow-\tau^{*}$ is $X_{\mathbf{r}}=\rho_{\mathbf{r}}(S)$ which is a combination of the modular symmetry transformation $S$ and the canonical CP transformation. All couplings in the Lagrangian would be real if the generalized CP symmetry is imposed.

We perform a systematical analysis of $S'_4$ modular models for lepton masses and mixing with/without generalized CP. We assume that the left-handed leptons transform as triplet of $S'_4$, and the right-handed charged leptons are assigned to singlets under $S'_4$, and we consider both the case where neutrino masses are described by the Weinberg operator and the case where neutrino masses arise from the type I seesaw mechanism. The charged lepton mass matrix can only take four possible forms in table~\ref{tab:charged lepton} and the forms of the neutrino mass matrices are summarized in table~\ref{tab:neutrino} if the weights of the involved modular forms are less than 4. The charged lepton masses $m_e$, $m_{\mu}$ and $m_{\tau}$ are in a one-to-one correspondence with the parameters $\alpha$, $\beta$ and $\gamma$ which can be taken real without loosing generality. We look for phenomenologically viable models with a minimal amount of free parameters. We find fifteen predictive lepton models which can describe the neutrino masses, mixing angles and CP violation phases in terms of five real parameters $|g_2/g_1|$, $\arg{(g_2/g_1)}$, $\texttt{Re}(\tau)$, $\texttt{Im}(\tau)$ and the overall scale $g^2_1v_u^2/\Lambda$. If generalized CP symmetry is imposed, the models have more predictive power and the phase $\arg{(g_2/g_1)}$ is restricted to be $0$ or $\pi$. Thus only four real input parameters $g_2/g_1$, $\texttt{Re}(\tau)$, $\texttt{Im}(\tau)$ and $g^2_1v_u^2/\Lambda$ are left in the neutrino sector, and we find seven out of the fifteen models
can fit the charged lepton masses and neutrino oscillation data very well,
as shown in table~\ref{tab:goodmodel_gCP}. The different observables are correlated with each other, as displayed in figures~\ref{fig:C1S5_I}-\ref{fig:C1S5_gCP}. A remarkable feature of these models is that the light neutrino masses can be very tiny, while the neutrino masses are typically quasi-degenerate in previous models based on $\Gamma_N$ modular group.

We have extended the $S'_4$ modular symmetry to quark sector, different possible assignments (triplet, the direct sum of a doublet and a singlet, or the direct sum of three singlets) of the quark fields under $S'_4$ are considered. For the first time we consider the generalized CP symmetry when constructing modular invariant models for quark masses and mixing. Because of the rich structure of the $S'_4$ modular group, we find many models can accommodate the observed patterns of quark masses and CKM mixing matrix. For illustration, we select eight benchmark models in which all the best fit values of observables fall in the $1\sigma$ experimental ranges. It is remarkable that the hierarchical quark masses, quark mixing angles and CP violation phase can be described very well by models with only nine real parameters including real and imaginary parts of the modulus $\tau$. Note that so far few predictive models use ten~\cite{Lu:2019vgm} or more free parameters~\cite{Okada:2018yrn,Okada:2019uoy,Okada:2020rjb}
to describe quark masses and CKM mixing matrix, consequently our benchmark models for quarks can be considered as the minimal ones.

Finally we present a quark-lepton unified model, this model predicting twenty-two observables is characterized by fifteen real parameters: thirteen real couplings $\alpha_{e, u,d}$, $\beta_{e, u,d}$, $\gamma_{e, u,d}$, $\delta_{u,d}$, $g_2/g_1$, $g_1^2v_u^2/\Lambda$ and the complex modulus $\tau$, and it is the most predictive modular invariant model for quark-lepton unification as we know. The masses and mixing of quarks and leptons can be explained simultaneously for a common value of the complex modulus $\tau$. The value of $\tau$ is mainly fixed by the precisely measured quark masses and mixing, then the entire neutrino sector including the three neutrino masses as well as the lepton mixing matrix only depends on two real parameters
$g_2/g_1$ and $g_1^2v_u^2/\Lambda$. We conclude that $S'_4$ modular symmetry is a promising framework to understand the flavor structure of quarks and leptons, and generalized CP can help to construct minimal and predictive models with modular symmetry. However, the bottom-up approach of modular invariance is less constrained so that a large number of phenomenologically viable models with few parameters could be constructed, as our present work shows. This drawback could potentially be overcome by embedding this approach in a more fundamental theory, for instance both modular weights and representation assignments of the matter fields are severely restricted in the eclectic flavor scheme~\cite{Nilles:2020nnc,Nilles:2020kgo} and thus model building with finite modular flavor symmetries becomes much more restrictive.

\vskip0.4in

\textbf{Note added}: During the final preparations of this work, a paper~\cite{Novichkov:2020eep} dealing with the same topic appeared on the arXiv. We use different representation basis of $S'_4$, the Clebsch-Gordan coefficients in our basis is simpler, and this basis is convenient to classify the $S'_4$ modular models. Modular forms of level $N=4$ up to weight 6 are constructed in this work, and higher weight modular forms until weight 10 are given in~\cite{Novichkov:2020eep}. The authors of~\cite{Novichkov:2020eep} present one Weinberg operator model and one type I seesaw model, and the right-handed charged leptons $E^c$ are assigned to a triplet of $S'_4$ in~\cite{Novichkov:2020eep}. We perform a systematical classification of modular $S'_4$ symmetry models for leptons with/without generalized CP symmetry, $E^c$ are assumed to transform as singlets under $S'_4$ in this work. We also apply the $S'_4$ modular symmetry to the quark sector, and construct a quark-lepton unified model. Our work significantly extends the model construction of~\cite{Novichkov:2020eep}.

\section*{Acknowledgements}

XGL and GJD are supported by the National Natural Science Foundation of China under Grant Nos~11975224, 11835013 and 11947301. CYY is supported in part by the Grants No.~NSFC-11975130, No.~NSFC-12035008, No.~NSFC-12047533, by the National Key Research and Development Program of China under Grant No.~2017YFA0402200 and the China Post-doctoral Science Foundation under Grant No.~2018M641621.

\newpage

\section*{Appendix}

\setcounter{equation}{0}
\renewcommand{\theequation}{\thesection.\arabic{equation}}

\begin{appendix}

\section{\label{sec:S4DC-app}Group theory of $S_4^{\prime}$}

The double covering group of $S_4$ has 48 elements, and it can be generated by three generators $S,T$ and $R$ satisfying the multiplication rules:
\begin{equation}
  S^2=R,\quad T^4=(ST)^3=1,\quad R^2=1,\quad RT=TR\,.
\end{equation}
After we input these multiplication rules in GAP~\cite{GAP}, its group ID can be determined as [48, 30]. Notice that $S_4$ is not a subgroup of $S'_4$, it is isomorphic to the quotient group of $S'_4$ over $Z_2^R$, i.e. $S_4\cong S_4^{\prime}/Z_2^R$, where $Z_2^R=\{1,R\}$ is a normal subgroup of $S_4^{\prime}$.
The homogeneous finite modular group $S'_4$ is isomorphic to $A_4\rtimes Z_4$. Therefore it can be expressed in terms of another set of generators $s$, $t$ and $r$ obeying the relations:
\begin{equation}
s^2=(st)^3=t^3=1,\quad r^4=1,\quad r s r^{-1}=s, \quad r t r^{-1} = (st)^2\,,
\end{equation}
where $s$ and $t$ generate a $A_4$ subgroup, $r$ generates a $Z_4$ subgroup, and the last two relations define the semidirect product "$\rtimes$". The generators $s,t$ and $r$ are related to $S$, $T$ and $R$ by
\begin{align}
\nonumber & s = T^2 R,\quad t=(ST)^2, \quad r=T \,,\\
& S= t^2 r^3, \quad T=r, \quad R=r^2 s\,.
\end{align}
All the elements of $S_4^{\prime}$ group can be divided into 10 conjugacy classes:
\begin{eqnarray}
\nonumber 1C_1&=&\{1\}\,, \\
\nonumber 1C_2&=&\{R\}=(1C_1)\cdot R\,, \\
\nonumber 3C_2&=&\left\{T^2,~ST^2S^3,~(ST^2)^2\right\}\,, \\
\nonumber3C_2'&=&\left\{T^2R,~ST^2S,~(ST^2)^2R\right\}=(3C_2)\cdot R\,, \\
\nonumber 8C_3&=&\left\{ST,~TS,~(ST)^2,~(TS)^2,~TS^3T^2,~T^2ST^3,~T^2S^3T,~T^3ST^2\right\}\,, \\
\nonumber6C_4&=&\left\{S,~TST^3,~T^2ST^2,~T^3ST,~TST^2S^3,~ST^2S^3T\right\}\,, \\
\nonumber 6C_4'&=&\left\{T,~ST^2,~T^2S,~T^3S^2,~TST,~STS^3\right\}\,, \\
\nonumber 6C_4''&=&\left\{SR,~TST^3R,~T^2ST^2R,~T^3STR,~TST^2S,~ST^2S^3TR\right\}=(6C_4)\cdot R\,, \\
\nonumber 6C_4'''&=&\left\{TR,~ST^2R,~T^2SR,~T^3,~TSTR,~STS\right\}=(6C_4')\cdot R \,, \\
\nonumber8C_6&=&\left\{STR,~TSR,~(ST)^2R,~(TS)^2R,~TS^3T^2R,~T^2ST^3R,~\right.\\
  && \left.  T^2S^3TR,~T^3ST^2R\right\}=(8C_3)\cdot R \,,
\end{eqnarray}
where $kC_n$ denotes a conjugacy class with $k$ elements of order $n$. Note that one half of these conjugacy classes can be written as the product of the other half with $R$. There are four one-dimensional irreducible representations $\mathbf{1},\mathbf{1}^{\prime},\mathbf{\hat{1}}$ and $\mathbf{\hat{1}}^{\prime}$, two two-dimensional irreducible representations $\mathbf{2}$ and $\mathbf{\hat{2}}$, and four three-dimensional irreducible representations $\mathbf{3},\mathbf{3}^{\prime},\mathbf{\hat{3}}$ and $\mathbf{\hat{3}}^{\prime}$. We have summarized the explicit matrix representations in table~\ref{tab:Rep_matrix}. In the representations $\mathbf{1}$, $\mathbf{1}^{\prime}$, $\mathbf{2}$, $\mathbf{3}$ and $\mathbf{3}^{\prime}$, the generator $R=1$ is identity matrix, the representation matrices of $S$ and $T$ coincide with those of $S_4$~\cite{Gui-JunDing:2019wap}, consequently $S'_4$ can not be distinguished from $S_4$ in these representations since they are represented by the same set of matrices. In the representations $\mathbf{\hat{1}}$, $\mathbf{\hat{1}}^{\prime}$, $\mathbf{\hat{2}}$, $\mathbf{\hat{3}}$ and $\mathbf{\hat{3}}^{\prime}$, the generator $R=-1$. The character table of $S_4^{\prime}$ can be obtained directly as shown in table~\ref{tab:character}. Moreover, the Kronecker products between all irreducible representations are given as follows:
\begin{eqnarray}
\nonumber&&\mathbf{1} \otimes \mathbf{1} = \mathbf{1}^{\prime} \otimes \mathbf{1}^{\prime} = \mathbf{\hat{1}} \otimes \mathbf{\hat{1}}^{\prime} = \mathbf{1}, \quad \mathbf{1} \otimes \mathbf{\hat{1}} = \mathbf{1}^{\prime} \otimes \mathbf{\hat{1}}^{\prime} = \mathbf{\hat{1}}  ,\\
\nonumber&&\mathbf{1} \otimes \mathbf{1}^{\prime} = \mathbf{\hat{1}} \otimes \mathbf{\hat{1}} = \mathbf{\hat{1}}^{\prime} \otimes \mathbf{\hat{1}}^{\prime} = \mathbf{1}^{\prime},\quad \mathbf{1} \otimes \mathbf{\hat{1}}^{\prime} = \mathbf{1}^{\prime} \otimes \mathbf{\hat{1}} = \mathbf{\hat{1}}^{\prime} ,\\
\nonumber&&\mathbf{1} \otimes \mathbf{2} = \mathbf{1}^{\prime} \otimes \mathbf{2} = \mathbf{\hat{1}} \otimes \mathbf{\hat{2}} = \mathbf{\hat{1}}^{\prime} \otimes \mathbf{\hat{2}} = \mathbf{2},\quad \mathbf{1} \otimes \mathbf{\hat{2}} = \mathbf{1}^{\prime} \otimes \mathbf{\hat{2}} = \mathbf{\hat{1}} \otimes \mathbf{2} = \mathbf{\hat{1}}^{\prime} \otimes \mathbf{2} = \mathbf{\hat{2}} ,\\
\nonumber&&\mathbf{1} \otimes \mathbf{3} = \mathbf{1}^{\prime} \otimes \mathbf{3}^{\prime} = \mathbf{\hat{1}} \otimes \mathbf{\hat{3}}^{\prime} = \mathbf{\hat{1}}^{\prime} \otimes \mathbf{\hat{3}} = \mathbf{3} ,\quad \mathbf{1} \otimes \mathbf{\hat{3}} = \mathbf{1}^{\prime} \otimes \mathbf{\hat{3}}^{\prime} = \mathbf{\hat{1}} \otimes \mathbf{3} = \mathbf{\hat{1}}^{\prime} \otimes \mathbf{3}^{\prime} = \mathbf{\hat{3}} ,\\
\nonumber&&\mathbf{1} \otimes \mathbf{3}^{\prime} = \mathbf{1}^{\prime} \otimes \mathbf{3} = \mathbf{\hat{1}} \otimes \mathbf{\hat{3}} = \mathbf{\hat{1}}^{\prime} \otimes \mathbf{\hat{3}}^{\prime} = \mathbf{3}^{\prime},\quad \mathbf{1} \otimes \mathbf{\hat{3}}^{\prime} = \mathbf{1}^{\prime} \otimes \mathbf{\hat{3}} = \mathbf{\hat{1}} \otimes \mathbf{3}^{\prime} = \mathbf{\hat{1}}^{\prime} \otimes \mathbf{3} = \mathbf{\hat{3}}^{\prime} ,\\
\nonumber&&\mathbf{2} \otimes \mathbf{2} = \mathbf{\hat{2}} \otimes \mathbf{\hat{2}} = \mathbf{1} \oplus \mathbf{1}^{\prime} \oplus \mathbf{2} ,\quad \mathbf{2} \otimes \mathbf{\hat{2}} = \mathbf{\hat{1}} \oplus \mathbf{\hat{1}}^{\prime} \oplus \mathbf{\hat{2}} ,\\
\nonumber&&\mathbf{2} \otimes \mathbf{3} = \mathbf{2} \otimes \mathbf{3}^{\prime} = \mathbf{\hat{2}} \otimes \mathbf{\hat{3}} = \mathbf{\hat{2}} \otimes \mathbf{\hat{3}}^{\prime} = \mathbf{3} \oplus \mathbf{3}^{\prime}, \quad \mathbf{2} \otimes \mathbf{\hat{3}} = \mathbf{2} \otimes \mathbf{\hat{3}}^{\prime} = \mathbf{\hat{2}} \otimes \mathbf{3} = \mathbf{\hat{2}} \otimes \mathbf{3}^{\prime} = \mathbf{\hat{3}} \oplus \mathbf{\hat{3}}^{\prime}  ,\\
\nonumber&&\mathbf{3} \otimes \mathbf{3} = \mathbf{3}^{\prime} \otimes \mathbf{3}^{\prime} = \mathbf{\hat{3}} \otimes \mathbf{\hat{3}}^{\prime} = \mathbf{1} \oplus \mathbf{2} \oplus \mathbf{3} \oplus \mathbf{3}^{\prime},\quad \mathbf{3} \otimes \mathbf{\hat{3}} = \mathbf{3}^{\prime} \otimes \mathbf{\hat{3}}^{\prime} = \mathbf{\hat{1}} \oplus \mathbf{\hat{2}} \oplus \mathbf{\hat{3}} \oplus \mathbf{\hat{3}}^{\prime}  ,\\
&&\mathbf{3} \otimes \mathbf{3}^{\prime} = \mathbf{\hat{3}} \otimes \mathbf{\hat{3}} = \mathbf{\hat{3}}^{\prime} \otimes \mathbf{\hat{3}}^{\prime} = \mathbf{1}^{\prime} \oplus \mathbf{2} \oplus \mathbf{3} \oplus \mathbf{3}^{\prime},\quad \mathbf{3} \otimes \mathbf{\hat{3}}^{\prime} = \mathbf{3}^{\prime} \otimes \mathbf{\hat{3}} = \mathbf{\hat{1}}^{\prime} \oplus \mathbf{\hat{2}} \oplus \mathbf{\hat{3}} \oplus \mathbf{\hat{3}}^{\prime} .
\end{eqnarray}

\begin{table}[!t]
\centering
\begin{tabular}{|c|c|c|c|}\hline\hline
 & $S$ & $T$ & $R$\\ \hline
    $\mathbf{1},\mathbf{1^{\prime}}$ & $\pm 1$ & $\pm 1$ & $1$ \\ \hline
    $\mathbf{\hat{1}},\mathbf{\hat{1}^{\prime}}$ & $\pm i$ & $\mp i$ & $-1$ \\ \hline
    $\mathbf{2}$ & $\begin{pmatrix}{0} & {1} \\ {1} & {0}\end{pmatrix}$ & $\begin{pmatrix}{0} & {\omega^{2}} \\ {\omega} & {0}\end{pmatrix}$ & $\begin{pmatrix}{1} & {0} \\ {0} & {1}\end{pmatrix}$ \\ \hline
    $\mathbf{\hat{2}}$ & $i\begin{pmatrix}{0} & {1} \\ {1} & {0}\end{pmatrix}$ & $-i\begin{pmatrix}{0} & {\omega^{2}} \\ {\omega} & {0}\end{pmatrix}$ & $-\begin{pmatrix}{1} & {0} \\ {0} & {1}\end{pmatrix}$ \\ \hline
    $\mathbf{3},\mathbf{3^{\prime}}$ & $\pm\dfrac{1}{3}\begin{pmatrix}{1} & {-2} & {-2} \\ {-2} & {-2} & {1} \\ {-2} & {1} & {-2}\end{pmatrix}$ & $\pm\dfrac{1}{3}\begin{pmatrix}{1} & {-2 \omega^{2}} & {-2 \omega} \\ {-2} & {-2 \omega^{2}} & {\omega} \\ {-2} & {\omega^{2}} & {-2 \omega}\end{pmatrix}$ & $\begin{pmatrix}{1} & {0} & {0} \\ {0} & {1} & {0} \\ {0} & {0} & {1}\end{pmatrix}$ \\ \hline
    $\mathbf{\hat{3}}, \mathbf{\hat{3}^{\prime}}$ & $\pm\dfrac{i}{3}\begin{pmatrix}{1} & {-2} & {-2} \\ {-2} & {-2} & {1} \\ {-2} & {1} & {-2}\end{pmatrix}$ & $\mp\dfrac{i}{3}\begin{pmatrix}{1} & {-2 \omega^{2}} & {-2 \omega} \\ {-2} & {-2 \omega^{2}} & {\omega} \\ {-2} & {\omega^{2}} & {-2 \omega}\end{pmatrix}$ & $-\begin{pmatrix}{1} & {0} & {0} \\ {0} & {1} & {0} \\ {0} & {0} & {1}\end{pmatrix}$ \\ \hline\hline
  \end{tabular}
\caption{The representation matrices of the generators $S, T$ and $R$ in the irreducible representations of $S_{4}^{\prime}$ in our working basis, where $\omega=e^{2 \pi i/3}$.}
\label{tab:Rep_matrix}
\end{table}

\begin{table}[t!]
  \centering
  \begin{tabular}{|c|c|c|c|c|c|c|c|c|c|c|}
    \hline\hline
    Classes & $1C_1$ & $1C_2$ & $3C_2$ & $3C_2'$ & $8C_3$ & $6C_4$ & $6C_4'$ & $6C_4''$ & $6C_4'''$ & $8C_6$\\ \hline
    $G$ & $1$ & $R$ & $T^2$ & $T^2R$ & $ST$ & $S$ & $T$ & $SR$ & $TR$ & $STR$\\ \hline
    $\mathbf{1}$ & $1$ & $1$ & $1$ & $1$ & $1$ & $1$ & $1$ & $1$ & $1$ & $1$\\ \hline
    $\mathbf{1^{\prime}}$ & $1$ & $1$ & $1$ & $1$ & $1$ & $-1$ & $-1$ & $-1$ & $-1$ & $1$\\ \hline
    $\mathbf{\hat{1}}$ & $1$ & $-1$ & $-1$ & $1$ & $1$ & $i$ & $-i$ & $-i$ & $i$ & $-1$\\ \hline
    $\mathbf{\hat{1}^{\prime}}$ & $1$ & $-1$ & $-1$ & $1$ & $1$ & $-i$ & $i$ & $i$ & $-i$ & $-1$\\ \hline
    $\mathbf{2}$ & $2$ & $2$ & $2$ & $2$ & $-1$ & $0$ & $0$ & $0$ & $0$ & $-1$\\ \hline
    $\mathbf{\hat{2}}$ & $2$ & $-2$ & $-2$ & $2$ & $-1$ & $0$ & $0$ & $0$ & $0$ & $1$\\ \hline
    $\mathbf{3}$ & $3$ & $3$ & $-1$ & $-1$ & $0$ & $-1$ & $1$ & $-1$ & $1$ & $0$\\ \hline
    $\mathbf{3^{\prime}}$ & $3$ & $3$ & $-1$ & $-1$ & $0$ & $1$ & $-1$ & $1$ & $-1$ & $0$\\ \hline
    $\mathbf{\hat{3}}$ & $3$ & $-3$ & $1$ & $-1$ & $0$ & $-i$ & $-i$ & $i$ & $i$ & $0$\\ \hline
    $\mathbf{\hat{3}^{\prime}}$ & $3$ & $-3$ & $1$ & $-1$ & $0$ & $i$ & $i$ & $-i$ & $-i$ & $0$\\ \hline\hline
  \end{tabular}
\caption{\label{tab:character}Character table of $S_4^{\prime}$, the representative element of each conjugacy class is given in the second row.}
\end{table}

Corresponding to the above direct product rule, we give the Clebsch-Gordan (CG) coefficients of $S_4^{\prime}$ one by one in our working basis. All CG coefficients can be expressed in the form of $\alpha\otimes\beta$, we use $\alpha_i(\beta_i)$ to denote the component of the left (right) basis vector $\alpha (\beta)$. For direct products involving singlet $\mathbf{\harp{\hat{1}}}^{(\prime)}$, their CG coefficients are as follows
\begin{equation}
\begin{array}{llll}
\mathbf{\harp{\hat{1}}}^{(\prime)} \otimes \mathbf{\harp{\hat{1}}}^{(\prime)}\rightarrow\mathbf{\harp{\hat{1}}}^{(\prime)}  & \left\{\begin{array}{c}
    ~\\[10.5ex]p=\mathrm{even}\\[10.5ex]~
  \end{array}\right. &
  \left.\begin{array}{l}
\mathbf{1}\otimes\mathbf{1}\to\mathbf{1}\\
\mathbf{1}\otimes\mathbf{1'}\to\mathbf{1'}\\
\mathbf{1}\otimes\mathbf{\hat{1}}\to\mathbf{\hat{1}}\\
\mathbf{1}\otimes\mathbf{\hat{1}'}\to\mathbf{\hat{1}'}\\
\mathbf{1'}\otimes\mathbf{1'}\to\mathbf{1}\\
\mathbf{1'}\otimes\mathbf{\hat{1}}\to\mathbf{\hat{1}'}\\
\mathbf{1'}\otimes\mathbf{\hat{1}'}\to\mathbf{\hat{1}}\\
\mathbf{\hat{1}}\otimes\mathbf{\hat{1}}\to\mathbf{1'}\\
\mathbf{\hat{1}}\otimes\mathbf{\hat{1}'}\to\mathbf{1}\\
\mathbf{\hat{1}'}\otimes\mathbf{\hat{1}'}\to\mathbf{1'}
  \end{array}\right\} &
\mathbf{\harp{\hat{1}}'}\sim\alpha\beta\\
    &&&\\
\mathbf{\harp{\hat{1}}}^{(\prime)} \otimes \mathbf{\harp{\hat{2}}}\rightarrow
    \mathbf{\harp{\hat{2}}}
    & \left\{\begin{array}{c}
    ~\\[2ex]p=\mathrm{even}\\[10ex]p=\mathrm{odd}\\[2ex]~
  \end{array}\right. &
  \left.\begin{array}{l}
    \mathbf{1} \otimes \mathbf{2}\rightarrow\mathbf{2}\\
    \mathbf{\hat{1}}^{\prime} \otimes \mathbf{\hat{2}}\rightarrow\mathbf{2}\\
    \mathbf{1} \otimes \mathbf{\hat{2}}\rightarrow\mathbf{\hat{2}}\\
    \mathbf{\hat{1}} \otimes \mathbf{2}\rightarrow\mathbf{\hat{2}}\\[2ex]
    \mathbf{1}^{\prime} \otimes \mathbf{2}\rightarrow\mathbf{2}\\
    \mathbf{\hat{1}} \otimes \mathbf{\hat{2}}\rightarrow\mathbf{2}\\
    \mathbf{1}^{\prime} \otimes \mathbf{\hat{2}}\rightarrow\mathbf{\hat{2}}\\
    \mathbf{\hat{1}}^{\prime} \otimes \mathbf{2}\rightarrow\mathbf{\hat{2}}\\
  \end{array}\right\} &
\mathbf{\harp{\hat{2}}}\sim\alpha\begin{pmatrix}
 (-1)^{p}\beta_1 \\
 \beta_2 \\
\end{pmatrix}\\
\end{array}\nonumber
\end{equation}

\begin{equation}
\begin{array}{llll}
    \mathbf{\harp{\hat{1}}}^{(\prime)} \otimes \mathbf{\harp{\hat{3}}}^{(\prime)} \rightarrow \mathbf{\harp{\hat{3}}}^{(\prime)}
    &
 \left\{\begin{array}{c}
    ~\\[19ex]p=\mathrm{even}\\[19ex]~
 \end{array}\right.
&
 \left.\begin{array}{l}
 \mathbf{1} \otimes \mathbf{3}\rightarrow\mathbf{3}\\
 \mathbf{1}^{\prime} \otimes \mathbf{3}^{\prime}\rightarrow\mathbf{3}\\
          \mathbf{\hat{1}} \otimes \mathbf{\hat{3}}^{\prime}\rightarrow\mathbf{3}\\
          \mathbf{\hat{1}}^{\prime} \otimes \mathbf{\hat{3}}\rightarrow\mathbf{3}\\
          \mathbf{1} \otimes \mathbf{3}^{\prime}\rightarrow\mathbf{3}^{\prime}\\
          \mathbf{1}^{\prime} \otimes \mathbf{3}\rightarrow\mathbf{3}^{\prime}\\
          \mathbf{\hat{1}} \otimes \mathbf{\hat{3}}\rightarrow\mathbf{3}^{\prime}\\
         \mathbf{\hat{1}}^{\prime} \otimes \mathbf{\hat{3}}^{\prime}\rightarrow\mathbf{3}^{\prime}\\
         \mathbf{1} \otimes \mathbf{\hat{3}}\rightarrow\mathbf{\hat{3}}\\
         \mathbf{1}^{\prime} \otimes \mathbf{\hat{3}}^{\prime}\rightarrow\mathbf{\hat{3}}\\
         \mathbf{\hat{1}} \otimes \mathbf{3}\rightarrow\mathbf{\hat{3}}\\
         \mathbf{\hat{1}}^{\prime} \otimes \mathbf{3}^{\prime}\rightarrow\mathbf{\hat{3}}\\
         \mathbf{1} \otimes \mathbf{\hat{3}}^{\prime}\rightarrow\mathbf{\hat{3}}^{\prime}\\
         \mathbf{1}^{\prime} \otimes \mathbf{\hat{3}}\rightarrow\mathbf{\hat{3}}^{\prime}\\
         \mathbf{\hat{1}} \otimes \mathbf{3}^{\prime}\rightarrow\mathbf{\hat{3}}^{\prime}\\
         \mathbf{\hat{1}}^{\prime} \otimes \mathbf{3}\rightarrow\mathbf{\hat{3}}^{\prime}\\
  \end{array}\right\}
      & \mathbf{\harp{\hat{3}}}^{(\prime)}\sim\alpha\begin{pmatrix}
 \beta_1 \\
 \beta_2 \\
 \beta_3 \\
\end{pmatrix}
\end{array}\nonumber
\end{equation}
where we have introduced the notation $p$ to distinguish between different products, it makes the results more compact. The CG coefficients for the direct product involving doublet $\mathbf{\harp{\hat{2}}}$ are as follows
\begin{equation}
\begin{array}{llll}
\mathbf{\harp{\hat{2}}} \otimes \mathbf{\harp{\hat{2}}}\rightarrow
    \mathbf{\harp{\hat{1}}}\oplus\mathbf{\harp{\hat{1}}}^{\prime}\oplus\mathbf{\harp{\hat{2}}}
    & \left\{\begin{array}{c}
    ~\\[-1ex]p=\mathrm{even}\\ \\[0.5ex]p=\mathrm{odd}\\
  \end{array}\right. &
  \left.\begin{array}{l}
          \mathbf{2} \otimes \mathbf{2} \rightarrow \mathbf{1} \oplus \mathbf{1^{\prime}} \oplus \mathbf{2}\\
          \mathbf{2} \otimes \mathbf{\hat{2}} \rightarrow \mathbf{\hat{1}} \oplus \mathbf{\hat{1}}^{\prime} \oplus \mathbf{\hat{2}}\\[2ex]
          \mathbf{\hat{2}} \otimes \mathbf{\hat{2}} \rightarrow \mathbf{1} \oplus \mathbf{1^{\prime}} \oplus \mathbf{2}
  \end{array}\right\} &
\begin{array}{l}
 \mathbf{\harp{\hat{1}}}\sim \alpha_1 \beta_2+(-1)^{p}\alpha_2 \beta_1 \\
 \mathbf{\harp{\hat{1}}}^{\prime}\sim \alpha_1 \beta_2-(-1)^{p}\alpha_2 \beta_1 \\
 \mathbf{\harp{\hat{2}}}\sim\begin{pmatrix}
 \alpha_2 \beta_2 \\
 (-1)^{p}\alpha_1 \beta_1 \\
\end{pmatrix} \\
\end{array}
  \\[2ex]
    &&&\\
\mathbf{\harp{\hat{2}}} \otimes \mathbf{\harp{\hat{3}}}^{(\prime)}\rightarrow
    \mathbf{\harp{\hat{3}}}\oplus\mathbf{\harp{\hat{3}}}^{\prime}
    & \left\{\begin{array}{c}
    ~\\[2ex] p=\mathrm{even}\\[11ex]p=\mathrm{odd}\\[4ex]
  \end{array}\right. &
  \left.\begin{array}{l}
          \mathbf{2} \otimes \mathbf{3} \rightarrow \mathbf{3} \oplus \mathbf{3}^{\prime}\\
          \mathbf{\hat{2}} \otimes \mathbf{\hat{3}}^{\prime} \rightarrow \mathbf{3} \oplus \mathbf{3}^{\prime}\\
          \mathbf{2} \otimes \mathbf{\hat{3}} \rightarrow \mathbf{\hat{3}} \oplus \mathbf{\hat{3}}^{\prime}\\
          \mathbf{\hat{2}} \otimes \mathbf{3} \rightarrow \mathbf{\hat{3}} \oplus \mathbf{\hat{3}}^{\prime}\\[2ex]
          \mathbf{2} \otimes \mathbf{3}^{\prime} \rightarrow \mathbf{3} \oplus \mathbf{3}^{\prime} \\
          \mathbf{\hat{2}} \otimes \mathbf{\hat{3}} \rightarrow \mathbf{3} \oplus \mathbf{3}^{\prime}\\
          \mathbf{2} \otimes \mathbf{\hat{3}}^{\prime} \rightarrow \mathbf{\hat{3}} \oplus \mathbf{\hat{3}}^{\prime}\\
          \mathbf{\hat{2}} \otimes \mathbf{3}^{\prime} \rightarrow \mathbf{\hat{3}} \oplus \mathbf{\hat{3}}^{\prime}\\
  \end{array}\right\} &
\begin{array}{l}
 \mathbf{\harp{\hat{3}}}\sim\begin{pmatrix}
 \alpha_2 \beta_3+(-1)^{p}\alpha_1 \beta_2 \\
 \alpha_2 \beta_1+(-1)^{p}\alpha_1 \beta_3 \\
 \alpha_2 \beta_2+(-1)^{p}\alpha_1 \beta_1 \\
\end{pmatrix}  \\[5ex]
 \mathbf{\harp{\hat{3}}}^{\prime}\sim\begin{pmatrix}
 \alpha_2 \beta_3-(-1)^{p}\alpha_1 \beta_2 \\
 \alpha_2 \beta_1-(-1)^{p}\alpha_1 \beta_3 \\
 \alpha_2 \beta_2-(-1)^{p}\alpha_1 \beta_1 \\
\end{pmatrix} \\
\end{array}
\end{array}\nonumber
\end{equation}
The last case involves the direct product of $\mathbf{\harp{\hat{3}}}^{(\prime)} \otimes \mathbf{\harp{\hat{3}}}^{(\prime)}$, the CG coefficients are given as follows
{\footnotesize
\begin{equation}
\arraycolsep=2.0pt
\begin{array}{llll}
\begin{array}{l}
    \mathbf{\harp{\hat{3}}}^{(\prime)} \otimes \mathbf{\harp{\hat{3}}}^{(\prime)}\rightarrow\\
    ~~~\mathbf{\harp{\hat{1}}}^{(\prime)}\oplus\mathbf{\harp{\hat{2}}}\oplus\mathbf{\harp{\hat{3}}}\oplus\mathbf{\harp{\hat{3}}}^{\prime}
\end{array}
    & \left\{\begin{array}{c}
    ~\\[2ex] p=\mathrm{even}\\[15ex]p=\mathrm{odd}\\[6ex]
  \end{array}\right. &
  \left.\begin{array}{l}
          \mathbf{3} \otimes \mathbf{3} \rightarrow \mathbf{1} \oplus \mathbf{2} \oplus \mathbf{3} \oplus \mathbf{3}^{\prime}\\
          \mathbf{3}^{\prime} \otimes \mathbf{3}^{\prime} \rightarrow \mathbf{1} \oplus \mathbf{2} \oplus \mathbf{3} \oplus \mathbf{3}^{\prime}\\
          \mathbf{\hat{3}} \otimes \mathbf{\hat{3}}^{\prime} \rightarrow \mathbf{1} \oplus \mathbf{2} \oplus \mathbf{3} \oplus \mathbf{3}^{\prime}\\
          \mathbf{3} \otimes \mathbf{\hat{3}} \rightarrow \mathbf{\hat{1}} \oplus \mathbf{\hat{2}} \oplus \mathbf{\hat{3}} \oplus \mathbf{\hat{3}}^{\prime}\\
          \mathbf{3}^{\prime} \otimes \mathbf{\hat{3}}^{\prime} \rightarrow \mathbf{\hat{1}} \oplus \mathbf{\hat{2}} \oplus \mathbf{\hat{3}} \oplus \mathbf{\hat{3}}^{\prime}\\[2ex]
          \mathbf{3} \otimes \mathbf{3}^{\prime} \rightarrow \mathbf{1}^{\prime} \oplus \mathbf{2} \oplus \mathbf{3} \oplus \mathbf{3}^{\prime}\\
          \mathbf{\hat{3}} \otimes \mathbf{\hat{3}} \rightarrow \mathbf{1}^{\prime} \oplus \mathbf{2} \oplus \mathbf{3} \oplus \mathbf{3}^{\prime}\\
          \mathbf{\hat{3}}^{\prime} \otimes \mathbf{\hat{3}}^{\prime}\rightarrow \mathbf{1}^{\prime} \oplus \mathbf{2} \oplus \mathbf{3} \oplus \mathbf{3}^{\prime}\\
          \mathbf{3} \otimes \mathbf{\hat{3}}^{\prime} \rightarrow \mathbf{\hat{1}}^{\prime} \oplus \mathbf{\hat{2}} \oplus \mathbf{\hat{3}} \oplus \mathbf{\hat{3}}^{\prime}\\
          \mathbf{3}^{\prime} \otimes \mathbf{\hat{3}} \rightarrow \mathbf{\hat{1}}^{\prime} \oplus \mathbf{\hat{2}} \oplus \mathbf{\hat{3}} \oplus \mathbf{\hat{3}}^{\prime}\\
  \end{array}\right\} &
  \begin{array}{l}
 \mathbf{\harp{\hat{1}}}^{(\prime)}\sim \alpha_1 \beta_1+\alpha_2 \beta_3+\alpha_3 \beta_2 \\[1ex]
 \mathbf{\harp{\hat{2}}}\sim\begin{pmatrix}
 (-1)^{p}(\alpha_1 \beta_3+\alpha_2 \beta_2+\alpha_3 \beta_1) \\
 \alpha_1 \beta_2+\alpha_2 \beta_1+\alpha_3 \beta_3 \\
\end{pmatrix} \\[2.5ex]
 \mathbf{\harp{\hat{3}}}\sim\begin{pmatrix}
 \alpha_1 \beta_1-\alpha_2 \beta_3+(-1)^{p}(\alpha_3 \beta_2-\alpha_1 \beta_1) \\
 \alpha_3 \beta_3-\alpha_1 \beta_2+(-1)^{p}(\alpha_2 \beta_1-\alpha_3 \beta_3) \\
 \alpha_2 \beta_2-\alpha_3 \beta_1+(-1)^{p}(\alpha_1 \beta_3-\alpha_2 \beta_2) \\
\end{pmatrix} \\[4.5ex]
 \mathbf{\harp{\hat{3}}}^{\prime}\sim\begin{pmatrix}
 \alpha_1 \beta_1-\alpha_2 \beta_3-(-1)^{p}(\alpha_3 \beta_2-\alpha_1 \beta_1) \\
 \alpha_3 \beta_3-\alpha_1 \beta_2-(-1)^{p}(\alpha_2 \beta_1-\alpha_3 \beta_3) \\
 \alpha_2 \beta_2-\alpha_3 \beta_1-(-1)^{p}(\alpha_1 \beta_3-\alpha_2 \beta_2) \\
\end{pmatrix} \\
  \end{array}
\end{array}\nonumber
\end{equation}
}

\section{\label{sec:higher weight-app}Higher weight modular forms of level $N=4$}
 In this Appendix, we present the explicit forms of the modular forms for higher weight $k=4,5,6$.

The weight 4 modular space has dimension $2\times4+1=9$, they arrange into the irreducible presentations $\mathbf{1}$, $\mathbf{2}$, $\mathbf{3}$ and $\mathbf{3'}$ of $S'_4$,
\begin{eqnarray}
\nonumber&& Y^{(4)}_{\mathbf{1}}=\left(Y^{(3)}_{\mathbf{\hat{3}}}Y^{(1)}_{\mathbf{\hat{3}'}}\right)_{\mathbf{1}}=4(Y_1^4-2Y_1(Y_2^3+Y_3^3)+3Y_2^2Y_3^2), \\
\nonumber&& Y^{(4)}_{\mathbf{2}}=\left(Y^{(3)}_{\mathbf{\hat{3}}}Y^{(1)}_{\mathbf{\hat{3}'}}\right)_{\mathbf{2}}=\begin{pmatrix}-2Y_3^4+4Y_1^3Y_3+4Y_2^3 Y_3-6Y_1^2Y_2^2 \\ -2Y_2^4+4Y_1^3Y_2+4Y_2Y_3^3-6Y_1^2Y_3^2\end{pmatrix}, \\
\nonumber&& \\[+0.05in]
\nonumber&& Y^{(4)}_{\mathbf{3}}=\left(Y^{(3)}_{\mathbf{\hat{3}}}Y^{(1)}_{\mathbf{\hat{3}'}}\right)_{\mathbf{3}}=\begin{pmatrix}6Y_1(-Y_2^3+Y_3^3) \\
6Y_1Y_3(Y_2^2-Y_1Y_3)+2Y_2(-2Y_1^3+Y_2^3+Y_3^3) \\
6Y_1Y_2(Y_1Y_2-Y_3^2)-2Y_3(-2Y_1^3+Y_2^3+Y_3^3)\end{pmatrix} ,\\
\nonumber&& \\[+0.05in]
&& Y^{(4)}_{\mathbf{3'}}=\left(Y^{(3)}_{\mathbf{\hat{3}}}Y^{(1)}_{\mathbf{\hat{3}'}}\right)_{\mathbf{3'}}=\begin{pmatrix}2(4Y_1^4-6Y_2^2Y_3^2+Y_1(Y_2^3+Y_3^3))\\ 2(Y_2^4-2Y_1^3Y_2 +7Y_2Y_3^3+3Y_1^2Y_3^2-9Y_1Y_2^2 Y_3) \\
2(Y_3^4-2Y_1^3Y_3+7Y_2^3Y_3+3Y_1^2Y_2^2-9Y_1Y_2Y_3^2)\end{pmatrix}.
\end{eqnarray}
The weight 5 modular forms of level 4 decompose as $\mathbf{\hat{2}}\oplus\mathbf{\hat{3}}\oplus\mathbf{\hat{3}'}\oplus\mathbf{\hat{3}'}$ under $S'_4$, and they are given by
\begin{eqnarray}
\nonumber&& Y^{(5)}_{\mathbf{\hat{2}}}=\left(Y^{(4)}_{\mathbf{3'}}Y^{(1)}_{\mathbf{\hat{3}'}}\right)_{\mathbf{\hat{2}}}=\begin{pmatrix}
2(Y_2^5+2Y_1^4Y_3+2Y_1Y_3^4+Y_2^2Y_3^3+Y_1^3Y_2^2-Y_1Y_2^3Y_3-6Y_1^2Y_2Y_3^2)\\ 2(Y_3^5+2Y_1^4Y_2+2Y_1Y_2^4+Y_1^3Y_3^2+Y_2^3Y_3^2-Y_1Y_2Y_3^3-6Y_1^2Y_2^2Y_3)
\end{pmatrix}, \\
\nonumber&& Y^{(5)}_{\mathbf{\hat{3}}}=\left(Y^{(4)}_{\mathbf{3}}Y^{(1)}_{\mathbf{\hat{3}'}}\right)_{\mathbf{\hat{3}}}=\begin{pmatrix}18 Y_1^2(-Y_2^3 + Y_3^3) \\ 4 Y_1^4 Y_2 + 4 Y_1(Y_2^4-5Y_2Y_3^3)+ 14 Y_1^3 Y_3^2 - 4 Y_3^2 (Y_2^3 + Y_3^3)+ 6 Y_1^2 Y_2^2 Y_3 \\ -4Y_1^4Y_3-4Y_1(Y_3^4-5Y_2^3Y_3)-14Y_1^3Y_2^2+4Y_2^2(Y_2^3+Y_3^3)-6Y_1^2Y_2Y_3^2\end{pmatrix}, \\
\nonumber&& \\[+0.05in]
\nonumber&& Y^{(5)}_{\mathbf{\hat{3}'},I}=\left(Y^{(4)}_{\mathbf{2}}Y^{(1)}_{\mathbf{\hat{3}'}}\right)_{\mathbf{\hat{3}'}}=\begin{pmatrix}8Y_1^3Y_2 Y_3-6Y_1^2(Y_2^3+Y_3^3)+2Y_2Y_3(Y_2^3+Y_3^3)\\
4Y_1^4Y_2-2Y_1Y_2^4-6Y_1^2Y_2^2Y_3 - 2Y_1^3 Y_3^2+4Y_2^3Y_3^2+4Y_1Y_2Y_3^3-2 Y_3^5\\ -2(Y_1^3Y_2^2+Y_2^5-2Y_1^4Y_3+3Y_1^2 Y_2 Y_3^2 -2Y_2^2Y_3^3+Y_1(-2Y_2^3 Y_3 + Y_3^4))\end{pmatrix} ,\\
\nonumber&& \\[+0.05in]
&& Y^{(5)}_{\mathbf{\hat{3}'},II}=\left(Y^{(4)}_{\mathbf{1}}Y^{(1)}_{\mathbf{\hat{3}'}}\right)_{\mathbf{\hat{3}'}}=\begin{pmatrix}4Y_1(Y_1^4+3Y_2^2Y_3^2-2 Y_1(Y_2^3+Y_3^3))\\ 4Y_2(Y_1^4 + 3Y_2^2 Y_3^2-2Y_1(Y_2^3+Y_3^3))\\ 4Y_3(Y_1^4+3 Y_2^2 Y_3^2 - 2 Y_1 (Y_2^3 + Y_3^3))\end{pmatrix}\,,
\end{eqnarray}
where $Y^{(5)}_{\mathbf{\hat{3}'},I}$ and $Y^{(5)}_{\mathbf{\hat{3}'},II}$ denote two weight 5 modular forms transforming as triplet $\mathbf{\hat{3}'}$ of $S'_4$, and they can also taken to be any two linearly independent combinations of $Y^{(5)}_{\mathbf{\hat{3}'},I}$ and $Y^{(5)}_{\mathbf{\hat{3}'},II}$. Finally there are 13 independent weight 6 modular forms of level 5, and they can be arranged into the following irreducible representations of $S'_4$,
\begin{small}
\begin{eqnarray}
\nonumber&& Y^{(6)}_{\mathbf{1'}}=\left(Y^{(5)}_{\mathbf{\hat{3}'},I}Y^{(1)}_{\mathbf{\hat{3}'}}\right)_{\mathbf{1'}} \\
\nonumber&&~~~~~~=-2(Y_2^6+Y_3^6-8Y_1^4Y_2Y_3+6Y_1^2Y_2^2Y_3^2-4Y_2^3Y_3^3+4Y_1^3(Y_2^3+Y_3^3)-2Y_1Y_2Y_3(Y_2^3+Y_3^3))\,,\\
\nonumber&& Y^{(6)}_{\mathbf{1}}=\left(Y^{(5)}_{\mathbf{\hat{3}}}Y^{(1)}_{\mathbf{\hat{3}'}}\right)_{\mathbf{1}}=4(Y_2^3-Y_3^3)(-8Y_1^3+Y_2^3+Y_3^3+6Y_1Y_2Y_3)\,,\\
\nonumber&& \\[+0.05in]
\nonumber&& Y^{(6)}_{\mathbf{2}}=\left(Y^{(5)}_{\mathbf{\hat{3}'},II}Y^{(1)}_{\mathbf{\hat{3}'}}\right)_{\mathbf{2}}=\begin{pmatrix}-4(Y_2^2+2Y_1Y_3)(Y_1^4+3 Y_2^2Y_3^2 -2Y_1(Y_2^3+Y_3^3))\\ 4(Y_3^2+2Y_1Y_2)(Y_1^4+3Y_2^2Y_3^2- 2Y_1(Y_2^3+ Y_3^3))\end{pmatrix} ,\\
\nonumber&& \\[+0.05in]
\nonumber&& Y^{(6)}_{\mathbf{3},I}=\left(Y^{(5)}_{\mathbf{\hat{3}'},I}Y^{(1)}_{\mathbf{\hat{3}'}}\right)_{\mathbf{3}} \\
\nonumber&& =\begin{pmatrix}2(Y_2^6 + Y_3^6 + 4 Y_1^4Y_2Y_3 + 6Y_1^2Y_2^2Y_3^2 - 4 Y_2^3 Y_3^3 - 5Y_1^3 (Y_2^3 + Y_3^3) + Y_1Y_2Y_3(Y_2^3 + Y_3^3))\\-2(2Y_1^5Y_2 - 5Y_1^4Y_3^2+ 3 Y_1^3Y_2^2Y_3 + 3Y_2^2Y_3(Y_2^3 - Y_3^3) + Y_1^2 (5Y_2Y_3^3-4Y_2^4) + Y_1 (Y_3^5-2Y_2^3 Y_3^2))\\-2(2Y_1^5Y_3-5Y_1^4Y_2^2+3Y_1^3Y_2Y_3^2+ 3 Y_2Y_3^2(Y_3^3-Y_2^3) + Y_1(Y_2^5-2Y_2^2Y_3^3) + Y_1^2 (5Y_2^3 Y_3 - 4Y_3^4))\end{pmatrix}\,, \\
\nonumber&& \\[+0.05in]
\nonumber&& Y^{(6)}_{\mathbf{3},II}=\left(Y^{(5)}_{\mathbf{\hat{3}'},II}Y^{(1)}_{\mathbf{\hat{3}'}}\right)_{\mathbf{3}}=\begin{pmatrix}8(Y_1^2-Y_2Y_3)(Y_1^4-2Y_1(Y_2^3+Y_3^3)+3Y_2^2 Y_3^2)\\ 8(Y_3^2-Y_1Y_2)(Y_1^4-2Y_1(Y_2^3+Y_3^3)+3Y_2^2Y_3^2)\\ 8(Y_2^2-Y_1Y_3)(Y_1^4-2Y_1(Y_2^3+Y_3^3)+3Y_2^2Y_3^2)\end{pmatrix} ,\\
\nonumber&& \\[+0.05in]
\nonumber&& Y^{(6)}_{\mathbf{3'}}=\left(Y^{(5)}_{\mathbf{\hat{3}'},I}Y^{(1)}_{\mathbf{\hat{3}'}}\right)_{\mathbf{3'}} \\
&& =\begin{pmatrix}-2(Y_2^3-Y_3^3)(Y_1^3+Y_2^3+Y_3^3-3Y_1Y_2Y_3)\\ 2(2Y_1^5 Y_2 - 7Y_1^3Y_2^2Y_3 - Y_1^4Y_3^2 - Y_2^2Y_3(Y_2^3 + Y_3^3) + Y_1^2 (2Y_2^4 + 5Y_2 Y_3^3) + Y_1(2Y_2^3Y_3^2 - Y_3^5))\\2(Y_1^4Y_2^2 + 7Y_1^3Y_2Y_3^2 - 2Y_1^5Y_3 + Y_2 Y_3^2(Y_2^3 + Y_3^3) - Y_1^2(5Y_2^3Y_3 + 2Y_3^4)+ Y_1(Y_2^5 - 2Y_2^2Y_3^3))\end{pmatrix}.
\end{eqnarray}
\end{small}
We note that the results of even weight modular forms obtained here coincide with those of our previous work~\cite{Gui-JunDing:2019wap} up to some overall constants. Specifically the following relations are fulfilled:
\begin{align}
\nonumber
&Y^{(2)}_\mathbf{2}=\frac{3}{2}(3i+\sqrt{3})\widetilde{Y}^{(2)}_\mathbf{2},~~ Y^{(2)}_\mathbf{3}=(3-3i\sqrt{3})\widetilde{Y}^{(2)}_\mathbf{3}\,,~~Y^{(4)}_\mathbf{1}=-27e^{3\pi i/4}(i+\sqrt{3})\widetilde{Y}^{(4)}_\mathbf{1},\\
\nonumber& Y^{(4)}_\mathbf{2}=27e^{3\pi i/4}(i+\sqrt{3})\widetilde{Y}^{(4)}_\mathbf{2},~~ Y^{(4)}_\mathbf{3}=18\sqrt{3}e^{\pi i/4}\widetilde{Y}^{(4)}_\mathbf{3},~~ Y^{(4)}_\mathbf{3'}=18\sqrt{3}e^{\pi i/4}\widetilde{Y}^{(4)}_\mathbf{3'}\,,\\
\nonumber
&Y^{(6)}_\mathbf{1}=-162\sqrt{6}(1+i)\widetilde{Y}^{(6)}_\mathbf{1},~~ Y^{(6)}_\mathbf{1'}=-81\sqrt{6}(1+i)\widetilde{Y}^{(6)}_\mathbf{1'},~~ Y^{(6)}_\mathbf{2}=81\sqrt{6}(1+i)\widetilde{Y}^{(6)}_\mathbf{2},\\
& Y^{(6)}_{\mathbf{3},I}=-162e^{7\pi i/12}\widetilde{Y}^{(6)}_{\mathbf{3},II},~~Y^{(6)}_{\mathbf{3},II}=-324e^{7\pi i/12}\widetilde{Y}^{(6)}_{\mathbf{3},I},~~Y^{(6)}_{\mathbf{3}'}=-162e^{7\pi i/12}\widetilde{Y}^{(6)}_{\mathbf{3}'}\,,
\end{align}
where the modular forms in~\cite{Gui-JunDing:2019wap} are denoted with a symbol ``$~\widetilde{}~$''.

\section{\label{sec:lepton sector-app}
Lepton models for triplet and doublet plus singlet assignment of right-handed charged leptons }
In order to accommodate the hierarchical charged lepton masses, the right-handed charged leptons are usually assumed to transform as singlets under modular flavor symmetry, and this type of assignment in the setup of $S'_4$ modular symmetry has been studied in section~\ref{sec:lepton-models}. However, the singlet assignment is not unique, they can also be assigned to a triplet or the direct sum of a doublet and a singlet under $S'_4$. If the three right-handed charged leptons are embedded into a triplet of $S'_4$, the
most general superpotential for the charged lepton masses is of the form
\begin{equation}
\mathcal{W}_e=\alpha (E^c L f_E(Y))_\mathbf{1}H_d\,,
\end{equation}
where the triplet assignment for $L$ is preserved, $f_E(Y)$ is a modular multiplet, and all possible modular invariant terms should be included. In the same way as we have done in section~\ref{sec:lepton-models}, the possible charged lepton models can be straightforwardly searched for. For illustration, we consider the modular forms with weight less than four and higher weight modular forms can be studied analogously. The predictions for the charged lepton mass matrix are summarized in table~\ref{tab:charged lepton_2}.

\begin{table}[th!]
\renewcommand{\tabcolsep}{0.58mm}
\centering
\resizebox{1.05\textwidth}{!}{
\begin{tabular}{|c|c|c|c|} \hline\hline
\multirow{2}{*}{\texttt{Cases}} ~&~ \texttt{rep} ~&~ \texttt{weights} &\multirow{2}{*}{\texttt{Charged lepton mass matrix}}\\
 ~&~ $(\rho_L,\rho_{E^c})$ ~&~ $k_L+k_{E^c}$ &  \\ \hline
 & & & \\[-0.15in]
$C_5$ & $\begin{cases}
(\mathbf{3},\mathbf{\hat{3}}) \\
(\mathbf{3'},\mathbf{\hat{3}'})  \\
(\mathbf{\hat{3}},\mathbf{3})  \\
(\mathbf{\hat{3}'},\mathbf{3'}) \end{cases}$
& $1$ &~~ $ M_e =\alpha \begin{pmatrix}
0 ~&~ -Y_3 ~&~ Y_2 \\
Y_3 ~&~ 0  ~&~ -Y_1 \\
 -Y_2 ~&~ Y_1 ~&~ 0 \\
 \end{pmatrix} v_d $~~ \\ \hline
 & & & \\[-0.15in]
 $C_6$ & $\begin{cases}
(\mathbf{3},\mathbf{\hat{3}'}) \\
(\mathbf{3'},\mathbf{\hat{3}})  \\
(\mathbf{\hat{3}'},\mathbf{3})  \\
(\mathbf{\hat{3}},\mathbf{3'}) \end{cases}$
& $1$ &~~ $ M_e =\alpha \begin{pmatrix}
2Y_1 ~&~ -Y_3 ~&~ -Y_2 \\
-Y_3 ~&~ 2Y_2  ~&~ -Y_1 \\
 -Y_2 ~&~ -Y_1 ~&~ 2Y_3 \\
 \end{pmatrix} v_d $~~ \\ \hline
 & & & \\[-0.15in]
$C_7$ & $\begin{cases}
(\mathbf{3},\mathbf{3}) \\
(\mathbf{3'},\mathbf{3'})  \\
(\mathbf{\hat{3}},\mathbf{\hat{3}'})  \\
(\mathbf{\hat{3}'},\mathbf{\hat{3}}) \end{cases}$
& $2$ &~~ $ M_e =\begin{pmatrix}
 0 ~&~ \alpha_1 Y^{(2)}_1- \alpha_2 Y^{(2)}_5 ~&~ \alpha_1 Y^{(2)}_2+ \alpha_2 Y^{(2)}_4 \\
\alpha_1 Y^{(2)}_1+ \alpha_2 Y^{(2)}_5 ~&~ \alpha_1 Y^{(2)}_2  ~&~
-\alpha_2 Y^{(2)}_3 \\
 \alpha_1 Y^{(2)}_2- \alpha_2 Y^{(2)}_4~&~ \alpha_2 Y^{(2)}_3  ~&~
 \alpha_1 Y^{(2)}_1 \\
 \end{pmatrix} v_d $~~ \\ \hline
 & & & \\[-0.15in]
$C_8$ & $\begin{cases}
(\mathbf{\hat{3}},\mathbf{\hat{3}}) \\
(\mathbf{\hat{3}'},\mathbf{\hat{3}'})  \\
(\mathbf{3},\mathbf{3'})  \\
(\mathbf{3'},\mathbf{3}) \end{cases}$
& $2$ &~~ $ M_e =\begin{pmatrix}
 2\alpha_2 Y^{(2)}_3 ~&~ \alpha_1 Y^{(2)}_1- \alpha_2 Y^{(2)}_5 ~&~
 -\alpha_1 Y^{(2)}_2 - \alpha_2 Y^{(2)}_4 \\
\alpha_1 Y^{(2)}_1 - \alpha_2 Y^{(2)}_5 ~&~ -\alpha_1 Y^{(2)}_2+2\alpha_2 Y^{(2)}_4  ~&~  -\alpha_2 Y^{(2)}_3 \\
 -\alpha_1 Y^{(2)}_2- \alpha_2 Y^{(2)}_4~&~ -\alpha_2 Y^{(2)}_3  ~&~
 \alpha_1 Y^{(2)}_1 + 2\alpha_2 Y^{(2)}_5\\
 \end{pmatrix} v_d $~~ \\ \hline
$C_9$ & $\begin{cases}
(\mathbf{3},\mathbf{\hat{3}}) \\
(\mathbf{3'},\mathbf{\hat{3}'})  \\
(\mathbf{\hat{3}},\mathbf{3})  \\
(\mathbf{\hat{3}'},\mathbf{3'}) \end{cases}$
& $3$ &~~ $ M_e =\begin{pmatrix}
 2\alpha_2 Y^{(3)}_2 + \alpha_1 Y^{(3)}_1 ~& -\alpha_3 Y^{(3)}_7-\alpha_2 Y^{(3)}_4 ~& \alpha_3 Y^{(3)}_6-\alpha_2 Y^{(3)}_3 \\
 \alpha_3 Y^{(3)}_7-\alpha_2 Y^{(3)}_4 ~& 2\alpha_2 Y^{(3)}_3 ~& -\alpha_3 Y^{(3)}_5-\alpha_2 Y^{(3)}_2 + \alpha_1 Y^{(3)}_1 \\
 -\alpha_3 Y^{(3)}_6-\alpha_2 Y^{(3)}_3 ~& \alpha_3 Y^{(3)}_5-\alpha_2 Y^{(3)}_2 + \alpha_1 Y^{(3)}_1 ~& 2\alpha_2 Y^{(3)}_4\\
 \end{pmatrix} v_d $~~ \\ \hline
 $C_{10}$ & $\begin{cases}
(\mathbf{3},\mathbf{\hat{3}'}) \\
(\mathbf{3'},\mathbf{\hat{3}})  \\
(\mathbf{\hat{3}'},\mathbf{3})  \\
(\mathbf{\hat{3}},\mathbf{3'}) \end{cases}$
& $3$ &~~ $ M_e =\begin{pmatrix}
 2\alpha_1 Y^{(3)}_5 & -\alpha_1 Y^{(3)}_7-\alpha_2 Y^{(3)}_4 & -\alpha_1 Y^{(3)}_6+\alpha_2 Y^{(3)}_3 \\
 -\alpha_1 Y^{(3)}_7+\alpha_2 Y^{(3)}_4 & 2\alpha_1 Y^{(3)}_6 & -\alpha_1 Y^{(3)}_5-\alpha_2 Y^{(3)}_2 \\
 -\alpha_1 Y^{(3)}_6-\alpha_2 Y^{(3)}_3 & -\alpha_1 Y^{(3)}_5+\alpha_2 Y^{(3)}_2 & 2\alpha_1 Y^{(3)}_7\\
 \end{pmatrix} v_d $~~ \\ \hline
\end{tabular} }
\caption{\label{tab:charged lepton_2}
The possible forms of the charged lepton mass matrix for the case that both $L$ and $E^c$ transform as triplets of $S'_4$, where integral weight and level 4 modular forms of weight less than 3 are considered.   }
\end{table}
If the right-handed charged leptons are assigned to the direct sum of a doublet and a singlet of $S'_4$, the most general superpotential for the charged lepton masses is given by
\begin{equation}
\mathcal{W}_e=\alpha (E_D^c L f_{E_D}(Y))_\mathbf{1}H_d + \alpha (E_3^c L f_{E_3}(Y))_\mathbf{1}H_d
\end{equation}
where $E^c_D=(E^c_1,\,E^c_2)^T$, and the modular multiplets $f_{E_D}(Y)$ and  $f_{E_3}(Y)$ have to be triplet of $S'_4$. As a consequence, the charged lepton mass matrices can be divided into two blocks as follow
\begin{equation}
\label{eq:doublet+singlet}M_e=\left(\begin{array}{c}
  \\[-0.1in]
 ~~~T~~~ \\
  \\[-0.15in]
\cdots \cdots \cdots  \\
 ~~~D~~~ \\
\end{array}\right)v_{d}\,,
\end{equation}
where $T$ and $D$ are $2\times 3$ and $1\times 3$ sub-matrices respectively.
Likewise we focus on modular forms of weight less than four, the charged lepton mass matrix can take 24 possible forms denoted as $T_i$-$D_j$ with $i=1,\ldots,6$ and $j=1,\ldots,4$, as shown in table~\ref{tab:charged-lepton-3}.
\begin{table}[th!]
\renewcommand{\tabcolsep}{0.58mm}
\centering
\resizebox{1.05\textwidth}{!}{
\begin{tabular}{|c|c|c|c|} \hline\hline
\multirow{2}{*}{\texttt{Cases}} ~&~ \texttt{rep} ~&~ \texttt{weights} &\multirow{2}{*}{\texttt{First two row of charged lepton mass matrix}}\\
 ~&~ $(\rho_L,\rho_{E_D^c})$ ~&~ $k_L+k_{E_D^c}$ &  \\ \hline
 & & & \\[-0.15in]
$T_1$ & $\begin{cases}
(\mathbf{\hat{3}},\mathbf{2}) \\
(\mathbf{3},\mathbf{\hat{2}})  \\ \end{cases}$
& $1$ &~~ $ T=\alpha \begin{pmatrix}
Y_2 ~&~ Y_1 ~&~ Y_3 \\
Y_3 ~&~ Y_2 ~&~ Y_1 \\
 \end{pmatrix}  $~~ \\ \hline
 & & & \\[-0.15in]
$T_2$ & $\begin{cases}
(\mathbf{\hat{3}'},\mathbf{2}) \\
(\mathbf{3'},\mathbf{\hat{2}})  \\ \end{cases}$
& $1$ &~~ $ T=\alpha \begin{pmatrix}
-Y_2 ~&~ -Y_1 ~&~ -Y_3 \\
Y_3 ~&~ Y_2 ~&~ Y_1 \\
 \end{pmatrix} $~~ \\ \hline
 & & & \\[-0.15in]
$T_3$ & $\begin{cases}
(\mathbf{3},\mathbf{2}) \\
(\mathbf{\hat{3}'},\mathbf{\hat{2}})  \\ \end{cases}$
& $2$ &~~ $T=\alpha \begin{pmatrix}
Y^{(2)}_4 ~&~ Y^{(2)}_3 ~&~ Y^{(2)}_5 \\
Y^{(2)}_5 ~&~ Y^{(2)}_4 ~&~ Y^{(2)}_3 \\
 \end{pmatrix} $~~ \\ \hline
 & & & \\[-0.15in]
$T_4$ & $\begin{cases}
(\mathbf{3'},\mathbf{2}) \\
(\mathbf{\hat{3}},\mathbf{\hat{2}})  \\ \end{cases}$
& $2$ &~~ $ T=\alpha \begin{pmatrix}
-Y^{(2)}_4 ~&~ -Y^{(2)}_3 ~&~ -Y^{(2)}_5 \\
Y^{(2)}_5 ~&~ Y^{(2)}_4 ~&~ Y^{(2)}_3 \\
 \end{pmatrix}  $~~ \\ \hline
$T_5$ & $\begin{cases}
(\mathbf{\hat{3}},\mathbf{2}) \\
(\mathbf{3},\mathbf{\hat{2}})  \\ \end{cases}$
& $3$ &~~ $ T= \begin{pmatrix}
\alpha_2 Y^{(3)}_6 - \alpha_1 Y^{(3)}_3 ~&~ \alpha_2 Y^{(3)}_5 - \alpha_1 Y^{(3)}_2 ~&~ \alpha_2 Y^{(3)}_7 - \alpha_1 Y^{(3)}_4 \\
\alpha_2 Y^{(3)}_7 + \alpha_1 Y^{(3)}_4 ~&~ \alpha_2 Y^{(3)}_6 + \alpha_1 Y^{(3)}_3 ~&~ \alpha_2 Y^{(3)}_5 + \alpha_1 Y^{(3)}_2 \\
 \end{pmatrix}  $~~ \\ \hline
$T_6$ & $\begin{cases}
(\mathbf{\hat{3}'},\mathbf{2}) \\
(\mathbf{3'},\mathbf{\hat{2}})  \\ \end{cases}$
& $3$ &~~ $  T=\begin{pmatrix}
-\alpha_2 Y^{(3)}_6 + \alpha_1 Y^{(3)}_3 ~&~ -\alpha_2 Y^{(3)}_5 + \alpha_1 Y^{(3)}_2 ~&~ -\alpha_2 Y^{(3)}_7 + \alpha_1 Y^{(3)}_4 \\
\alpha_2 Y^{(3)}_7 + \alpha_1 Y^{(3)}_4 ~&~ \alpha_2 Y^{(3)}_6 + \alpha_1 Y^{(3)}_3 ~&~ \alpha_2 Y^{(3)}_5 + \alpha_1 Y^{(3)}_2 \\
 \end{pmatrix} $~~ \\ \hline \hline

\multirow{2}{*}{\texttt{Cases}} ~&~ \texttt{rep} ~&~ \texttt{weights} &\multirow{2}{*}{\texttt{Third row of charged lepton mass matrix}}\\
 ~&~ $(\rho_L,\rho_{E_3^c})$ ~&~ $k_L+k_{E_3^c}$ &  \\ \hline
 & & & \\[-0.15in]
$D_1$ & $\begin{cases}
(\mathbf{3},\mathbf{\hat{1}}) \\
(\mathbf{3'},\mathbf{\hat{1}'})  \\
(\mathbf{\hat{3}},\mathbf{1})  \\
(\mathbf{\hat{3}'},\mathbf{1'})  \\ \end{cases}$
& $1$ &~~ $ D= \beta \begin{pmatrix}
Y_1 ~&~ Y_3 ~&~ Y_2 \\
 \end{pmatrix}$~~ \\ \hline
 & & & \\[-0.15in]
$D_2$ & $\begin{cases}
(\mathbf{3},\mathbf{1}) \\
(\mathbf{3'},\mathbf{1'})  \\
(\mathbf{\hat{3}},\mathbf{\hat{1}'})  \\
(\mathbf{\hat{3}'},\mathbf{\hat{1}})  \\ \end{cases}$
& $2$ &~~ $ D=\beta \begin{pmatrix}
Y^{(2)}_3 ~&~ Y^{(2)}_5 ~&~ Y^{(2)}_4 \\
 \end{pmatrix}$~~ \\ \hline
 & & & \\[-0.15in]
$D_3$ & $\begin{cases}
(\mathbf{3},\mathbf{\hat{1}'}) \\
(\mathbf{3'},\mathbf{\hat{1}})  \\
(\mathbf{\hat{3}},\mathbf{1'})  \\
(\mathbf{\hat{3}'},\mathbf{1})  \\ \end{cases}$
& $3$ &~~ $ D=\beta \begin{pmatrix}
Y^{(3)}_2 ~&~ Y^{(3)}_4 ~&~ Y^{(3)}_3 \\
 \end{pmatrix}$~~ \\ \hline
 & & & \\[-0.15in]
$D_4$ & $\begin{cases}
(\mathbf{3},\mathbf{\hat{1}}) \\
(\mathbf{3'},\mathbf{\hat{1}'})  \\
(\mathbf{\hat{3}},\mathbf{1})  \\
(\mathbf{\hat{3}'},\mathbf{1'})  \\ \end{cases}$
& $3$ &~~ $ D=\beta \begin{pmatrix}
Y^{(3)}_5 ~&~ Y^{(3)}_7 ~&~ Y^{(3)}_6 \\
 \end{pmatrix}$~~ \\ \hline \hline
\end{tabular} }
\caption{\label{tab:charged-lepton-3} The possible forms of the charged lepton mass matrix for the doublet plus singlet assignment of the right-handed charged leptons, where $T$ and $D$ defined in Eq.~\eqref{eq:doublet+singlet} denote the first two rows and the last row of the charged lepton mass matrix respectively. Hence the charged lepton mass matrix can take 24 possible forms denoted as $T_1$-$D_1$,$\dots$, $T_6$-$D_4$. }
\end{table}

Combining the possible new constructions $C_{5,6,7,8,9,10}$ and $T_i$-$D_j$ in the charged lepton sector with the 18 possible neutrino models summarized in table~\ref{tab:neutrino}, we can obtain totally $540$ lepton models. The transformation properties of matter fields under $S'_4$ and their modular weights can be straightforwardly read out. We numerically scan the parameter space of each model to estimate whether the measured values of lepton masses and mixing angles can be accommodated. The generalized CP symmetry is included to make the model more predictive, all couplings constants are enforced to be real. In the following, we present 5 typical models compatible with the current experimental data. These models contain only 7 real free parameters including $\texttt{Re}(\tau)$ and $\texttt{Im}(\tau)$ at low energy, the best fit values of the input parameters and the corresponding predictions for neutrino masses and mixing parameters are listed in table~\ref{tab:goodmodel2_para_obs}. Note that the correct charged lepton masses can be reproduced and consequently they are not shown in this table.

\begin{table}[ht!]
\centering
\renewcommand{\arraystretch}{1.1}
\begin{tabular}{c}
\resizebox{1.0\textwidth}{!}{
\begin{tabular}{|c|c|c|c|c|c|c|c|c|c|c|c|}  \hline\hline
 Models  & \multicolumn{7}{c|}{ Best fit values of the input parameters for NO} &\multirow{2}{*}{$\chi^2_{\mathrm{min}}$}  \\ \cline{2-8}
with gCP & $\texttt{Re}\langle \tau \rangle$ &$\texttt{Im}\langle \tau \rangle$  &$\alpha_2/\alpha_1$ &$\alpha_3/\alpha_1(\gamma/\alpha)$  &$g_2/g_1$ ($\Lambda_2/\Lambda_1$) & $\alpha_1 v_d$/MeV  & $\dfrac{g^2_1v_u^2}{\Lambda}/$meV & \\ \hline
 $C_{9}$-$S_{5}$ & 0.1018 & 1.0158 & 0.4996 & $-1.3198$ & $-0.0117$ & 10.3363 & 0.3496 & 5.9620 \\ \hline
$C_{9}$-$S_{14}$  & 0.0001 & 1.2710 & 0.5007 & $-1.6978$ & $-2.6304$ & 12.1858 & 0.0414 & 10.6475 \\ \hline
$T_5$-$D_3$-$S_{2}$ & 0.0262 & 1.6776 & $-1.1567$ & 30.7289 & $-1.1794$
& 0.9848 & 0.4933 & 12.2994 \\ \hline
$T_5$-$D_3$-$S_{5}$ & $-0.1996$ & 0.9969 & $-1.0008$ & 0.1076 &
$-0.0066$ & 13.3618 & 0.3712 & 5.6448 \\ \hline
$T_5$-$D_4$-$S_{16}$ & 0.2728 & 1.0610 & 1.0111 & 34.3734 & 0.3230
& 0.9356 & 12.0795 & 1.7592 \\ \hline \hline
\end{tabular}}\\
  \resizebox{1.0\textwidth}{!}{
  \begin{tabular}{|c|c|c|c|c|c|c|c|c|c|c|c|}
 Models & \multicolumn{10}{c|}{Predictions for mixing parameters and neutrino masses at best fit point}  \\ \cline{2-11}
with gCP  & $\sin^2\theta_{12}$ &$\sin^2\theta_{13}$ &$\sin^2\theta_{23}$&$\delta^l_{CP}/\pi$ &$\alpha_{21}/\pi$  &$\alpha_{31}/\pi$ & $m_1$/meV & $m_2$/meV & $m_3$/meV & $|m_{ee}|$/meV  \\ \hline
$C_{9}$-$S_{5}$ & 0.3108 & 0.02236 & 0.5046 & 1.6383 & 0.3533 \
& 1.2565 & 12.2470 & 14.9629 & 51.7726 & 12.1087 \\ \hline
$C_{9}$-$S_{14}$ & 0.3322 & 0.02266 & 0.4972 & 0.9997 & 0.9997 & 1.9997 &
19.3821 & 21.2030 & 53.8872 & 6.9868 \\ \hline
$T_5$-$D_3$-$S_{2}$ & 0.3099 & 0.02246 & 0.4789 & 1.4599 & 1.8344 & \
0.8841 & 51.3018 & 52.0171 & 71.8306 & 50.5241 \\ \hline
$T_5$-$D_3$-$S_{5}$ & 0.3105 & 0.02239 & 0.5060 & 1.4595 & 1.9142 \
& 0.9431 & 14.6563 & 16.9914 & 52.3797 & 16.0717 \\ \hline
$T_5$-$D_4$-$S_{16}$& 0.3039 & 0.02180 & 0.5637 & 1.8261 & \
0.6232 & 0.6580 & 11.8968 & 14.6777 & 51.0478 & 6.6853 \\ \hline\hline
\end{tabular}}
\end{tabular}
\caption{\label{tab:goodmodel2_para_obs}
Fit results of the models in which the right-handed charged leptons are assigned to transform as a triplet or the direct sum of a doublet and a singlet of $S'_4$, and the generalized CP symmetry is imposed on the models. Notice in the CP dual point $\tau\rightarrow-\tau^{*}$, the signs of Dirac and Majorana CP phases are reversed while the predictions for lepton mixing angles and neutrino masses are unchanged. }
\end{table}

\end{appendix}

\clearpage
\newpage

\providecommand{\href}[2]{#2}\begingroup\raggedright\endgroup

\end{document}